\documentclass[fleqn,usenatbib]{mnras}

\usepackage{newtxtext,newtxmath}
\usepackage{array}
\usepackage{afterpage}

\usepackage[T1]{fontenc}

\DeclareRobustCommand{\VAN}[3]{#2}
\let\VANthebibliography\thebibliography
\def\thebibliography{\DeclareRobustCommand{\VAN}[3]{##3}\VANthebibliography}

\usepackage{graphicx}	% Including figure files
\usepackage{amsmath}	% Advanced maths commands
\newcommand\cm{{\rm\thinspace cm}}

\newcommand\erg{{\rm\thinspace erg}}

\newcommand\K{{\rm\thinspace K}}
\newcommand\keV{{\rm\thinspace keV}}
\newcommand\km{{\rm\thinspace km}}
\newcommand\kpc{{\rm\thinspace kpc}}

\newcommand\Mpc{{\rm\thinspace Mpc}}
\newcommand\Msun{\hbox{$\rm\thinspace M_{\odot}$}}

\newcommand\s{{\rm\thinspace s}}
\newcommand\yr{{\rm\thinspace yr}}
\newcommand\Gyr{{\rm\thinspace Gyr}}

\newcommand\cmsq{\hbox{$\cm^2\,$}}

\newcommand\pcmcuK{\hbox{$\cm^{-3}\K\,$}}

\newcommand\ergps{\hbox{$\erg\s^{-1}\,$}}

\newcommand\kmps{\hbox{$\km\s^{-1}\,$}}
\newcommand\kmpspmpc{\hbox{$\km\,\s^{-1}\Mpc^{-1}\,$}}

\newcommand\Msunpyr{\hbox{$\Msun\yr^{-1}\,$}}

\newcommand\pcmsq{\hbox{$\cm^{-2}\,$}}

\newcommand\psqcm{\hbox{$\cm^{-2}\,$}}

\title[Hidden Cooling Flows III]{Hidden Cooling Flows in  Clusters of Galaxies III: Accretion onto the Central Black Hole}

\author[A. C. Fabian et al.]{
A. C. Fabian,$^{1}$\thanks{E-mail: acf@ast.cam.ac.uk },  J.S. Sanders$^{2}$, G.J. Ferland$^{3}$, B.R. McNamara$^{4}$, C. Pinto$^{5}$ and S.A. Walker$^6$
\\
$^{1}$Institute of Astronomy, University of Cambridge, Madingley Road, Cambridge CB3 0HA, UK\\
$^{2} $Max-Planck-Institut fur extraterrestrische Physik, Giessenbachstrasse 1, 85748 Garching, Germany\\
$^{3} $Department of Physics, University of Kentucky, Lexington KY 40506, USA\\
$^{4} $Department of Physics and Astronomy, University of Waterloo, 200 University Avenue West, Waterloo, ON N2L 3G1, Canada\\
$^{5} $INAF-IASF Palermo, Via U. La Malfa 153, I-90146 Palermo, Italy\\
$^6$ Department of Physics and Astronomy, The University of Alabama in Huntsville, Huntsville, AL 35899, USA\\ }
\date{Accepted XXX. Received YYY; in original form ZZZ}

\pubyear{2022}

\begin{document}
\label{firstpage}
\pagerange{\pageref{firstpage}--\pageref{lastpage}}
\maketitle

\begin{abstract}
Recently, we have uncovered Hidden Cooling Flows (HCF) in the X-ray spectra of the  central Brightest Galaxies of 11 clusters, 1 group and 2 elliptical galaxies. Here we report such flows in a further 15 objects, consisting of 8 clusters, 3 groups, 3 ellipticals and 1 Red Nugget. The mass cooling rates are about $1\Msunpyr$ in the ellipticals, 2 to $20\Msunpyr$ in the groups and 20 to $100\Msunpyr$ in regular clusters. The Red Nugget, MRK 1216, has an HCF of $10\Msunpyr$. We review the fate of the cooled gas  and investigate how some of it might accrete onto the central black hole. The gas is likely to be very cold and to have fragmented into low mass stars and smaller objects before being swallowed whole, with little luminous output. If such a scenario is correct and operates at a few $\Msunpyr$ then such objects may host the fastest growing black holes in the low redshift Universe. We briefly discuss the relevance of HCF to the growth of early galaxies and black holes. 
\end{abstract}


%%%%%%%%%%%%%%%%%%%%%%%%%%%%%%%%%%%%%%%%%%%%%%%%%%

%%%%%%%%%%%%%%%%% BODY OF PAPER %%%%%%%%%%%%%%%%%%

\section{Introduction}
We have recently found Hidden Cooling Flows in clusters and groups of galaxies, as well as a couple of nearby elliptical galaxies (HCFI and HCFII) \citep{Fabian22, Fabian23}, using spectra from the XMM Reflection Grating Spectrometer (RGS). These soft X-ray-emitting flows are hidden within photoelectrically-absorbing cold clouds and dust near the centres of the central brightest galaxies. They represent the cooler inner parts of larger, wider-scale cooling flows. AGN feedback acts to reduce the main cooling flow in the larger body of these objects but the inner parts drop from direct view behind cold absorbing clouds. The total mass cooling rates can be 20 to 50 per cent or more of the unabsorbed rates inferred earlier from X-ray imaging studies. 

The findings again raise the "cooling flow problem" of what happens to the cooled gas? HCF mass cooling flow rates of tens of Solar masses per year in regular clusters and  $1\Msunpyr$ in early-type galaxies lasting $\sim8\Gyr$  (since  redshift $z=1$)  means  almost $10^{11}\Msun$ and $ 10^{10}\Msun$, respectively, of accumulated cooled gas. Where does it go?
The issue is not new\footnote{We do not repeat here the history of absorption studies in cooling flows, which is discussed in HCFI and II.} but has largely been ignored for the past 2 decades, even at the low rates allowed without absorption (see \cite{Liu2019} and Section 2).  

We have proposed and discussed several possibilities, namely that a) the gas cools to invisibility (i.e. so cold that it radiates little), b) the cooled gas  fragments into low mass stars and substellar objects, c) cooled gas is dragged out from the centre by the bubbling action of AGN feedback. a) and b) mean that there is increasing unseen mass of gas and/or low mass stars at the centres of these objects. c) may be consistent with observed metal abundance profiles. These possibilities are not of course mutually exclusive. 

Here we investigate how much cooled gas can end up in the central black holes. Many of the most massive black holes at low redshift lie in Brightest Cluster Galaxies \citep{McConnell2013,Bogdan2018}, and we include a couple here including Holm 15A, the central galaxy of A85, which has a black hole of mass $4\times 10 ^{10}$ \citep{Mehrgan2019}. There is some evidence that the black hole to galaxy stellar mass ratio of early-type galaxies has increased significantly from $z=1$ to the present day \citep{Farrah2023}. Since massive black holes can swallow stars whole, such accretion need not be luminous. 

We now search for hidden cooling flows in 8  cool core clusters, 3 X-ray luminous groups and 4 relatively isolated elliptical galaxy including a Red Nugget. They are found in all objects and have the typical mass cooling rates found in HCFI and II. One is the very X-ray luminous cluster, ZW3146, at medium redshift $z\sim 0.3$, the results for which compare well with other high luminosity clusters found at similar and higher redshift. A significant part of its high cooling rate of $\sim1000\Msunpyr$, goes into observed normal star formation but it is unlikely to be  a long-lived  situation. As discussed in HCFII \citep{Fabian23}, rapid accretion onto the central black hole has the potential to turn it into a luminous quasar, as seen in the Phoenix cluster, perhaps ending in a massive outburst such as has occurred in MS 0735+7421 \citep{Mcnamara2005}. 

We then speculate whether hidden accretion is taking place into the central black holes of HCFs. Hidden in the sense of unobserved because the infalling  matter consists of low mass stars etc which are swallowed whole without emitting radiation. 
Finally we speculate on high pressure star formation, which occurs in HCF, and discuss its relevance to early galaxy formation and in particular to the origin of "red nuggets".

\begin{figure}
    \centering    
\includegraphics[width=0.48\textwidth]{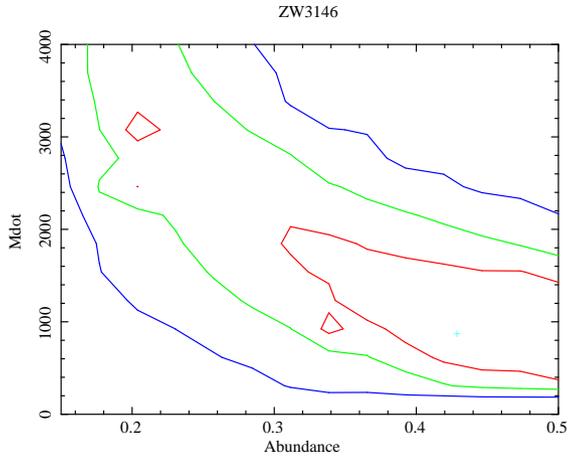} 
    \caption{ RGS spectrum of ZW3146 with HCF component shown in red and \textsc{McCall} component dotted, Mass cooling rate in $\Msunpyr$ versus  total column density in units of $10^{22}\cmsq$, Mass cooling rate versus Covering Fraction of the HCF component. Contours at 68\% (red), 90\% (green) and 99\% (blue). }
\end{figure}

\begin{figure}
    \centering    
\includegraphics[width=0.48\textwidth]{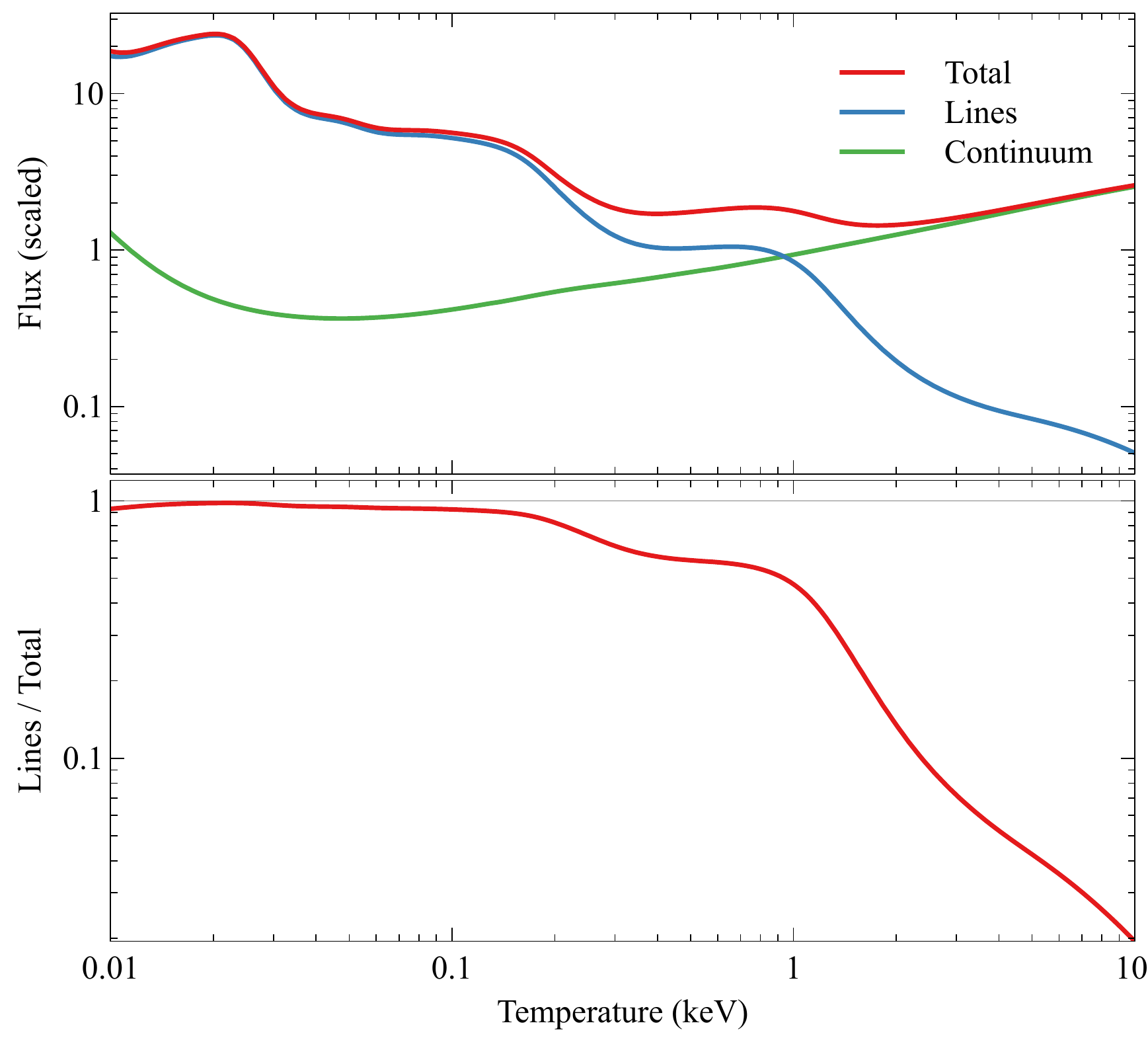} 
    \caption{ Flux from cooling flow emerging in lines (blue) and continuum (green) (top panel); as a fraction (bottom panel).  }
\end{figure}

\begin{figure}
    \centering    
\includegraphics[width=0.48\textwidth]{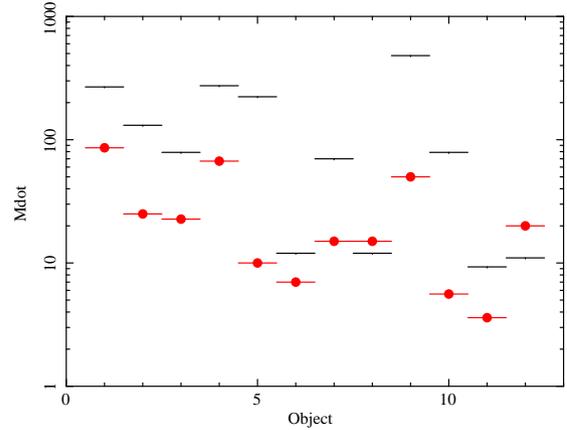} 
    \caption{Mass cooling rates, classical imaging rate from \citep{Hudson2010} (black), if available,  and  spectroscopic HCF rate (red). Objects: 1) 2A0335; 2) A85; 3) A496; 4) A2597; 5) S159, 6) A262, 7) A2052; 8) Cen; 9) Per, 10) A2199 11) NGC1550 and 12) NGC5044. The average ratio of red (HCF) to black (classical) is 0.45.}
\end{figure}

\begin{table*}
  \caption{Observed targets, giving the used source position (deg; J2000), observation identifiers, and average cleaned exposure of the RGS cameras.}

  \begin{tabular}{lrr>{\raggedright}p{5cm}rrr}
  Target & RA & Dec & OBSIDs & Exposure (ks) \\ \hline
  2A0335 & $54.6691$ & $9.9697$ & 0109870101 0109870201 0147800201 & $145$ \\
  A85 & $10.4601$ & $-9.3031$ & 0065140101 0723802101 0723802201 & $215$ \\
  A496 & $68.4074$ & $-13.2619$ & 0135120201 0506260301 0506260401 & $162$ \\
  A2597 & $351.3321$ & $-12.1243$ & 0108460201 0147330101 0723801601 0723801701 & $257$ \\
  A2199 & $247.1594$ & $39.5512$ & 0008030201 0008030301 0008030601 0723801101 0723801201 & $137$ \\
  M87 & $187.7059$ & $12.3911$ & 0114120101 0200920101 0803670501 0803670601 0803671001 0803671101 & $430$ \\
  NGC1399 & $54.6210$ & $-35.4505$ & 0012830101 0400620101 & $139$ \\
  NGC720 & $28.2519$ & $-13.7387$ & 0112300101 0602010101 & $121$ \\
  NGC1550 & $64.9080$ & $2.4101$ & 0152150101 0723800401 0723800501 & $200$ \\
  NGC1600 & $67.9156$ & $-5.0861$ & 0400490101 0400490201 & $81$ \\
  NGC3091 & $150.0591$ & $-19.6364$ & 0041180301 0041180701 & $30$ \\
  NGC5813 & $225.2969$ & $1.7019$ & 0302460101 0554680201 0554680301 0554680401 & $170$ \\
  NGC5846 & $226.6223$ & $1.6048$ & 0021540101 0021540501 0045340101 0723800101 0723800201 & $228$ \\
  MRK1216 & $127.1964$ & $-6.9402$ & 0822960101 0822960201 & $235$ \\
  ZW3146 & $155.9147$ & $4.1866$ & 0108670101 0108670401 0605540201 0605540301 & $240$ \\
  \hline
  \end{tabular}
\end{table*}

\begin{table*}
	\centering
	\caption{Spectral Fitting Results. The units of column density $N_{\rm H}$ are $10^{22}\pcmsq$, the temperature $kT$ of the \textsc{apec} and maximum of \textsc{mkcflow} component $kT$ (which are the same) is in $\keV$, $Z$ is  abundance  relative to Solar and $\dot M$ is in $\Msunpyr$. $\dot M_{\rm u}$ is the uncovered rate (i.e. with no absorption).  (f) means that a parameter is fixed. All uncertainties correspond to the 90\% confidence level.}
	\begin{tabular}{lcccccccccr} 
		\hline
		Cluster & $N_{\rm H}$ &  $kT$ & $Z$ &$z$ & $Norm$ &  $CFrac$ & ${N_{\rm H}}^{'}$ &  $\dot M$ & $\chi^2/{\rm dof}$ &$\dot M_{\rm u}$\\
		\hline
		 & $10^{22}\cmsq$ & $\keV$ & $Z_{\odot}$ & & &  & $10^{22}\cmsq$ & $\Msunpyr$ & $\Msunpyr$\\
		 \hline
     2A0335 &0.26 & $1.85^{+0.02}_{-0.01}$ & $0.32\pm 0.01$ &0.035 &4.9e-2  & 1& 0.32& $86^{+9}_{-7}$ & 1359/1224  & $48\pm4$\\
   A85 &0.029 & $3.6\pm0.4$ & $0.29^{+0.07}_{-0.003}$ & 0.056&2e-2 & 0.95&1 &$25^{+60}_{-20}$ &1406/1327  & $14.4^{+3}_{-5}$\\
   A496 &0.05 & 2.49& 0.32&0.033 &2.1e-2 &0.95 &2.2 &$22.7^{+7.5}_{-10}$ &1406/1299 & $<2.1$\\
   A2597 & 0.023 & $3.3\pm0.13$ & $0.35\pm0.03$ & 0.082&1.5e-2 &0.71 &5 & $67^{+139}_{-19}$  &1457/1364 & $19.5^{+6}_{-4}$ \\
   A2199 &0.008&$2.96^{+0.16}_{-0.08}$&$0.29^{+0.14}_{-0.02}$&0.0296&2.42e-2&0.95&0.46&$5.6^{+4.5}_{2.3}$&1409/1312&$2.3\pm1.5$ \\
   M87 &0.018&1.47&0.23&4.28e-3&7.3e-2&1.&0.2&$0.8^{+0.2}_{-0.15}$&1875/992&$0.5\pm0.04$\\
   NGC1399 &0.014&$1.08\pm0.01$&$0.27\pm0.02$&5.5e-3&3.2e-3&1.&2.67&$3.3^{+3.4}_{-0.7}$&1086/766&$0.15\pm0.03$\\
   NGC720 &0.14 & $0.61\pm0.02$ &$0.11\pm0.02$ & 0.0065& 7.9e-4&0.95 & 1.13&$1.3^{+2.6}_{-0.95}$ & 232/214 & $0.25\pm0.1$\\
   NGC1550 &0.114 & $1.26\pm0.01$& $0.26\pm0.02$& 0.0132&7.5e-3&1. & 1.33&$3.6^{+2.4}_{-1.0}$ & 1115/1051 & $0.55\pm0.2$\\
   NGC1600& 0.04 & $1.33^{+0.02}_{-0.07}$  & $0.12^{+0.07}_{-0.04}$&0.0163 &1.1e-3 &0.65 &1.3 &$0.81^{+7}_{-0.7}$ &161/146 & $<0.6$\\
   NGC3091 & 0.013 & $0.01\pm0.03$&$0.09\pm0.02$ & 0.013&1.37e-3 &0.95 &3.16 &$8.5\pm6.6$ & 92/78 & $<1$ \\
   NGC5813 & 0.043 & $0.72\pm0.01$& $0.48\pm0.08$ & 0.0063& 1.8e-3& 1& 0.5&$2.0^{+5}_{-1}$  & 1227/881 & $2.1$\\
   NGC5846 & 0.043 & $0.83\pm 0.01$ & $0.33\pm 0.02$ & 0.0053& 1.5e-3&1 & 3.2&$1.29^{+0.3}_{-0.15}$ & 839/617 & $0.8\pm0.5$\\
   MRK1216 & 0.04 & $0.7\pm0.14$ &$0.33\pm0.1$ &0.021 & 4.e-4&0.95 & 2&$9.7\pm 2.7$ &336/336 & $1.3\pm0.4$\\
    ZW3146 & 0.024& $5.7\pm 0.45$& $0.4f$ &0.288 & 8.8e-3& 0.8& 3&$895^{+1000}_{-300}$ & 1196/1169 & $230^{+50}_{-80}$\\
		\hline
	\end{tabular}
\end{table*}

\begin{table*}
	\centering
	\caption{Relevant Cluster Properties. See subsections of Appendix A for individual object references. A dash indicates lack of data.}
	\label{tab:properties_table}
	\begin{tabular}{lcccccr} 
		\hline
		Cluster & $L({\rm FIR})$ & $L_a$ & $\dot M$ & $L({\rm H}\alpha)$ & $M_{\rm CO}$ &$M_{\rm BH}$ \\
		\hline
		 & $\ergps$ &$\ergps$& $\Msunpyr$ & $\ergps$ & $\Msun$ & $\Msun$ \\
		 \hline
   \hline
   2A0335 & 4e43 & 2.1e43 & $86$ & 8e41 & 1.1e9 & - \\
   A85 & 2.8e43 & 9.9e42 &$23$ & -& - & 4e10\\
   A496 &- &9.6e42 & $23$ & 5e40& - & - \\
   A2597 & 6.5e43 &2.1e43 &$67$  & 3e42 & 2.3e9 & - \\
   A2199&-&1.5e42&5.6&3.5e40&-&4e9\\
   M87&5.0e41&1.6e41&0.8&1.9e40&-&6.5e9\\
   NGC1399&-&7.4e41&3.3&1e39&&$1e9$\\
   NGC720 & -& 1.5e41& $1.0$ & -& 1.1e7 & - \\
   NGC1550 & -& 8.7e41& $1.5$ & -& - & 4.5e9\\
   NGC1600& - &1.3e41 & $0.8$ &4e39 &-  & 1.7e10\\
   NGC3091 & - & 1.6e42& $8.5$ &- & - & 3.6e9 \\
   NGC5813 &1.1e42 &5.9e41 &$2.0$  & 1.6e40   & - & - \\
   NGC5846 & 6.2e41 & 2.0e41& $1.3$ & 2.5e40 & 2e6  & - \\
   MRK1216 & - & 1.3e41& $9.7$ & -& - & 4.9e9 \\
    ZW3146 &1.0e45 & 6.3e44& $1570$ &6e42 & 5e10& - \\
   \hline
		NGC5044 & 3.0e42 & 3.6e42 & 20 &  7.0e40  &1.5e8 \\
		Sersic 159 & 7.3e42 & 2.5e42 & 10  & 2.0e41  & 1.1e9\\
		 A262 & 8.0e42 &2.1e42 & 7 & 9.4e40  &4.0e8\\
		 A2052 & 8.3e42 & 4.4e42 & 15 & 6e40 & 2.8e8\\
		 RXJ0821 & 4.5e44 & 7.8e42 & 40 &3.0e41   &3.9e10\\
		 RXJ1532 & 2.3e45 & 2.0e44 &1000 &3e42  &8.7e10\\
		 MACS1931 & 5.6e45 & 4.6e44 &1000 & 2e42 &9.0e10\\ 
		 Phoenix Cluster &3.7e46&3.3e44&2000& 8.5e43& 2e10\\
		 M84 &1.0e42&3.3e41&2.0&4.0e39&<1.8e7\\
		 M49 &1.2e42&2.0e41&1.0&5.8e39&<1.4e7\\
		 \hline
		 Centaurus &3.2e42 &3.6e42&15&1.7e40&1.0e8\\
		 Perseus &5.6e44&5.8e42&50&3.2e42&2.0e10\\
		 A1835 &3.2e45&5.2e43&400&4.4e42&5.0e10\\
		 \hline
		 RXJ1504 &-&1.9e44&520&3.2e43&1.9e10\\
		 \hline
	
		\hline
	\end{tabular}
\end{table*}

\section{Spectral Analysis}

The objects and data used are listed in Table 1. SAS 20.0 was used for the data reduction.
The spectra were extracted using \textsc{rgsproc} with a 95\% extraction in PSF width  (corresponding to 1.7 arcmin) and 95\% in pulse-height distribution. To create Good Time Intervals (GTIs), light curves were created for each RGS instrument from CCD 9, with events with flag values of 8 or 16, extracted with a cross-dispersion angle of greater than $1.5\times 10^{-4}$, in time bins of 200s. The GTIs were created when the rate was below 0.3 cts/s. Background spectra were created with \textsc{rgsbkgmodel}. The spectra and background spectra from RGS1 and RGS2 were combined using \textsc{rgscombine} before spectral fitting.
The spectra were then analysed using \textsc{xspec} \cite{Arnaud1996} over the energy range of $8-22$A, where background is minimised. 

The spectral model used is \textsc {tbabs(gsmooth*apec+gsmooth(partcov*mlayerz)mkcflow)}. The intrinsic absorption model \textsc{mlayerz} (see HCFII for details) represents a sequence of interleaved emission and absorption layers with a total column density $N_{\rm H}$ listed in Table 2. \textsc{TBABS} is the Galactic absorption in the direction of the target \textsc{APEC} is a constant temperature thermal emission model which represents the outer cluster gas. Its temperature is also used as the hotter temperature in the cooling flow \textsc{mkcflow} model. \textsc{partcov} enables the measurement of the  total mass cooling rate of both unabsorbed and absorbed components. 
A covering fraction of one means that all the cooling flow component is absorbed and if zero then none is absorbed. The model assumes no particular geometry for the absorbed and unabsorbed components. It does assume that all absorbed components are identical. The minimum temperature of the cooling flow model is set at 0.1 keV. 

The spectra are shown  in the  Appendix as Figs B1 to B15,  together with contour plots  of  absorbed mass cooling rate ($\dot M_{\rm a}$) versus intrinsic column density $N_{\rm H}$ and  covering fraction.

Since the RGS is a slitless spectrometer \citep{denHerder2001} there is some blurring of the energy scale associated with extended sources. This is included  in the spectral model by smoothing the spectral components with separate gaussian kernels for the outer \textsc{APEC} component and  inner HCF. When making the contour plots for the less bright objects we often needed to freeze the smoothing parameters to their best fit values in order to have convergence.   
Detailed spectral results are given in Table 2 and are compared with data from other wavebands in Table 3.

\section{The Spectral Results}

As noted in HCFII, $\chi^2$-space for the HCF model is often corrugated  which can lead to complex contour plots.  We are using a very simplistic model and a real hidden cooling flow is expected to be far more complicated in both space and column density. RGS spectra provide no more than a rough average over the inner arcmin of the target source.

A source like ZW3146, where there is a large continuum fraction, can have  a very uncertain abundance $Z$, with it anticorrelating with the  mass cooling rate (see Fig 1).  In this case we fix it at $Z=0.4$. 

Of the 15 sources studied  here, all but 3 require a best-fit covering fraction of 0.95 or more. This emphasises that they are indeed "hidden". The intrinsic column densities range from $2\times 10^{21}$ to $3\times 10^{22}\pcmsq$.

We also refit the spectra with the Covering Fraction set to zero, in order to determine the mass cooling rate if there is no absorption, $\dot M_{\rm u}$. This is listed in the last column of Table 2. As expected, it is generally very low, but quite large for 2A0335. The lowest $\chi^2$ value for this no absorption case is however 14 above that for the best fit HCF \textsc{mlayerz} model, which is therefore the statistically preferred one. The value in the case of A2597 is about what is expected from the HCF model where the Covering Fraction is about 70 per cent.  

When the temperature of the gas is above about 0.4 keV the fraction of the energy emerging in continuum is about 50 per cent and drops below 10 per cent below 0.2 keV (Fig 2). Most of the flux below 0.4 keV is absorbed away in our HCF fits, meaning the continuum shape plays a significant role in our spectral fit results.

We reduced the energy band   of the spectra of several lower temperature objects to 12--20A due to broad excess residuals around 10A. These are likely due to the \textsc{APEC} component having a (small) spread in temperature.

The absorbed luminosities ($L_{\rm a}$, Table 3) are all less than the Far Infrared luminosities, where available. This indicates that the energy lost in the cooling flow to absorption is energetically capable of emerging as radiation from dust in the absorbing gas. 

Fig 3 shows the HCF mass cooling rates (in red) compared (where available) with the "classical" rates from X-ray imaging listed by \citep{Hudson2010} (in black). The mean ratio of Hidden to classical rates is 45 per cent, with a range from 4 to 180 per cent. (Most lie between 17 and 58 percent.) The HCF rates are about $1\Msunpyr$ for elliptical galaxies, 2 to $20\Msunpyr$ for Brightest Group Galaxies (BGG) and about 10 to $100\Msunpyr$ for regular Brightest Cluster Galaxies (BCG). There are then a group of more distant, exceptionally X-ray luminous, BCGs with 400 to $>1000\Msunpyr$. 

We suspect that the last group of rare objects may be highly time variable, with peak luminosity followed by a quasar eruption. The regular clusters and elliptical galaxies generally have low luminosity nuclei, with  radio emission from jets that blow bubbles in the intracluster medium. The bubbles and related activity generally lie outside the inner kpc studied here. 

\section{The accumulation of cooled gas}

Over a billion years $10^9\Msun$ of gas will have cooled in a typical elliptical, up to $10^{10}\Msun$ in a BGG and up to $10^{11}\Msun$ in a BCG. These are large values, the higher end of which exceeds the cold molecular masses observed via CO emission in BCGs \citep{Russell2019,Olivares2019}. It is possible that the mass of molecular gas has been underestimated due to low abundance and an unseen diffuse component, but this is unlikely to  make a very large  difference.  

In HCFI we considered the following possibilities: a) continued cooling to invisibility at 3K, b) fragmentation and collapse into substellar objects since the Jeans mass is less than $0.1\Msun$, c)
outward  dragging of cooled clouds by the bubbling process or d) cold front formation. The gas and dark matter peaks may be  offset by a kpc or more. 

We also flagged the similarity in conditions (e.g. gas pressure) of cooled dusty molecular clouds of a BCG core to those in the Crab Nebula. More detailed observational comparisons are warranted.

It is likely that the dominant process is a combination of c) and d) in which the cooled material is spread over the innermost few kpc of the core. Clear evidence of the dragging out of dust-enriched material from the centre is provided by the peaks in metal abundance seen $\sim10\kpc$ from the centre of low redshift clusters \citep{Panagoulia2015,Lakhchaura2019, Liu2019}.

Detailed measurements of the mass profile of each separate component in a cool core (black hole, dark matter, stars, gas etc) will be invaluable in sorting the possibilities out. We now consider whether some small fraction of the  very cold clouds and substellar objects can be swallowed by the central  black hole in the next section.

\subsection{Accretion of fragmented cold matter by the central black hole}

We showed in HCFI \citep{Fabian22} that, under the high pressure conditions of an HCF ($nT\sim 10^{6.5} - 10^{7.5}\pcmcuK$) and no heating, the gas cools rapidly (timescale of tens of years) to $\sim 3K$. The Jeans mass is below about $0.1\Msun$ \citep{Jura1977,Ferland1994} and the gas expected to clump and fragment into low mass stars, brown dwarfs etc, some of which will fall into the black hole emitting little radiation. Exactly how large a fraction will be swallowed depends on how angular momentum is transported outward. The turbulent viscosity of a luminous accretion disc is absent here and a possible path is that the innermost cooled gas forms a thick disc of low mass stars and cold gas clouds around the black hole. Dynamical gravitational instabilities such as spiral waves\footnote{A spiral feature is seen at the centre of the Centaurus cluster, see Fig 6 in HCFII.} and bars within bars transport angular momentum outward in non-spherical systems so that some of the matter falls inward \citep{Shlosman1989,Hopkins2011,Gualandris2017} to be swallowed directly by the central black hole without a standard accretion disc forming.  
\footnote{\cite{King2016} discusses an upper mass limit for a black hole to have a luminous accretion disc of $\approx5\times 10^{10}\Msun$ . Above that limit, the innermost parts of a luminous turbulent disc would be gravitationally unstable to fragmentation, so preventing the existence of any luminous gaseous disc. In our case we consider that the infalling matter has already fragmented and collapsed at larger radii, irrespective of the black hole mass.}

A very crude estimate of the mass inflow rate may be obtained from an isothermal Bondi flow. This of course assumes the matter is a fluid and ignores rotation but does give some idea of the rate at which matter comes under the gravitational influence of the black hole.  This simple rate is 
\begin{equation}
    \dot M = 4.5 \pi \frac{G^2 M^2}{c_s^3} \rho,
\end{equation}
where $M, c_s$ and $\rho$ are the black hole mass $M_9=10^9\Msun$, the speed of sound (or random motions) and density of the surrounding gas. Taking $c_s=300\kmps$ and $\rho$ equal to the mass density if the medium has a mass in units of  $10^9\Msun$ per sphere of radius $1\kpc$, we obtain $\dot M\approx 4 M_9^2\Msunpyr$ and an accretion radius of $\sim50$ pc.

\cite{Hopkins2011} give an analytical estimate of the accretion rate from gravitational torques which agrees with their numerical simulations and find $1\Msunpyr$ in the middle of the range. The predicted rate has a weak black hole mass dependence of $M^{1/6}$. Using observations and analytical work, \cite{Genzel2023} show that such torques operating in disc galaxies at $z\sim 2$ lead to largescale inflow on about 10 dynamical times. 

In the case of the elliptical and brightest group galaxies studied here, the accretion rate could  exceed the HCF mass cooling rate, which would then become the determining rate. We conclude that rates of a few $\Msunpyr$ may be possible.  We are of course assuming a high efficiency with which the cold matter is swallowed by the black hole.  

The possibility thus emerges that the mass of black holes in low redshift Elliptical Galaxies  is increasing due to inflow from HCF at a  rate of  several $\Msunpyr$. Angular momentum transfer is due to gravitational torques.  The black hole mass can thus increase by up to $\sim 10^{10}\Msun$ since $z=1$ and possibly even more for the most massive objects in the BCGs of the  most massive clusters. Such objects need not have a luminous AGN, although an ADAF due to a weak gaseous inflow may persist and power jets thus a radio source in these objects. They would be the highest accretion rate black holes in the low redshift Universe. If the accretion rate continues for several Gyr then this would lead to the most massive black holes appearing now. Examples in our sample include NGC1600 and the central galaxy of A85, Holm 15A, with $1.7\times 10^{10}\Msun$  and $4\times 10^{10}\Msun$ black holes, respectively. Other BCGs have very high mass black hole including that of the BCG of A1201 for which a gravitationally lensed arc reveals a central mass of $3\times 10^{10}\Msun$ \citep{Nightingale2023}.

\subsection{Relevance to galaxy formation}

The standard model of galaxy formation involves gas falling into dark matter haloes and heated by shocks and compression, gas that can cool quickly (on a dynamical time or less) then leads to star formation enriching the gas with metals and dust and the core of a new galaxy \citep{white1978}. Further accretion, mergers and feedback later build the  outer galaxy. The gas which has a longer cooling time than the local dynamical time, but shorter that the age of the Universe at the time, can form a cooling flow \citep{Nulsen1995}. If the conditions such as metal and dust enrichment and especially the pressure are high ($nT>10^6\pcmcuK$) then they may resemble the nearby Hidden Cooling Flows discussed here. If at high redshift, the higher temperature of the Cosmic Microwave Background will in turn require a higher Jeans mass. If cloud collapse does lead to large populations of low mass stars and brown dwarfs then early supermassive black holes can grow by swallowing such fragments whole, independent of the Eddington limit.   

\subsection{Red Nuggets and MRK 1216}
A population of compact Early-Type Galaxies (ETG) have been identified at redshifts of 2 and above which may be examples of galaxies that did not progress beyond the early core formation galaxy stage \citep{Daddi2005}. These are known as "Red Nuggets" \citep{Damjanov2009} and have stellar masses of $1-2\times 10^{11}\Msun$ and effective radii of only 1-2 kpc. Later some examples have been identified at low redshifts, e.g. NGC1277 \citep{Trujillo2014}  a galaxy  unable to grow larger  by mergers, or by accretion of cold gas, since it lies in the core of the rich Perseus Cluster. More recently further examples have been found \citep{Ferre015} including the isolated rotating ETG MRK 1216 \citep{Ferre2017} which lies at a distance of 94 Mpc and hosts a black hole of mass $4.5\times 10^9\Msun$ \citep{walsh2017}. 

\cite{Werner2018} noted that MRK 1216 might lie in a halo of mass up to $10^{13}\Msun$ and so have an X-ray halo.  They indeed found extended thermal emission with an X-ray luminosity $L_{\rm X}=7\times 10^{41}\ergps.$ \cite{Buote2019} found that its dark matter halo has a high concentration, implying  early formation.  We have included MRK1216 in our sample and find a significant HCF of $9.7\pm2.7 \Msunpyr$, larger than the rate of typical ETGs. \cite{Ferre2017} show it to have a very bottom-heavy IMF which is consistent with a significant accumulation of low mass stars and brown dwarfs.

MRK1216 could provide the nearest link between low and high redshift HCF and clearly merits deeper study. 

\subsection{Observational Possibilities}
Further observations at the time of writing are limited. Hopefully, XRISM will be launched soon and provide new high resolution, non-dispersive, X-ray spectra of the inner regions of clusters, groups and ETG. Its Field of View is larger than that of the RGS so can see how any HCF region matches into the rest of the cluster.  High spatial resolution X-ray studies await next generation telescopes such as AXIS. As well as resolving the expected irregular appearance of HCF due to the absorption, it will be particularly helpful for examining the immediate surroundings of the central black hole. The X-IFU of Athena will  spectroscopically map HCFs in great detail, as will the Light Element Emission Mapper Probe.  JWST may open up the inner regions in the near IR. Since most of the flow of cooled gas takes place at very low temperatures below $10\K$, the bulk of the flow will be inaccessible, except to absorption measurements.

ALMA has opened up molecular {\em absorption} studies of cool BCGs using the  central radio source as a backlight \citep{David2014, Tremblay2016a, Rose2019, Rose2023}. Four objects in the last study show molecular gas moving towards the central source at $200-300\kmps$, each plausibly forming part of an inward cold accretion flow.  

\section{Conclusion} 
We find that significant cooling flows, closely linked with cold absorbing gas, are common in the brightest galaxies of cool core clusters and groups as well as large elliptical galaxies. The mass cooling rates range from 1 to over $1000\Msunpyr$. In most cases they are reduced by AGN Feedback to a factor  2 to 3 times lower than the simple cooling rates derived from X-ray imaging. The gas in the central hidden/absorbed part can cool to below 10K, collapsing and fragmenting  into low mass stars, brown dwarfs etc., most of which are dragged outward by the bubbling and cold front processes. We speculate that some matter within the inner tens pc may fall into the black hole, with a rate of a few $\Msunpyr$ being plausible. Such accretion  emits little radiation, although it is likely that some thin plasma is present, possibly in the form of a low luminosity ADAF, to power the jets usually seen in radio images.  If cooled collapsed matter does fall in, then the mass accretion rate can be among the highest  in the low redshift Universe.

\section{Acknowledgements}   BRM acknowledges the Natural Sciences and Engineering Research Council for their support. We thank the referee for a prompt report. 

\section{Data Availability} All data used here are available from ESA's XMM-Newton Science Archive.

\bibliographystyle{mnras}
\bibliography{cool_core_Mdot} % if your bibtex file is called example.bi

\begin{thebibliography}{}
\makeatletter
\relax
\def\mn@urlcharsother{\let\do\@makeother \do\$\do\&\do\#\do\^\do\_\do\%\do\~}
\def\mn@doi{\begingroup\mn@urlcharsother \@ifnextchar [ {\mn@doi@}
  {\mn@doi@[]}}
\def\mn@doi@[#1]#2{\def\@tempa{#1}\ifx\@tempa\@empty \href
  {http://dx.doi.org/#2} {doi:#2}\else \href {http://dx.doi.org/#2} {#1}\fi
  \endgroup}
\def\mn@eprint#1#2{\mn@eprint@#1:#2::\@nil}
\def\mn@eprint@arXiv#1{\href {http://arxiv.org/abs/#1} {{\tt arXiv:#1}}}
\def\mn@eprint@dblp#1{\href {http://dblp.uni-trier.de/rec/bibtex/#1.xml}
  {dblp:#1}}
\def\mn@eprint@#1:#2:#3:#4\@nil{\def\@tempa {#1}\def\@tempb {#2}\def\@tempc
  {#3}\ifx \@tempc \@empty \let \@tempc \@tempb \let \@tempb \@tempa \fi \ifx
  \@tempb \@empty \def\@tempb {arXiv}\fi \@ifundefined
  {mn@eprint@\@tempb}{\@tempb:\@tempc}{\expandafter \expandafter \csname
  mn@eprint@\@tempb\endcsname \expandafter{\@tempc}}}

\bibitem[\protect\citeauthoryear{{Arnaud}}{{Arnaud}}{1996}]{Arnaud1996}
{Arnaud} K.~A.,  1996, in {Jacoby} G.~H.,  {Barnes} J.,  eds,  Astronomical
  Society of the Pacific Conference Series Vol. 101, Astronomical Data Analysis
  Software and Systems V. p.~17

\bibitem[\protect\citeauthoryear{{Bogd{\'a}n}, {Lovisari}, {Volonteri}  \&
  {Dubois}}{{Bogd{\'a}n} et~al.}{2018}]{Bogdan2018}
{Bogd{\'a}n} {\'A}.,  {Lovisari} L.,  {Volonteri} M.,   {Dubois} Y.,  2018,
  \mn@doi [\apj] {10.3847/1538-4357/aa9ab5}, \href
  {https://ui.adsabs.harvard.edu/abs/2018ApJ...852..131B} {852, 131}

\bibitem[\protect\citeauthoryear{{Buote} \& {Barth}}{{Buote} \&
  {Barth}}{2019}]{Buote2019}
{Buote} D.~A.,  {Barth} A.~J.,  2019, \mn@doi [\apj]
  {10.3847/1538-4357/ab1008}, \href
  {https://ui.adsabs.harvard.edu/abs/2019ApJ...877...91B} {877, 91}

\bibitem[\protect\citeauthoryear{{Buote}, {Jeltema}, {Canizares}  \&
  {Garmire}}{{Buote} et~al.}{2002}]{Buote2002}
{Buote} D.~A.,  {Jeltema} T.~E.,  {Canizares} C.~R.,   {Garmire} G.~P.,  2002,
  \mn@doi [\apj] {10.1086/342158}, \href
  {https://ui.adsabs.harvard.edu/abs/2002ApJ...577..183B} {577, 183}

\bibitem[\protect\citeauthoryear{{Crawford} \& {Fabian}}{{Crawford} \&
  {Fabian}}{1992}]{Crawford1992}
{Crawford} C.~S.,  {Fabian} A.~C.,  1992, \mn@doi [\mnras]
  {10.1093/mnras/259.2.265}, \href
  {https://ui.adsabs.harvard.edu/abs/1992MNRAS.259..265C} {259, 265}

\bibitem[\protect\citeauthoryear{{Daddi} et~al.,}{{Daddi}
  et~al.}{2005}]{Daddi2005}
{Daddi} E.,  et~al., 2005, \mn@doi [\apj] {10.1086/430104}, \href
  {https://ui.adsabs.harvard.edu/abs/2005ApJ...626..680D} {626, 680}

\bibitem[\protect\citeauthoryear{{Damjanov} et~al.,}{{Damjanov}
  et~al.}{2009}]{Damjanov2009}
{Damjanov} I.,  et~al., 2009, \mn@doi [\apj] {10.1088/0004-637X/695/1/101},
  \href {https://ui.adsabs.harvard.edu/abs/2009ApJ...695..101D} {695, 101}

\bibitem[\protect\citeauthoryear{{David} et~al.,}{{David}
  et~al.}{2014}]{David2014}
{David} L.~P.,  et~al., 2014, \mn@doi [\apj] {10.1088/0004-637X/792/2/94},
  \href {https://ui.adsabs.harvard.edu/abs/2014ApJ...792...94D} {792, 94}

\bibitem[\protect\citeauthoryear{{Donahue}, {Sun}, {O'Dea}, {Voit}  \&
  {Cavagnolo}}{{Donahue} et~al.}{2007}]{Donahue2007}
{Donahue} M.,  {Sun} M.,  {O'Dea} C.~P.,  {Voit} G.~M.,   {Cavagnolo} K.~W.,
  2007, \mn@doi [\aj] {10.1086/518230}, \href
  {https://ui.adsabs.harvard.edu/abs/2007AJ....134...14D} {134, 14}

\bibitem[\protect\citeauthoryear{{Dupke} \& {White}}{{Dupke} \&
  {White}}{2003}]{Dupke2003}
{Dupke} R.,  {White} Raymond~E. I.,  2003, \mn@doi [\apjl] {10.1086/367824},
  \href {https://ui.adsabs.harvard.edu/abs/2003ApJ...583L..13D} {583, L13}

\bibitem[\protect\citeauthoryear{{Egami} et~al.,}{{Egami}
  et~al.}{2006}]{Egami2006}
{Egami} E.,  et~al., 2006, \mn@doi [\apj] {10.1086/504519}, \href
  {https://ui.adsabs.harvard.edu/abs/2006ApJ...647..922E} {647, 922}

\bibitem[\protect\citeauthoryear{{Event Horizon Telescope Collaboration}
  et~al.,}{{Event Horizon Telescope Collaboration} et~al.}{2019}]{EHT2019}
{Event Horizon Telescope Collaboration} et~al., 2019, \mn@doi [\apjl]
  {10.3847/2041-8213/ab0ec7}, \href
  {https://ui.adsabs.harvard.edu/abs/2019ApJ...875L...1E} {875, L1}

\bibitem[\protect\citeauthoryear{{Fabian}, {Ferland}, {Sanders}, {McNamara},
  {Pinto}  \& {Walker}}{{Fabian} et~al.}{2022}]{Fabian22}
{Fabian} A.~C.,  {Ferland} G.~J.,  {Sanders} J.~S.,  {McNamara} B.~R.,  {Pinto}
  C.,   {Walker} S.~A.,  2022, \mn@doi [\mnras] {10.1093/mnras/stac2003}, \href
  {https://ui.adsabs.harvard.edu/abs/2022MNRAS.515.3336F} {515, 3336}

\bibitem[\protect\citeauthoryear{{Fabian}, {Sanders}, {Ferland}, {McNamara},
  {Pinto}  \& {Walker}}{{Fabian} et~al.}{2023}]{Fabian23}
{Fabian} A.~C.,  {Sanders} J.~S.,  {Ferland} G.~J.,  {McNamara} B.~R.,  {Pinto}
  C.,   {Walker} S.~A.,  2023, \mn@doi [\mnras] {10.1093/mnras/stad507}, \href
  {https://ui.adsabs.harvard.edu/abs/2023MNRAS.521.1794F} {521, 1794}

\bibitem[\protect\citeauthoryear{{Farrah} et~al.,}{{Farrah}
  et~al.}{2023}]{Farrah2023}
{Farrah} D.,  et~al., 2023, \mn@doi [\apj] {10.3847/1538-4357/acac2e}, \href
  {https://ui.adsabs.harvard.edu/abs/2023ApJ...943..133F} {943, 133}

\bibitem[\protect\citeauthoryear{{Ferland}, {Fabian}  \& {Johnstone}}{{Ferland}
  et~al.}{1994}]{Ferland1994}
{Ferland} G.~J.,  {Fabian} A.~C.,   {Johnstone} R.~M.,  1994, \mn@doi [\mnras]
  {10.1093/mnras/266.2.399}, \href
  {https://ui.adsabs.harvard.edu/abs/1994MNRAS.266..399F} {266, 399}

\bibitem[\protect\citeauthoryear{{Ferr{\'e}-Mateu}, {Mezcua}, {Trujillo},
  {Balcells}  \& {van den Bosch}}{{Ferr{\'e}-Mateu} et~al.}{2015}]{Ferre015}
{Ferr{\'e}-Mateu} A.,  {Mezcua} M.,  {Trujillo} I.,  {Balcells} M.,   {van den
  Bosch} R. C.~E.,  2015, \mn@doi [\apj] {10.1088/0004-637X/808/1/79}, \href
  {https://ui.adsabs.harvard.edu/abs/2015ApJ...808...79F} {808, 79}

\bibitem[\protect\citeauthoryear{{Ferr{\'e}-Mateu}, {Trujillo},
  {Mart{\'\i}n-Navarro}, {Vazdekis}, {Mezcua}, {Balcells}  \&
  {Dom{\'\i}nguez}}{{Ferr{\'e}-Mateu} et~al.}{2017}]{Ferre2017}
{Ferr{\'e}-Mateu} A.,  {Trujillo} I.,  {Mart{\'\i}n-Navarro} I.,  {Vazdekis}
  A.,  {Mezcua} M.,  {Balcells} M.,   {Dom{\'\i}nguez} L.,  2017, \mn@doi
  [\mnras] {10.1093/mnras/stx171}, \href
  {https://ui.adsabs.harvard.edu/abs/2017MNRAS.467.1929F} {467, 1929}

\bibitem[\protect\citeauthoryear{{Genzel} et~al.,}{{Genzel}
  et~al.}{2023}]{Genzel2023}
{Genzel} R.,  et~al., 2023, \mn@doi [arXiv e-prints]
  {10.48550/arXiv.2305.02959}, \href
  {https://ui.adsabs.harvard.edu/abs/2023arXiv230502959G} {p. arXiv:2305.02959}

\bibitem[\protect\citeauthoryear{{Grossov{\'a}} et~al.,}{{Grossov{\'a}}
  et~al.}{2022}]{Grossova2022}
{Grossov{\'a}} R.,  et~al., 2022, \mn@doi [\apjs] {10.3847/1538-4365/ac366c},
  \href {https://ui.adsabs.harvard.edu/abs/2022ApJS..258...30G} {258, 30}

\bibitem[\protect\citeauthoryear{{Gualandris}, {Read}, {Dehnen}  \&
  {Bortolas}}{{Gualandris} et~al.}{2017}]{Gualandris2017}
{Gualandris} A.,  {Read} J.~I.,  {Dehnen} W.,   {Bortolas} E.,  2017, \mn@doi
  [\mnras] {10.1093/mnras/stw2528}, \href
  {https://ui.adsabs.harvard.edu/abs/2017MNRAS.464.2301G} {464, 2301}

\bibitem[\protect\citeauthoryear{{Hopkins} \& {Quataert}}{{Hopkins} \&
  {Quataert}}{2011}]{Hopkins2011}
{Hopkins} P.~F.,  {Quataert} E.,  2011, \mn@doi [\mnras]
  {10.1111/j.1365-2966.2011.18542.x}, \href
  {https://ui.adsabs.harvard.edu/abs/2011MNRAS.415.1027H} {415, 1027}

\bibitem[\protect\citeauthoryear{{Hudson}, {Mittal}, {Reiprich}, {Nulsen},
  {Andernach}  \& {Sarazin}}{{Hudson} et~al.}{2010}]{Hudson2010}
{Hudson} D.~S.,  {Mittal} R.,  {Reiprich} T.~H.,  {Nulsen} P.~E.~J.,
  {Andernach} H.,   {Sarazin} C.~L.,  2010, \mn@doi [\aap]
  {10.1051/0004-6361/200912377}, \href
  {https://ui.adsabs.harvard.edu/abs/2010A&A...513A..37H} {513, A37}

\bibitem[\protect\citeauthoryear{{Ichinohe}, {Werner}, {Simionescu}, {Allen},
  {Canning}, {Ehlert}, {Mernier}  \& {Takahashi}}{{Ichinohe}
  et~al.}{2015}]{Ichinohe2015}
{Ichinohe} Y.,  {Werner} N.,  {Simionescu} A.,  {Allen} S.~W.,  {Canning}
  R.~E.~A.,  {Ehlert} S.,  {Mernier} F.,   {Takahashi} T.,  2015, \mn@doi
  [\mnras] {10.1093/mnras/stv217}, \href
  {https://ui.adsabs.harvard.edu/abs/2015MNRAS.448.2971I} {448, 2971}

\bibitem[\protect\citeauthoryear{{Jura}}{{Jura}}{1977}]{Jura1977}
{Jura} M.,  1977, \mn@doi [\apj] {10.1086/155085}, \href
  {https://ui.adsabs.harvard.edu/abs/1977ApJ...212..634J} {212, 634}

\bibitem[\protect\citeauthoryear{{King}}{{King}}{2016}]{King2016}
{King} A.,  2016, \mn@doi [\mnras] {10.1093/mnrasl/slv186}, \href
  {https://ui.adsabs.harvard.edu/abs/2016MNRAS.456L.109K} {456, L109}

\bibitem[\protect\citeauthoryear{{Kolokythas} et~al.,}{{Kolokythas}
  et~al.}{2020}]{Kolokythas2020}
{Kolokythas} K.,  et~al., 2020, \mn@doi [\mnras] {10.1093/mnras/staa1506},
  \href {https://ui.adsabs.harvard.edu/abs/2020MNRAS.496.1471K} {496, 1471}

\bibitem[\protect\citeauthoryear{{Lakhchaura} et~al.,}{{Lakhchaura}
  et~al.}{2018}]{Lakhchaura2018}
{Lakhchaura} K.,  et~al., 2018, \mn@doi [\mnras] {10.1093/mnras/sty2565}, \href
  {https://ui.adsabs.harvard.edu/abs/2018MNRAS.481.4472L} {481, 4472}

\bibitem[\protect\citeauthoryear{{Lakhchaura}, {Mernier}  \&
  {Werner}}{{Lakhchaura} et~al.}{2019}]{Lakhchaura2019}
{Lakhchaura} K.,  {Mernier} F.,   {Werner} N.,  2019, \mn@doi [\aap]
  {10.1051/0004-6361/201834755}, \href
  {https://ui.adsabs.harvard.edu/abs/2019A&A...623A..17L} {623, A17}

\bibitem[\protect\citeauthoryear{{Liu}, {Pinto}, {Fabian}, {Russell}  \&
  {Sanders}}{{Liu} et~al.}{2019}]{Liu2019}
{Liu} H.,  {Pinto} C.,  {Fabian} A.~C.,  {Russell} H.~R.,   {Sanders} J.~S.,
  2019, \mn@doi [\mnras] {10.1093/mnras/stz456}, \href
  {https://ui.adsabs.harvard.edu/abs/2019MNRAS.485.1757L} {485, 1757}

\bibitem[\protect\citeauthoryear{{Machacek}, {Jerius}, {Kraft}, {Forman},
  {Jones}, {Randall}, {Giacintucci}  \& {Sun}}{{Machacek}
  et~al.}{2011}]{Machacek2011A}
{Machacek} M.~E.,  {Jerius} D.,  {Kraft} R.,  {Forman} W.~R.,  {Jones} C.,
  {Randall} S.,  {Giacintucci} S.,   {Sun} M.,  2011, \mn@doi [\apj]
  {10.1088/0004-637X/743/1/15}, \href
  {https://ui.adsabs.harvard.edu/abs/2011ApJ...743...15M} {743, 15}

\bibitem[\protect\citeauthoryear{{McConnell} \& {Ma}}{{McConnell} \&
  {Ma}}{2013}]{McConnell2013}
{McConnell} N.~J.,  {Ma} C.-P.,  2013, \mn@doi [\apj]
  {10.1088/0004-637X/764/2/184}, \href
  {https://ui.adsabs.harvard.edu/abs/2013ApJ...764..184M} {764, 184}

\bibitem[\protect\citeauthoryear{{McNamara} et~al.,}{{McNamara}
  et~al.}{2001}]{McNamara2001b}
{McNamara} B.~R.,  et~al., 2001, \mn@doi [\apjl] {10.1086/338326}, \href
  {https://ui.adsabs.harvard.edu/abs/2001ApJ...562L.149M} {562, L149}

\bibitem[\protect\citeauthoryear{{McNamara}, {Nulsen}, {Wise}, {Rafferty},
  {Carilli}, {Sarazin}  \& {Blanton}}{{McNamara} et~al.}{2005}]{Mcnamara2005}
{McNamara} B.~R.,  {Nulsen} P.~E.~J.,  {Wise} M.~W.,  {Rafferty} D.~A.,
  {Carilli} C.,  {Sarazin} C.~L.,   {Blanton} E.~L.,  2005, \mn@doi [\nat]
  {10.1038/nature03202}, \href
  {https://ui.adsabs.harvard.edu/abs/2005Natur.433...45M} {433, 45}

\bibitem[\protect\citeauthoryear{{Mehrgan}, {Thomas}, {Saglia}, {Mazzalay},
  {Erwin}, {Bender}, {Kluge}  \& {Fabricius}}{{Mehrgan}
  et~al.}{2019}]{Mehrgan2019}
{Mehrgan} K.,  {Thomas} J.,  {Saglia} R.,  {Mazzalay} X.,  {Erwin} P.,
  {Bender} R.,  {Kluge} M.,   {Fabricius} M.,  2019, \mn@doi [\apj]
  {10.3847/1538-4357/ab5856}, \href
  {https://ui.adsabs.harvard.edu/abs/2019ApJ...887..195M} {887, 195}

\bibitem[\protect\citeauthoryear{{Morris} \& {Fabian}}{{Morris} \&
  {Fabian}}{2005}]{Morris2005}
{Morris} R.~G.,  {Fabian} A.~C.,  2005, \mn@doi [\mnras]
  {10.1111/j.1365-2966.2005.08822.x}, \href
  {https://ui.adsabs.harvard.edu/abs/2005MNRAS.358..585M} {358, 585}

\bibitem[\protect\citeauthoryear{{Nightingale} et~al.,}{{Nightingale}
  et~al.}{2023}]{Nightingale2023}
{Nightingale} J.~W.,  et~al., 2023, \mn@doi [\mnras] {10.1093/mnras/stad587},
  \href {https://ui.adsabs.harvard.edu/abs/2023MNRAS.521.3298N} {521, 3298}

\bibitem[\protect\citeauthoryear{{Nulsen} \& {Fabian}}{{Nulsen} \&
  {Fabian}}{1995}]{Nulsen1995}
{Nulsen} P.~E.~J.,  {Fabian} A.~C.,  1995, \mn@doi [\mnras]
  {10.1093/mnras/277.2.561}, \href
  {https://ui.adsabs.harvard.edu/abs/1995MNRAS.277..561N} {277, 561}

\bibitem[\protect\citeauthoryear{{Nulsen} et~al.,}{{Nulsen}
  et~al.}{2013}]{Nulsen2013}
{Nulsen} P. E.~J.,  et~al., 2013, \mn@doi [\apj] {10.1088/0004-637X/775/2/117},
  \href {https://ui.adsabs.harvard.edu/abs/2013ApJ...775..117N} {775, 117}

\bibitem[\protect\citeauthoryear{{Oegerle}, {Cowie}, {Davidsen}, {Hu},
  {Hutchings}, {Murphy}, {Sembach}  \& {Woodgate}}{{Oegerle}
  et~al.}{2001}]{0egerle2001}
{Oegerle} W.~R.,  {Cowie} L.,  {Davidsen} A.,  {Hu} E.,  {Hutchings} J.,
  {Murphy} E.,  {Sembach} K.,   {Woodgate} B.,  2001, \mn@doi [\apj]
  {10.1086/322246}, \href
  {https://ui.adsabs.harvard.edu/abs/2001ApJ...560..187O} {560, 187}

\bibitem[\protect\citeauthoryear{{Oldham} \& {Auger}}{{Oldham} \&
  {Auger}}{2018}]{Oldham2018}
{Oldham} L.,  {Auger} M.,  2018, \mn@doi [\mnras] {10.1093/mnras/stx2969},
  \href {https://ui.adsabs.harvard.edu/abs/2018MNRAS.474.4169O} {474, 4169}

\bibitem[\protect\citeauthoryear{{Olivares} et~al.,}{{Olivares}
  et~al.}{2019}]{Olivares2019}
{Olivares} V.,  et~al., 2019, \mn@doi [\aap] {10.1051/0004-6361/201935350},
  \href {https://ui.adsabs.harvard.edu/abs/2019A&A...631A..22O} {631, A22}

\bibitem[\protect\citeauthoryear{{Panagoulia}, {Sanders}  \&
  {Fabian}}{{Panagoulia} et~al.}{2015}]{Panagoulia2015}
{Panagoulia} E.~K.,  {Sanders} J.~S.,   {Fabian} A.~C.,  2015, \mn@doi [\mnras]
  {10.1093/mnras/stu2469}, \href
  {https://ui.adsabs.harvard.edu/abs/2015MNRAS.447..417P} {447, 417}

\bibitem[\protect\citeauthoryear{{Randall} et~al.,}{{Randall}
  et~al.}{2015}]{Randall2015b}
{Randall} S.~W.,  et~al., 2015, \mn@doi [\apj] {10.1088/0004-637X/805/2/112},
  \href {https://ui.adsabs.harvard.edu/abs/2015ApJ...805..112R} {805, 112}

\bibitem[\protect\citeauthoryear{{Romero} et~al.,}{{Romero}
  et~al.}{2020}]{Romero2020}
{Romero} C.~E.,  et~al., 2020, \mn@doi [\apj] {10.3847/1538-4357/ab6d70}, \href
  {https://ui.adsabs.harvard.edu/abs/2020ApJ...891...90R} {891, 90}

\bibitem[\protect\citeauthoryear{{Rose} et~al.,}{{Rose}
  et~al.}{2019}]{Rose2019}
{Rose} T.,  et~al., 2019, \mn@doi [\mnras] {10.1093/mnras/stz2138}, \href
  {https://ui.adsabs.harvard.edu/abs/2019MNRAS.489..349R} {489, 349}

\bibitem[\protect\citeauthoryear{{Rose} et~al.,}{{Rose}
  et~al.}{2023}]{Rose2023}
{Rose} T.,  et~al., 2023, \mn@doi [\mnras] {10.1093/mnras/stac3194}, \href
  {https://ui.adsabs.harvard.edu/abs/2023MNRAS.518..878R} {518, 878}

\bibitem[\protect\citeauthoryear{{Runge} \& {Walker}}{{Runge} \&
  {Walker}}{2021}]{Runge2021}
{Runge} J.,  {Walker} S.~A.,  2021, \mn@doi [\mnras] {10.1093/mnras/stab444},
  \href {https://ui.adsabs.harvard.edu/abs/2021MNRAS.502.5487R} {502, 5487}

\bibitem[\protect\citeauthoryear{{Russell} et~al.,}{{Russell}
  et~al.}{2019}]{Russell2019}
{Russell} H.~R.,  et~al., 2019, \mn@doi [\mnras] {10.1093/mnras/stz2719}, \href
  {https://ui.adsabs.harvard.edu/abs/2019MNRAS.490.3025R} {490, 3025}

\bibitem[\protect\citeauthoryear{{Sanders}, {Fabian}  \& {Taylor}}{{Sanders}
  et~al.}{2009}]{Sanders2009b}
{Sanders} J.~S.,  {Fabian} A.~C.,   {Taylor} G.~B.,  2009, \mn@doi [\mnras]
  {10.1111/j.1365-2966.2009.14892.x}, \href
  {https://ui.adsabs.harvard.edu/abs/2009MNRAS.396.1449S} {396, 1449}

\bibitem[\protect\citeauthoryear{{Shlosman}, {Frank}  \& {Begelman}}{{Shlosman}
  et~al.}{1989}]{Shlosman1989}
{Shlosman} I.,  {Frank} J.,   {Begelman} M.~C.,  1989, \mn@doi [\nat]
  {10.1038/338045a0}, \href
  {https://ui.adsabs.harvard.edu/abs/1989Natur.338...45S} {338, 45}

\bibitem[\protect\citeauthoryear{{Su} et~al.,}{{Su} et~al.}{2019}]{Su2019}
{Su} Y.,  et~al., 2019, \mn@doi [\aj] {10.3847/1538-3881/ab1d51}, \href
  {https://ui.adsabs.harvard.edu/abs/2019AJ....158....6S} {158, 6}

\bibitem[\protect\citeauthoryear{{Tamura}, {Bleeker}, {Kaastra}, {Ferrigno}  \&
  {Molendi}}{{Tamura} et~al.}{2001}]{Tamura2001}
{Tamura} T.,  {Bleeker} J.~A.~M.,  {Kaastra} J.~S.,  {Ferrigno} C.,   {Molendi}
  S.,  2001, \mn@doi [\aap] {10.1051/0004-6361:20011317}, \href
  {https://ui.adsabs.harvard.edu/abs/2001A&A...379..107T} {379, 107}

\bibitem[\protect\citeauthoryear{{Temi}, {Brighenti}  \& {Mathews}}{{Temi}
  et~al.}{2007}]{Temi07}
{Temi} P.,  {Brighenti} F.,   {Mathews} W.~G.,  2007, \mn@doi [\apj]
  {10.1086/520123}, \href
  {https://ui.adsabs.harvard.edu/abs/2007ApJ...666..222T} {666, 222}

\bibitem[\protect\citeauthoryear{{Temi}, {Amblard}, {Gitti}, {Brighenti},
  {Gaspari}, {Mathews}  \& {David}}{{Temi} et~al.}{2018}]{Temi2018}
{Temi} P.,  {Amblard} A.,  {Gitti} M.,  {Brighenti} F.,  {Gaspari} M.,
  {Mathews} W.~G.,   {David} L.,  2018, \mn@doi [\apj]
  {10.3847/1538-4357/aab9b0}, \href
  {https://ui.adsabs.harvard.edu/abs/2018ApJ...858...17T} {858, 17}

\bibitem[\protect\citeauthoryear{{Thomas}, {Ma}, {McConnell}, {Greene},
  {Blakeslee}  \& {Janish}}{{Thomas} et~al.}{2016}]{Thomas2016}
{Thomas} J.,  {Ma} C.-P.,  {McConnell} N.~J.,  {Greene} J.~E.,  {Blakeslee}
  J.~P.,   {Janish} R.,  2016, \mn@doi [\nat] {10.1038/nature17197}, \href
  {https://ui.adsabs.harvard.edu/abs/2016Natur.532..340T} {532, 340}

\bibitem[\protect\citeauthoryear{{Tremblay} et~al.,}{{Tremblay}
  et~al.}{2012a}]{Tremblay2012a}
{Tremblay} G.~R.,  et~al., 2012a, \mn@doi [\mnras]
  {10.1111/j.1365-2966.2012.21281.x}, \href
  {https://ui.adsabs.harvard.edu/abs/2012MNRAS.424.1026T} {424, 1026}

\bibitem[\protect\citeauthoryear{{Tremblay} et~al.,}{{Tremblay}
  et~al.}{2012b}]{Tremblay2012b}
{Tremblay} G.~R.,  et~al., 2012b, \mn@doi [\mnras]
  {10.1111/j.1365-2966.2012.21278.x}, \href
  {https://ui.adsabs.harvard.edu/abs/2012MNRAS.424.1042T} {424, 1042}

\bibitem[\protect\citeauthoryear{{Tremblay} et~al.,}{{Tremblay}
  et~al.}{2016a}]{Tremblay2016a}
{Tremblay} G.~R.,  et~al., 2016a, \mn@doi [\nat] {10.1038/nature17969}, \href
  {https://ui.adsabs.harvard.edu/abs/2016Natur.534..218T} {534, 218}

\bibitem[\protect\citeauthoryear{{Tremblay} et~al.,}{{Tremblay}
  et~al.}{2016b}]{Tremblay2016b}
{Tremblay} G.~R.,  et~al., 2016b, \mn@doi [\nat] {10.1038/nature17969}, \href
  {https://ui.adsabs.harvard.edu/abs/2016Natur.534..218T} {534, 218}

\bibitem[\protect\citeauthoryear{{Trinchieri} \& {di Serego
  Alighieri}}{{Trinchieri} \& {di Serego Alighieri}}{1991}]{Trinchieri1991}
{Trinchieri} G.,  {di Serego Alighieri} S.,  1991, \mn@doi [\aj]
  {10.1086/115794}, \href
  {https://ui.adsabs.harvard.edu/abs/1991AJ....101.1647T} {101, 1647}

\bibitem[\protect\citeauthoryear{{Trujillo}, {Ferr{\'e}-Mateu}, {Balcells},
  {Vazdekis}  \& {S{\'a}nchez-Bl{\'a}zquez}}{{Trujillo}
  et~al.}{2014}]{Trujillo2014}
{Trujillo} I.,  {Ferr{\'e}-Mateu} A.,  {Balcells} M.,  {Vazdekis} A.,
  {S{\'a}nchez-Bl{\'a}zquez} P.,  2014, \mn@doi [\apjl]
  {10.1088/2041-8205/780/2/L20}, \href
  {https://ui.adsabs.harvard.edu/abs/2014ApJ...780L..20T} {780, L20}

\bibitem[\protect\citeauthoryear{{Vantyghem} et~al.,}{{Vantyghem}
  et~al.}{2016}]{Vantyghem2016}
{Vantyghem} A.~N.,  et~al., 2016, \mn@doi [\apj] {10.3847/0004-637X/832/2/148},
  \href {https://ui.adsabs.harvard.edu/abs/2016ApJ...832..148V} {832, 148}

\bibitem[\protect\citeauthoryear{{Vantyghem} et~al.,}{{Vantyghem}
  et~al.}{2021}]{Vantyghem2021}
{Vantyghem} A.~N.,  et~al., 2021, \mn@doi [\apj] {10.3847/1538-4357/abe306},
  \href {https://ui.adsabs.harvard.edu/abs/2021ApJ...910...53V} {910, 53}

\bibitem[\protect\citeauthoryear{{Vaughan}, {Davies}, {Zieleniewski}  \&
  {Houghton}}{{Vaughan} et~al.}{2018}]{Vaughan2018}
{Vaughan} S.~P.,  {Davies} R.~L.,  {Zieleniewski} S.,   {Houghton} R. C.~W.,
  2018, \mn@doi [\mnras] {10.1093/mnras/sty1434}, \href
  {https://ui.adsabs.harvard.edu/abs/2018MNRAS.479.2443V} {479, 2443}

\bibitem[\protect\citeauthoryear{{Walsh}, {van den Bosch}, {Gebhardt},
  {Y{\i}ld{\i}r{\i}m}, {G{\"u}ltekin}, {Husemann}  \& {Richstone}}{{Walsh}
  et~al.}{2017}]{walsh2017}
{Walsh} J.~L.,  {van den Bosch} R. C.~E.,  {Gebhardt} K.,  {Y{\i}ld{\i}r{\i}m}
  A.,  {G{\"u}ltekin} K.,  {Husemann} B.,   {Richstone} D.~O.,  2017, \mn@doi
  [\apj] {10.3847/1538-4357/835/2/208}, \href
  {https://ui.adsabs.harvard.edu/abs/2017ApJ...835..208W} {835, 208}

\bibitem[\protect\citeauthoryear{{Werner} et~al.,}{{Werner}
  et~al.}{2013}]{Werner2013}
{Werner} N.,  et~al., 2013, \mn@doi [\apj] {10.1088/0004-637X/767/2/153}, \href
  {https://ui.adsabs.harvard.edu/abs/2013ApJ...767..153W} {767, 153}

\bibitem[\protect\citeauthoryear{{Werner}, {Lakhchaura}, {Canning}, {Gaspari}
  \& {Simionescu}}{{Werner} et~al.}{2018}]{Werner2018}
{Werner} N.,  {Lakhchaura} K.,  {Canning} R.~E.~A.,  {Gaspari} M.,
  {Simionescu} A.,  2018, \mn@doi [\mnras] {10.1093/mnras/sty862}, \href
  {https://ui.adsabs.harvard.edu/abs/2018MNRAS.477.3886W} {477, 3886}

\bibitem[\protect\citeauthoryear{{White} \& {Rees}}{{White} \&
  {Rees}}{1978}]{white1978}
{White} S.~D.~M.,  {Rees} M.~J.,  1978, \mn@doi [\mnras]
  {10.1093/mnras/183.3.341}, \href
  {https://ui.adsabs.harvard.edu/abs/1978MNRAS.183..341W} {183, 341}

\bibitem[\protect\citeauthoryear{{den Herder} et~al.,}{{den Herder}
  et~al.}{2001}]{denHerder2001}
{den Herder} J.~W.,  et~al., 2001, \mn@doi [\aap] {10.1051/0004-6361:20000058},
  \href {https://ui.adsabs.harvard.edu/abs/2001A&A...365L...7D} {365, L7}

\makeatother
\end{thebibliography}

\appendix

\section{Notes on Objects}
\subsection{2A0335-096}
The Chandra X-ray data \citep{Sanders2009b} of 2A0335 show that the central arcmin consists of 6 or so blobs of emission. Some are softer than the others. Cavities are seen and in the optical a bright H$\alpha$ nebula is found. ALMA CO data \citep{Vantyghem2016} of the core reveals $10^9\Msun$ of cold molecular gas. The paper also discusses the patchy dust extinction reported earlier by \citep{Donahue2007}. We use the column density of \citep{Sanders2009b} in the spectral fitting of this object.  

\subsection{A85}
Holm 15 A, the BCG of A85, hosts one of the most massive black holes known, at $4\times 10^{10}\Msun$ \citep{Mehrgan2019}. The cluster is merging with a subcluster seen to the South. Cold fronts and a cavity are seen in Chandra X-ray data presented by \citep{Ichinohe2015}.

\subsection{A496}
The RGS data have been studied by \citep{Tamura2001} in which no evidence for X-ray emitting gas cooler than $1\keV$ was reported. (No additional absorption was considered.) Chandra X-ray images showing cold fronts have been presented by \citep{Dupke2003}. Optical emission lines spectra have been studied by \citep{Crawford1992}.

\subsection{A2597}
A2597 has consistently shown evidence for residual cooling, starting with OVI UV emission  detected with FUSE \citep{0egerle2001}, together with CIII]. The OVI line originates from thermal gas at around  $7\times 10^5\K$ and could be due to a cooling flow of $20\pm15\Msunpyr$, or  $75\Msunpyr$ when corrected for dust extinction.

\cite{Morris2005} analysed XMM data, including the RGS, and found a possible cooling flow rate of $\sim45\Msunpyr$ (corrected to $H_0=70\kmpspmpc$). \cite{McNamara2001b} found ghost bubbles in the Chandra X-ray images and \cite{Tremblay2012a,Tremblay2012b,Tremblay2016a,Tremblay2016b} found evidence for some residual cooling as well as CO absorption from ALMA data. They argue for a multiphase outflow or fountain in the cluster core. 

\subsection{A2199} 
\cite{Nulsen2013} presented the Chandra data for A2199. It shows signs of sloshing and has a complex radio source 3C338. 

\subsection{M87} 
M87 is the second most luminous galaxy but is situated at the centre of   the nearest cluster to us, the  Virgo cluster. Its central black hole has a mass of $6.5\times 10^9\Msun$  as measured by the Event Horizon Telescope \citep{EHT2019}. \citep{Werner2013} studied the Chandra emission in the core of M87, showing that it is highly multiphase and fitting an intrinsic absorption model.
\citep{Temi07} report the Spitzer FIR flux. A powerlaw component  with photon index of 2 was included in the fit to represent the nucleus and jet.
\cite{Oldham2018} Found that M87 has a botton-heavy IMF. 

\subsection{NGC1399}
This is the BCG of the Fornax cluster which is the next most distant cluster after Virgo. \cite{Su2019} present analysis of the Chandra X-ray data, revealing a pair of X-ray cavities coincident with radio lobes. Optical spectroscopy studied by \citep{Vaughan2018} shows that IMF in the central region is bottom heavy. They also comment on a filament of ionized gas.

\subsection{NGC720}
NGC720 is an isolated nearby elliptical galaxy. Its Chandra X-ray data have been studied by \cite{Buote2002}.

\subsection{NGC1550}
NGC1550 is the brightest galaxy in an X-ray bright group. It has been studied recently with Chandra by \cite{Kolokythas2020}. It shows signs of activity near its core due to interactions with its radio source.

\subsection{NGC1600}
NGC1600 is the brightest  galaxy in a small group. It hosts a very massive black hole of $1.6\times 10^{10}\Msun$, as determined by stellar velocity measurements by \cite{Thomas2016}. \cite{Runge2021} present  the Chandra data on NGC1600 at the Bondi radius. They find evidence for a multiphase (2-temperature) X-ray emitting gas. H${\alpha}$ emission is reported by \cite{Trinchieri1991}

\subsection{NGC3091}
NGC3091 is a bright elliptical galaxy in the relatively isolated compact group, Hickson 42. It has been selected for not being otherwise special. Our results show that it hosts a modest HCF, 

\subsection{NGC5813}
NGC5813 is the brightest galaxy in a subgroup of the larger NGC5846 group. They constitute the third nearest massive cluster/group assembly after the Virgo and Fornax clusters. The Chandra X-ray data of NGC5813 reveal a series of bubbles extending from the BGG \citep{Randall2015b}. 
ALMA CO and other data are reported by \cite{Temi2018}.

\subsection{NGC5846}
Chandra X-ray data on NGC5846 have been studied by \cite{Machacek2011A}.
Further X-ray analysis and multiband data on NGC5846 and many other elliptical galaxies are given by \citep{Lakhchaura2018}. Radio images of this and some of the above galaxies can be found in \citep{Grossova2022}. 

\subsection{MRK1216}
The Red Nugget MRK1216 is discussed in Subsection 3.2.

\subsection{ZW3146}
ZW3146 is a massive X-ray luminous cluster at redshift 0.29.
X-ray and ALMA CO and other data are discussed by \cite{Vantyghem2021}. The centroid of the $5\times 10^{10}\Msun$ of molecular gas is offset by 2.6 kpc from the central AGN. The FIR detection is presented by \cite{Egami2006}.  MUSTANG-2 SZ results are reported by \citep{Romero2020}.

\section{Spectra of Objects}
\begin{figure}
    \centering    \includegraphics[width=0.48\textwidth]{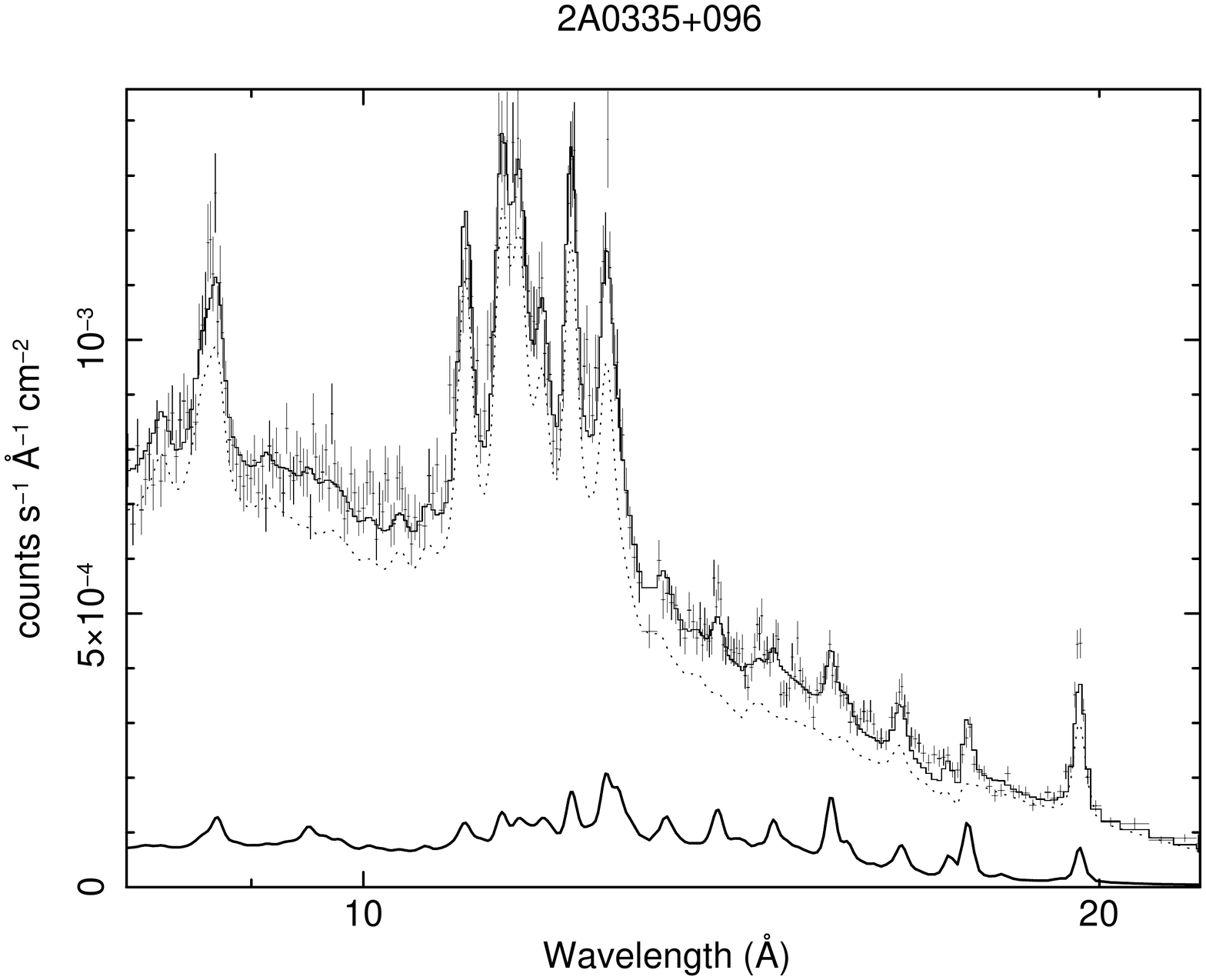}
    \hspace{-0.85cm}\includegraphics[width=0.48\textwidth]
    {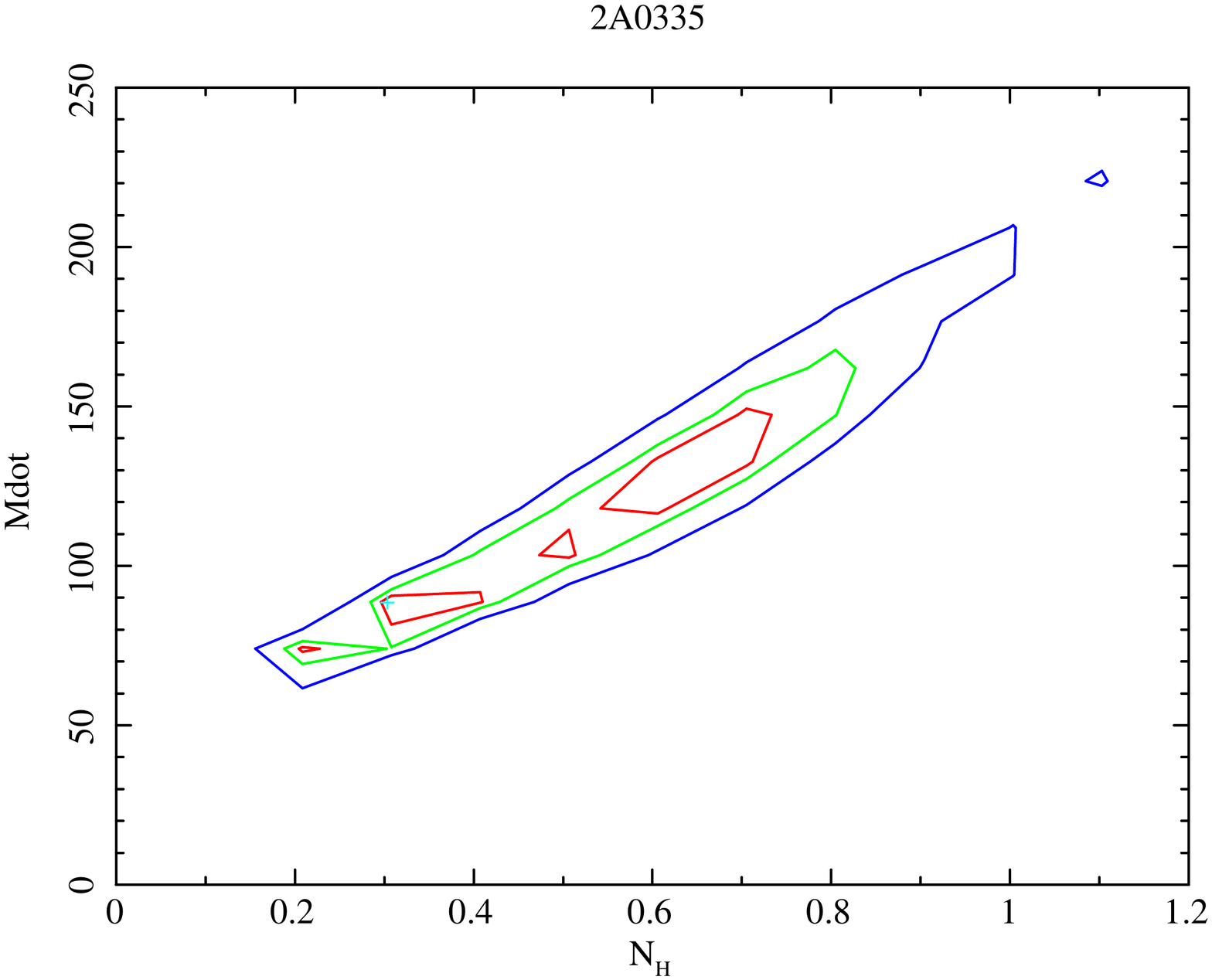}
      \hspace{-0.85cm} \includegraphics[width=0.48\textwidth]{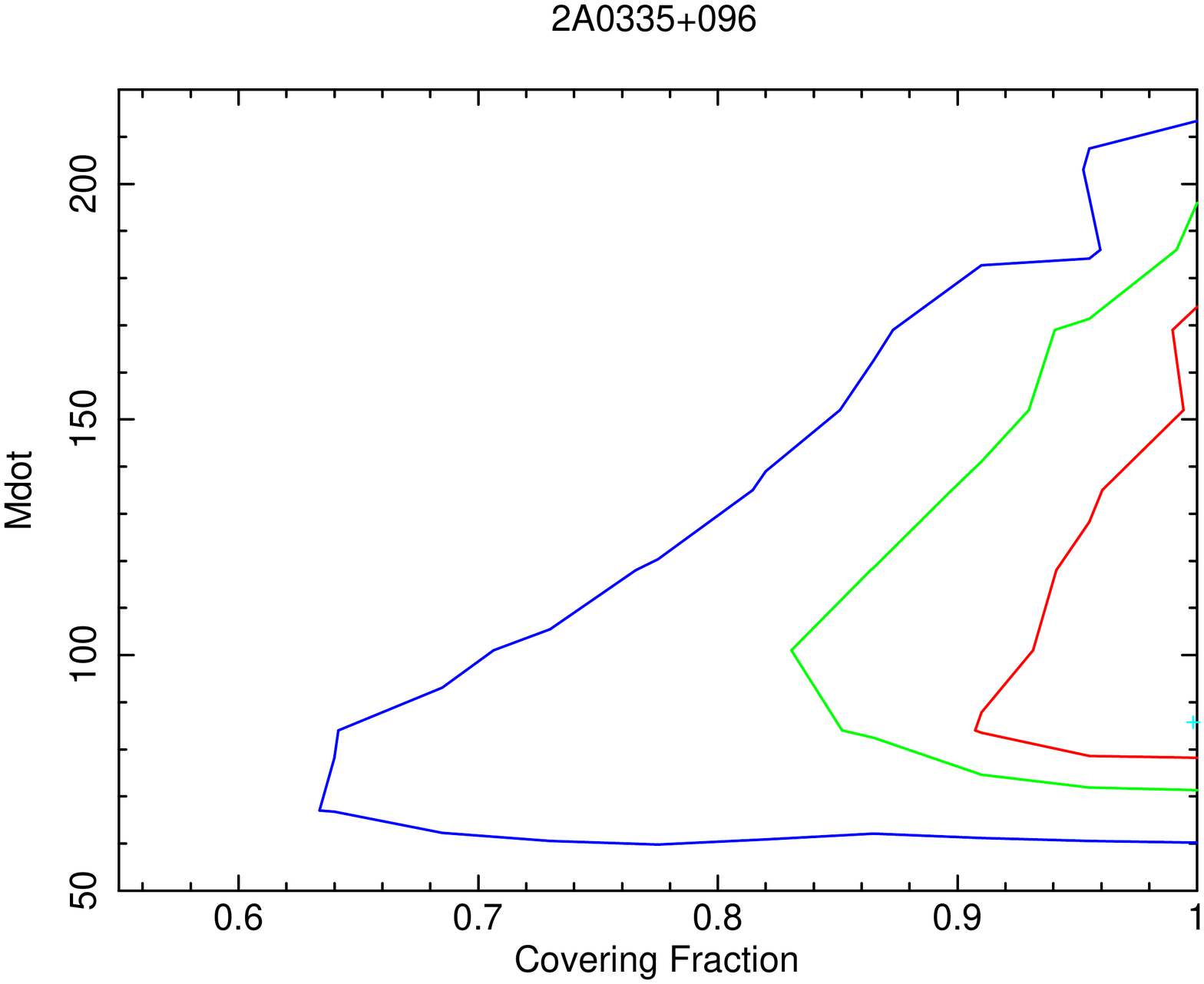}
    \caption{Top to Bottom: RGS spectrum of 2A0335+096 with HCF component shown in red and \textsc{mkcflow} component dotted, Mass cooling rate in $\Msunpyr$ versus  total column density in units of $10^{22}\cmsq$, Mass cooling rate versus Covering Fraction of the HCF component. Contours at 68\% (red), 90\% (green) and 99\% (blue). }
\end{figure}

\begin{figure}
    \centering    \includegraphics[width=0.48\textwidth]{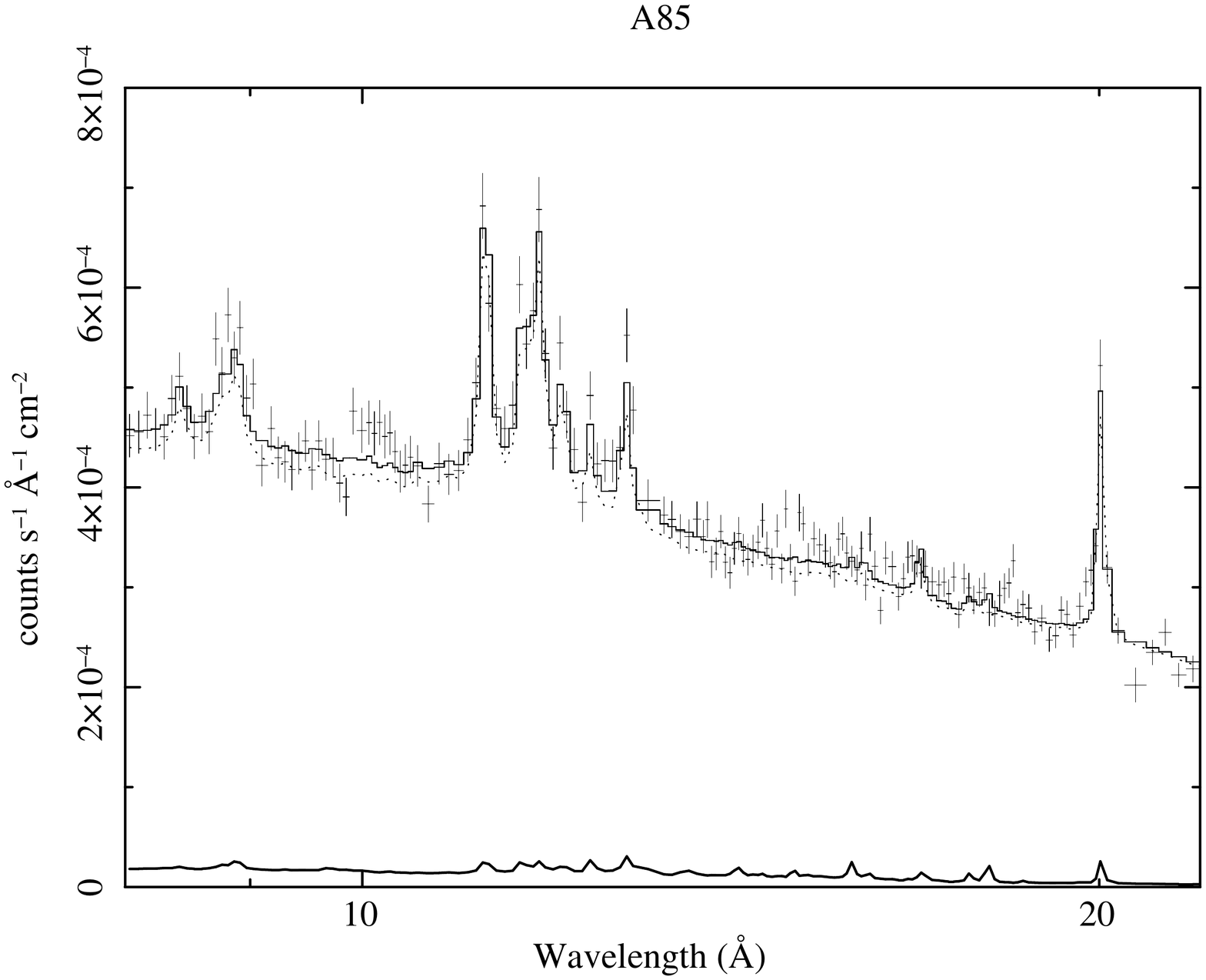}
    \hspace{-0.85cm}\includegraphics[width=0.48\textwidth]
    {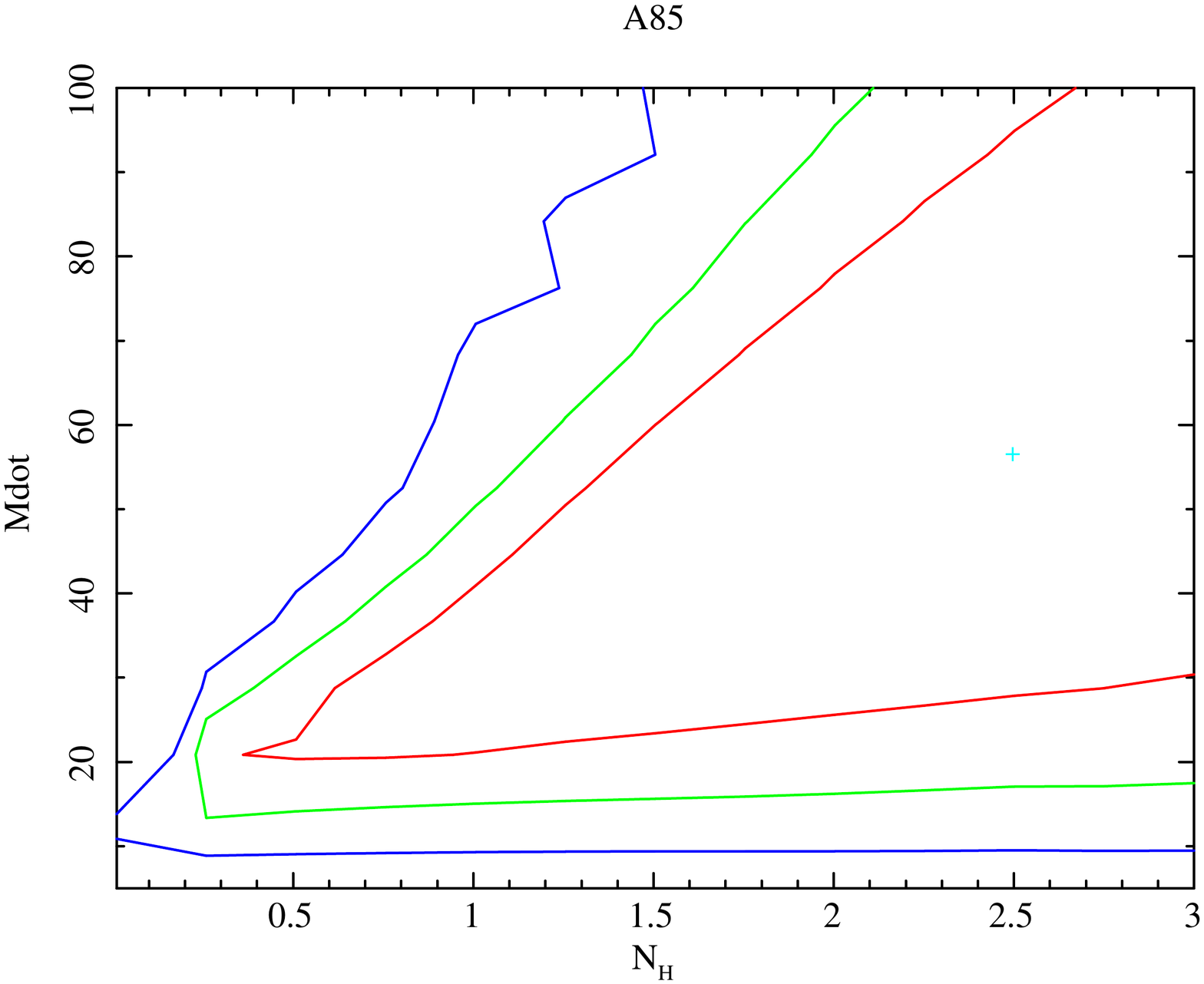}
      \hspace{-0.85cm} \includegraphics[width=0.48\textwidth]{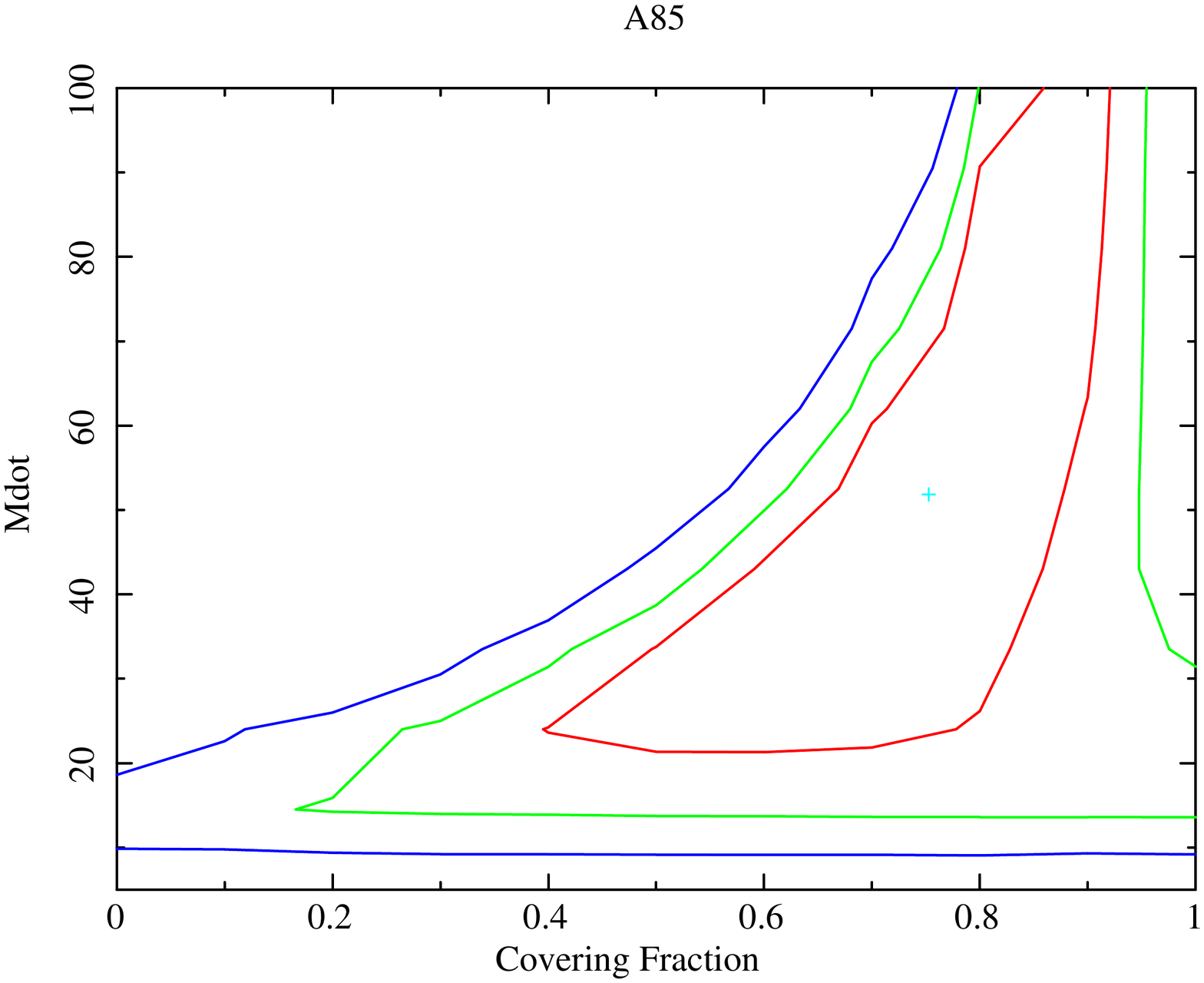}
    \caption{A85, with details as in Fig B2.}
\end{figure}

\begin{figure}
    \centering    \includegraphics[width=0.48\textwidth]{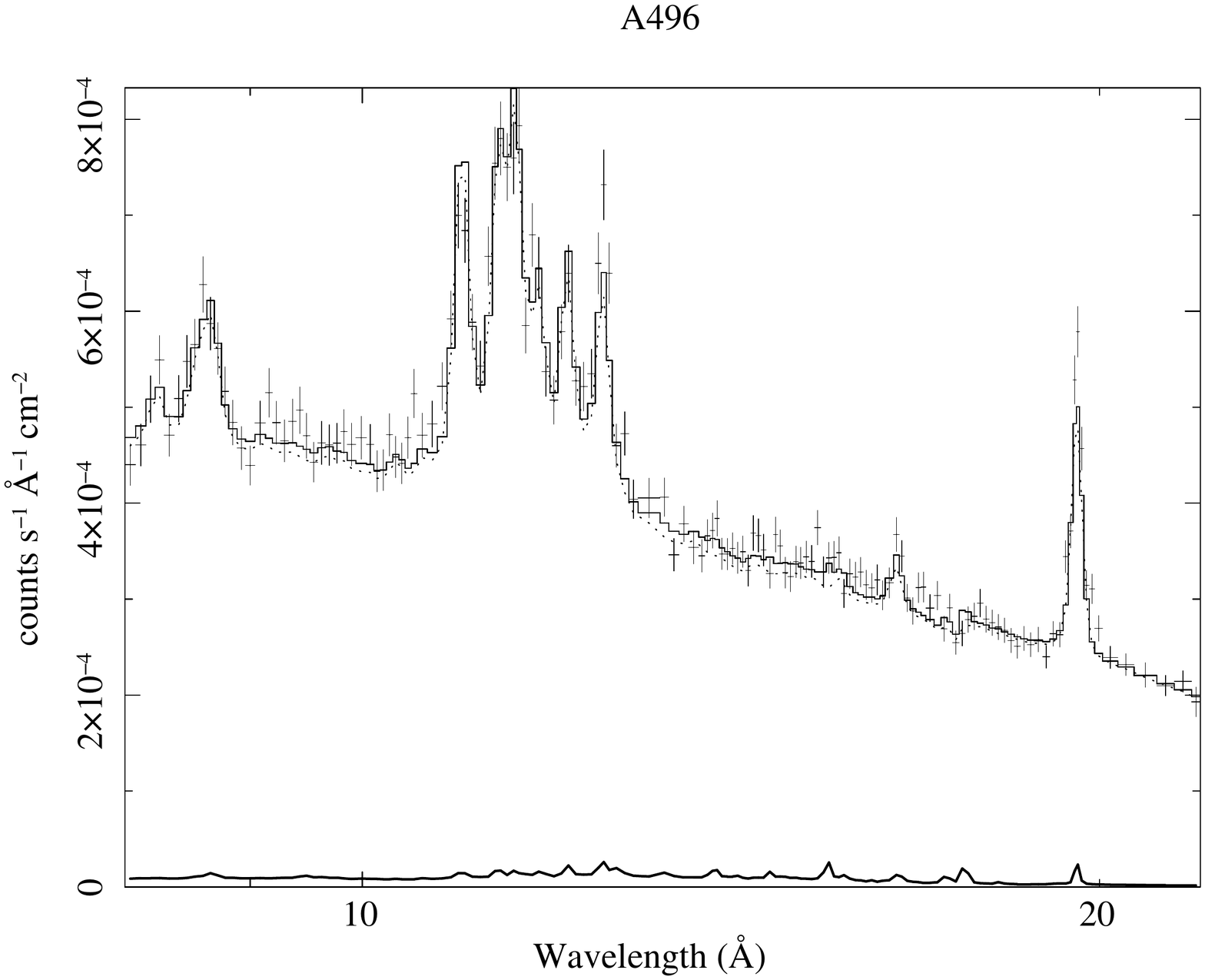}
    \hspace{-0.85cm}\includegraphics[width=0.48\textwidth]
    {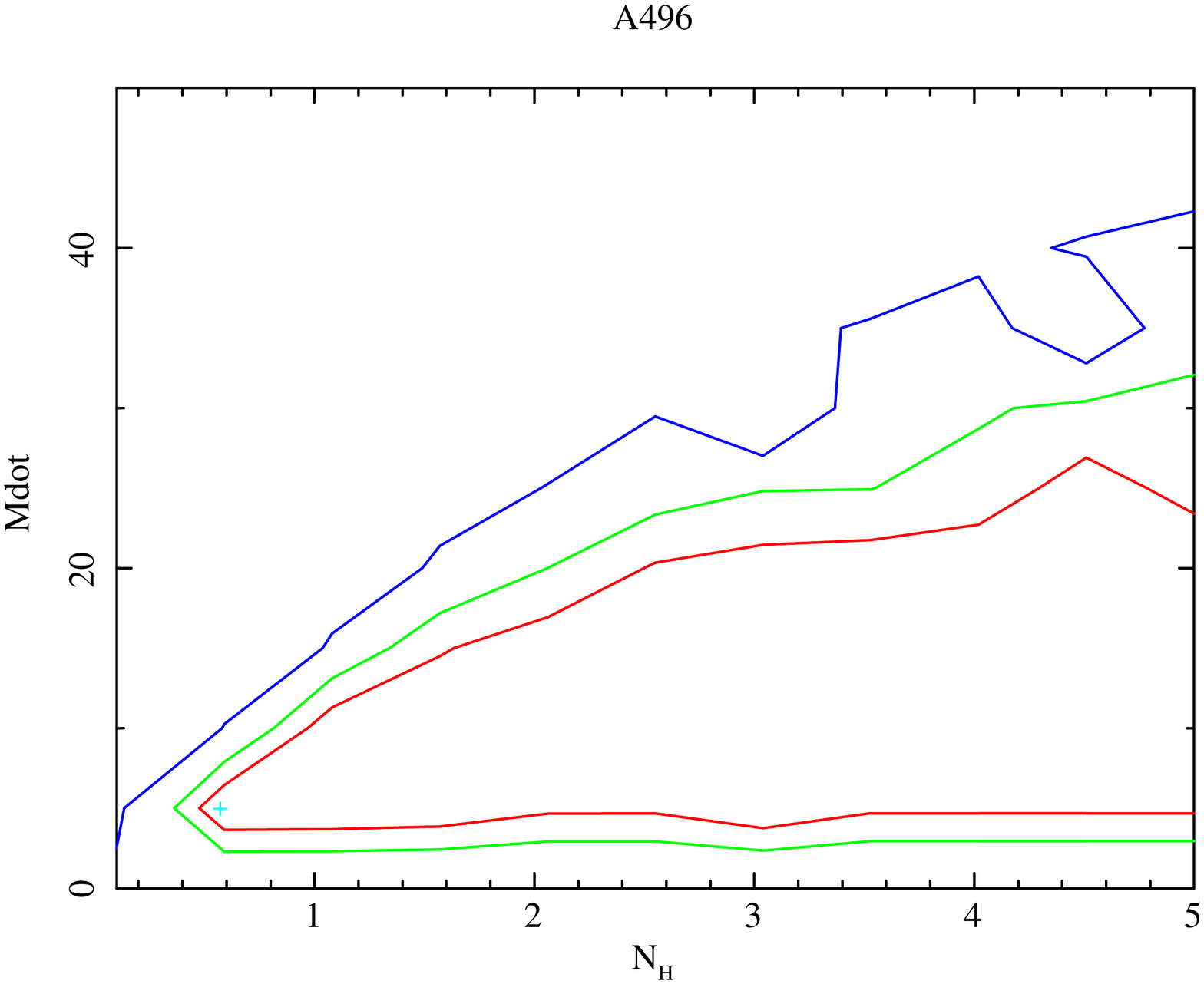}
      \hspace{-0.85cm} \includegraphics[width=0.48\textwidth]{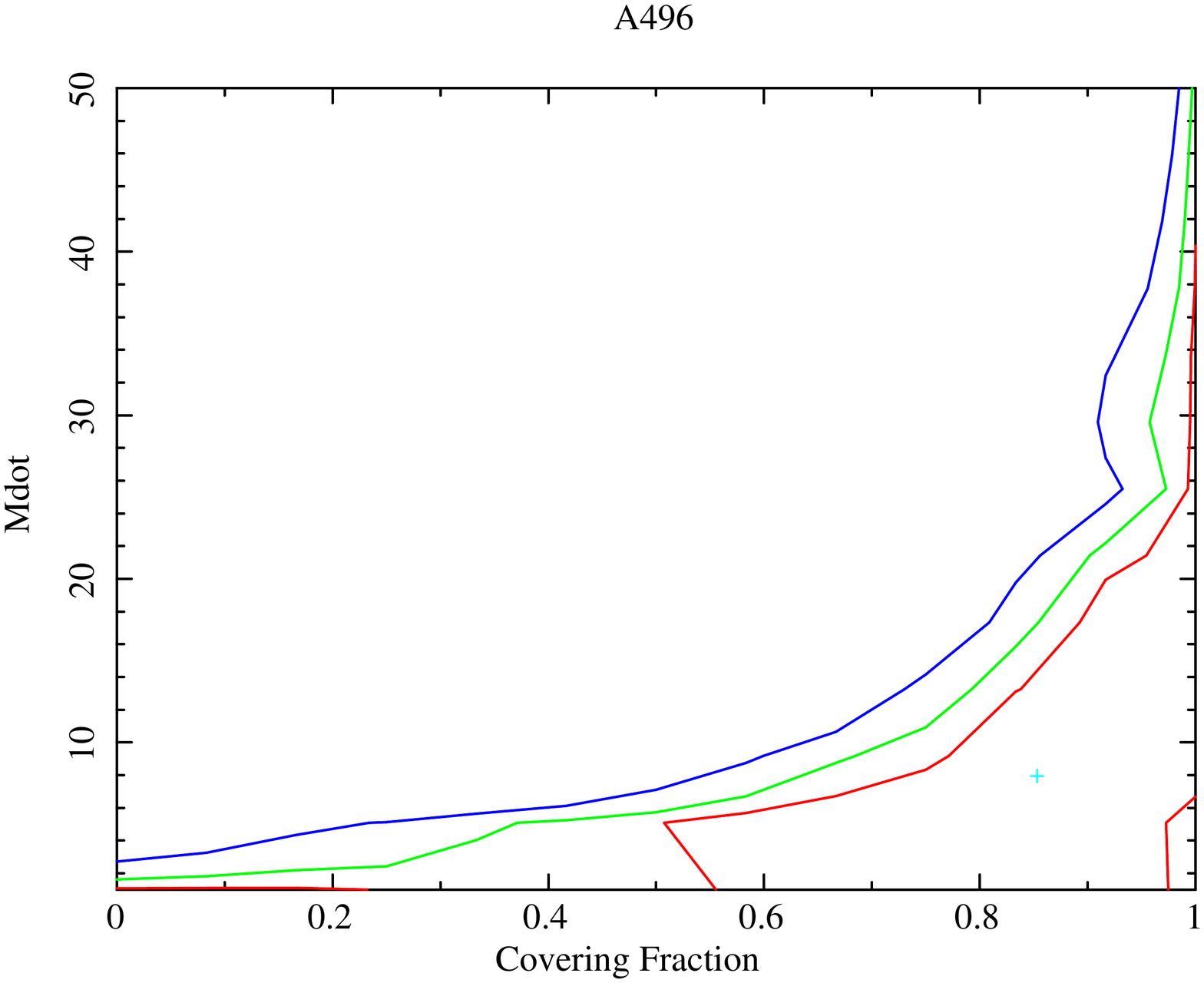}
    \caption{A496,  with details as in Fig B2.}
\end{figure}

\begin{figure}
    \centering    \includegraphics[width=0.48\textwidth]{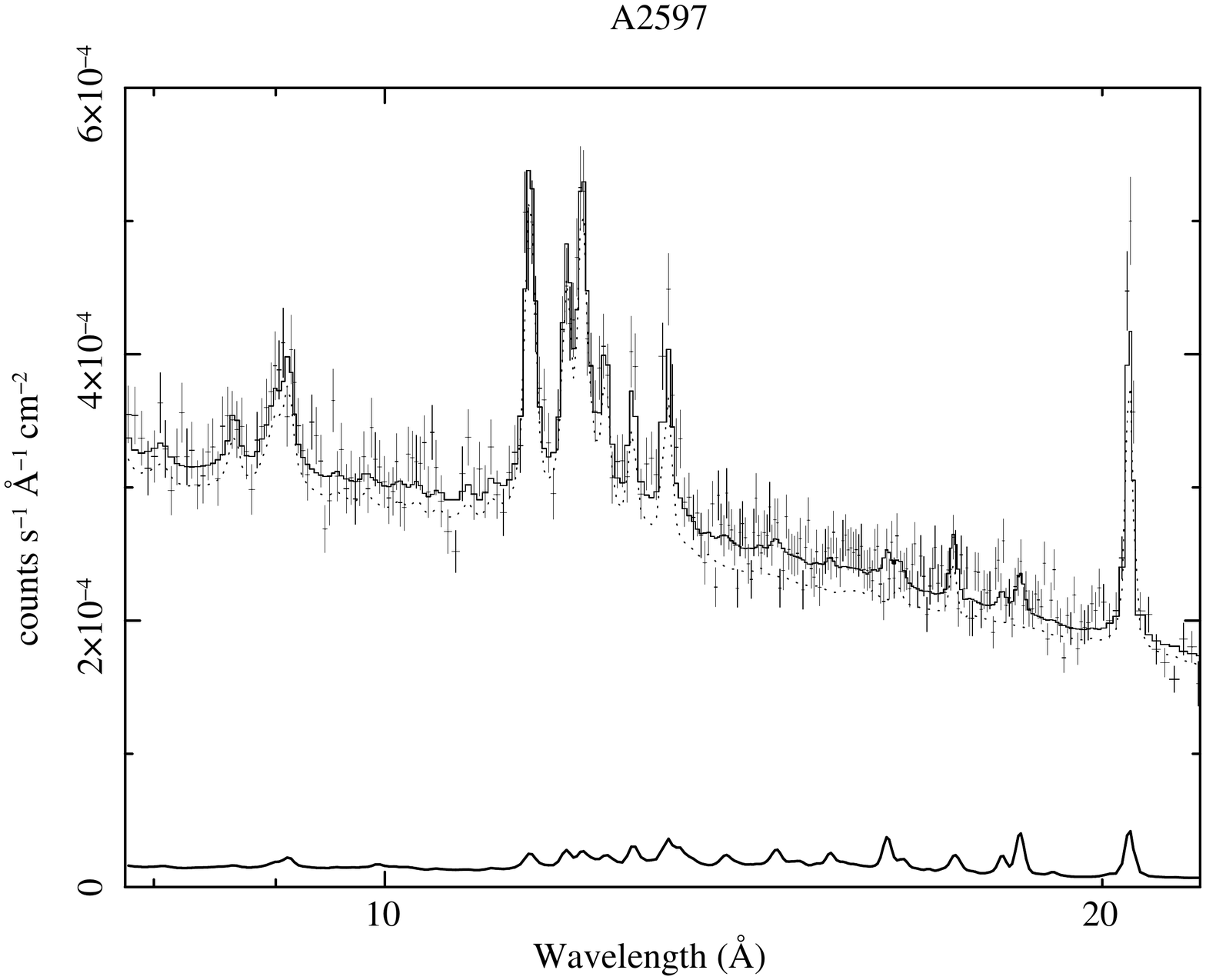}
   \hspace{-0.85cm}\includegraphics[width=0.48\textwidth]   {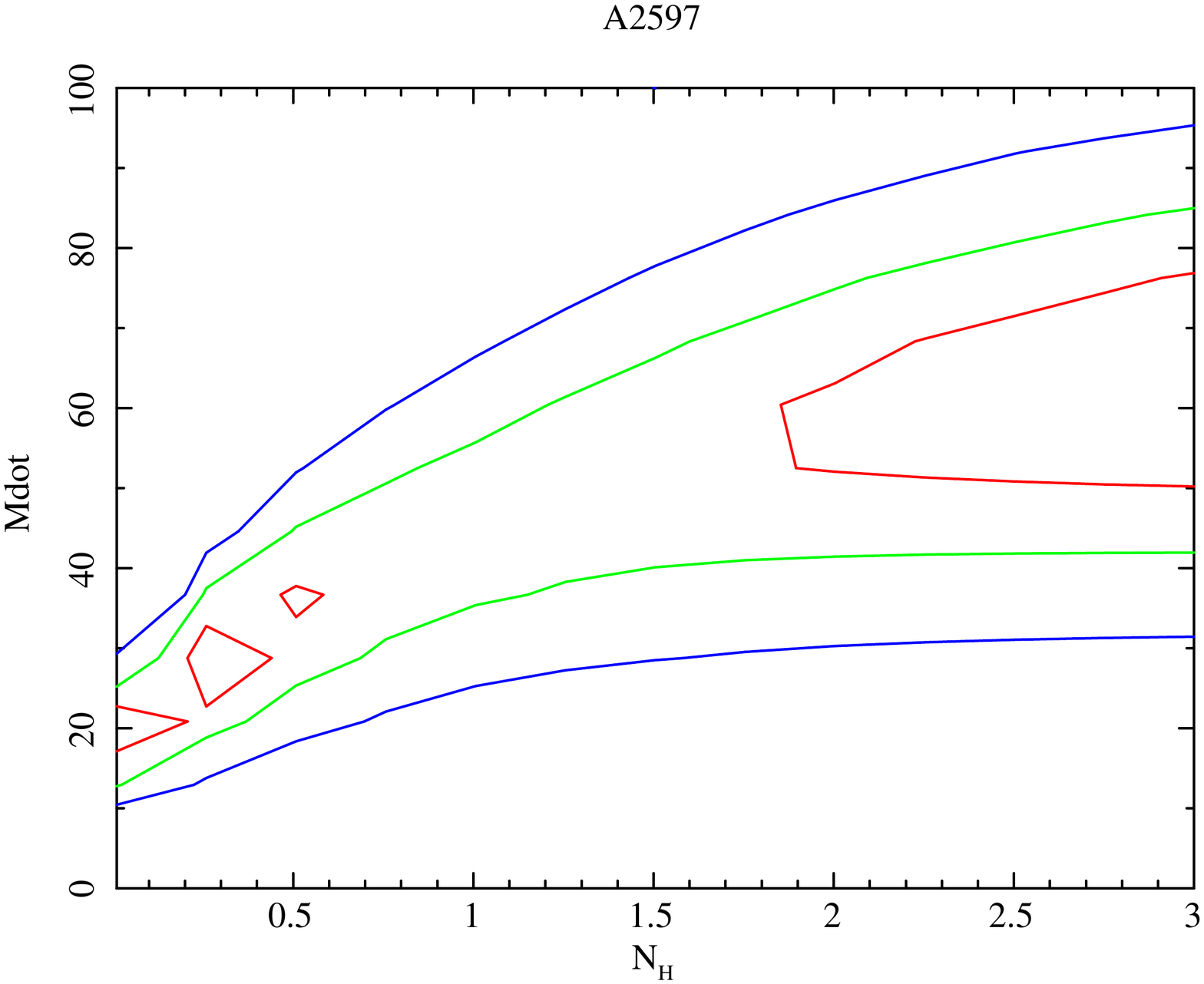}
   \hspace{-0.85cm} \includegraphics[width=0.48\textwidth]{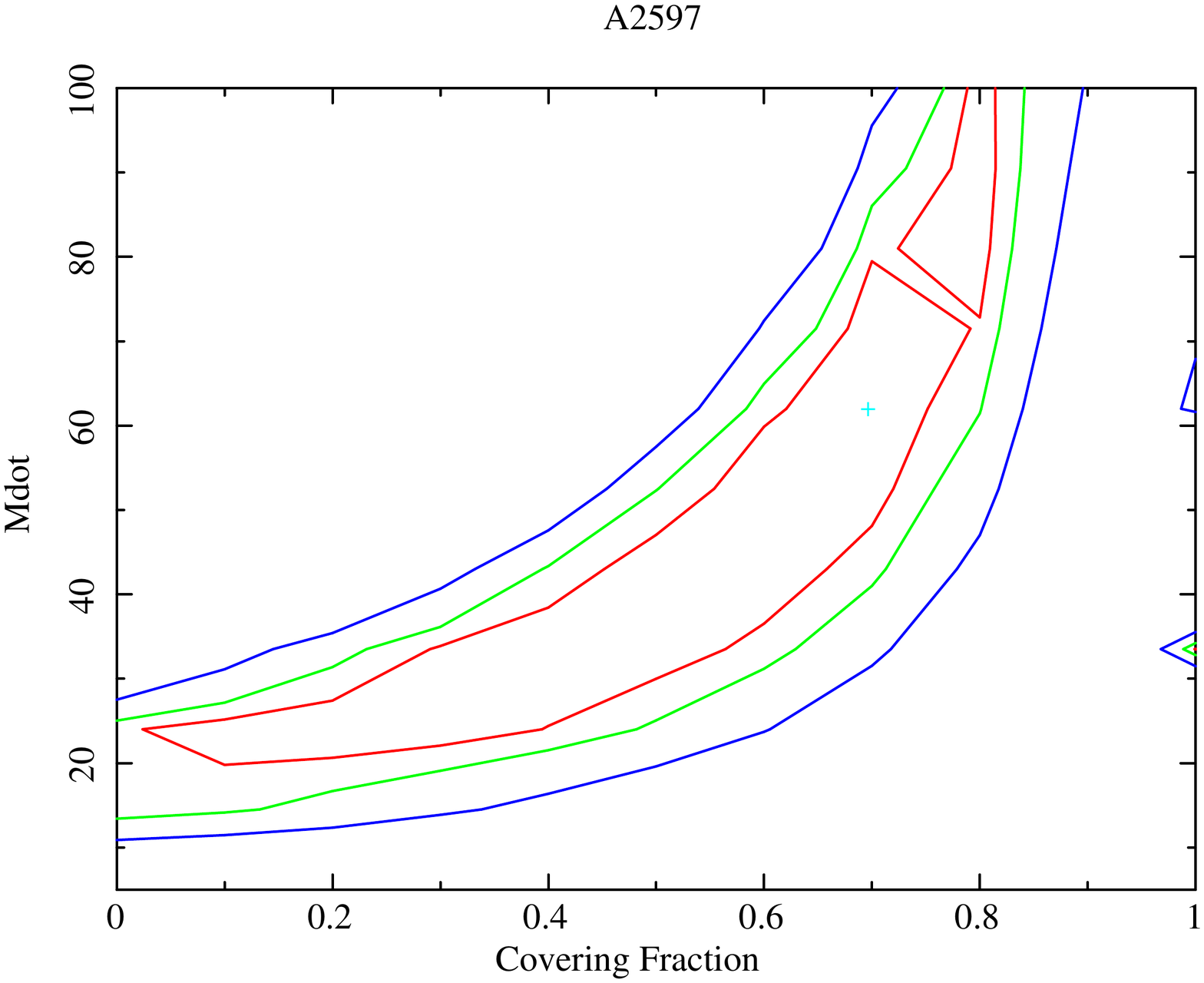}
    \caption{A2597,  with details as in Fig B2.}
\end{figure}

\begin{figure}
    \centering    \includegraphics[width=0.48\textwidth]{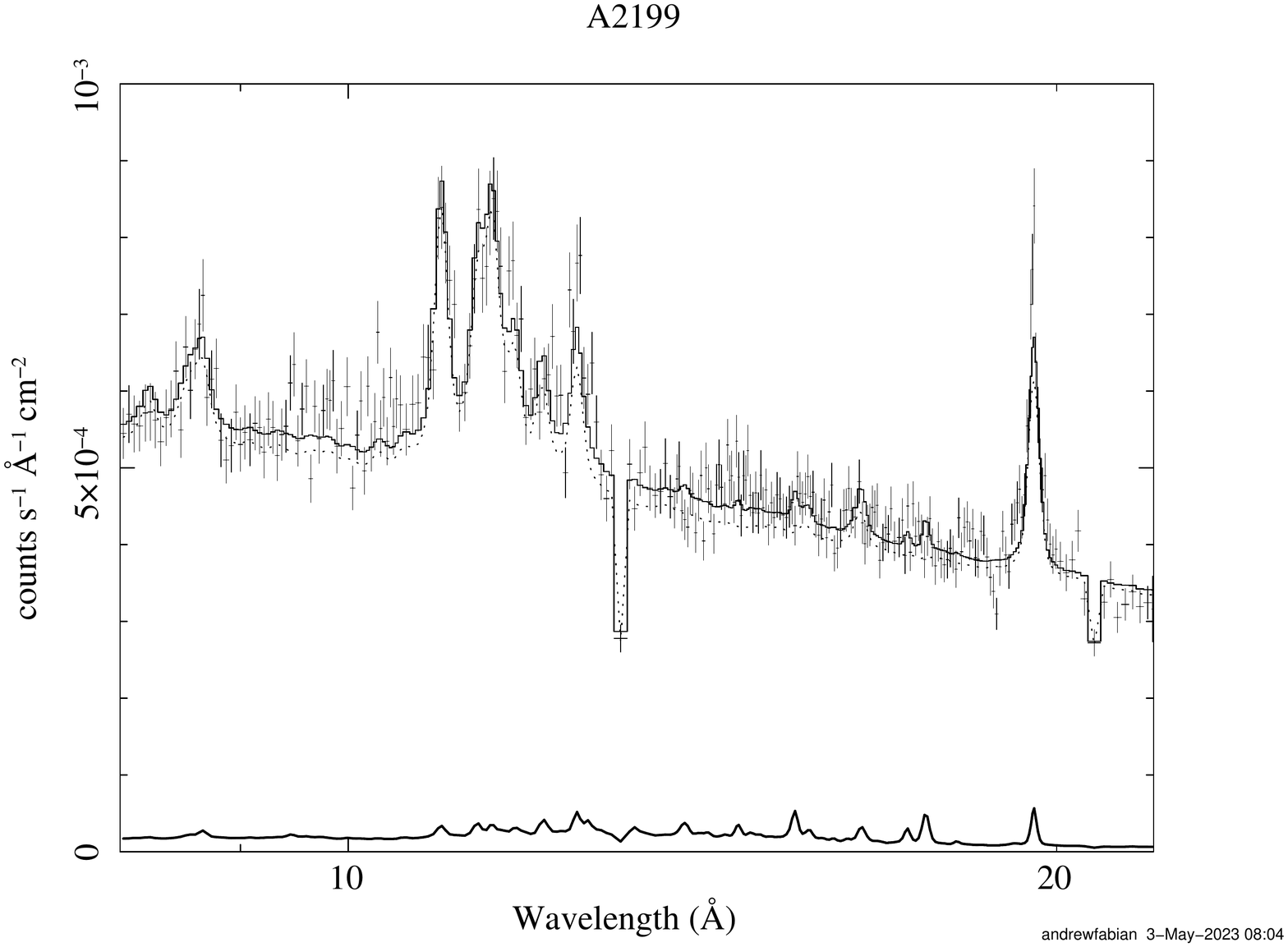}
    \hspace{-0.85cm}\includegraphics[width=0.48\textwidth]
    {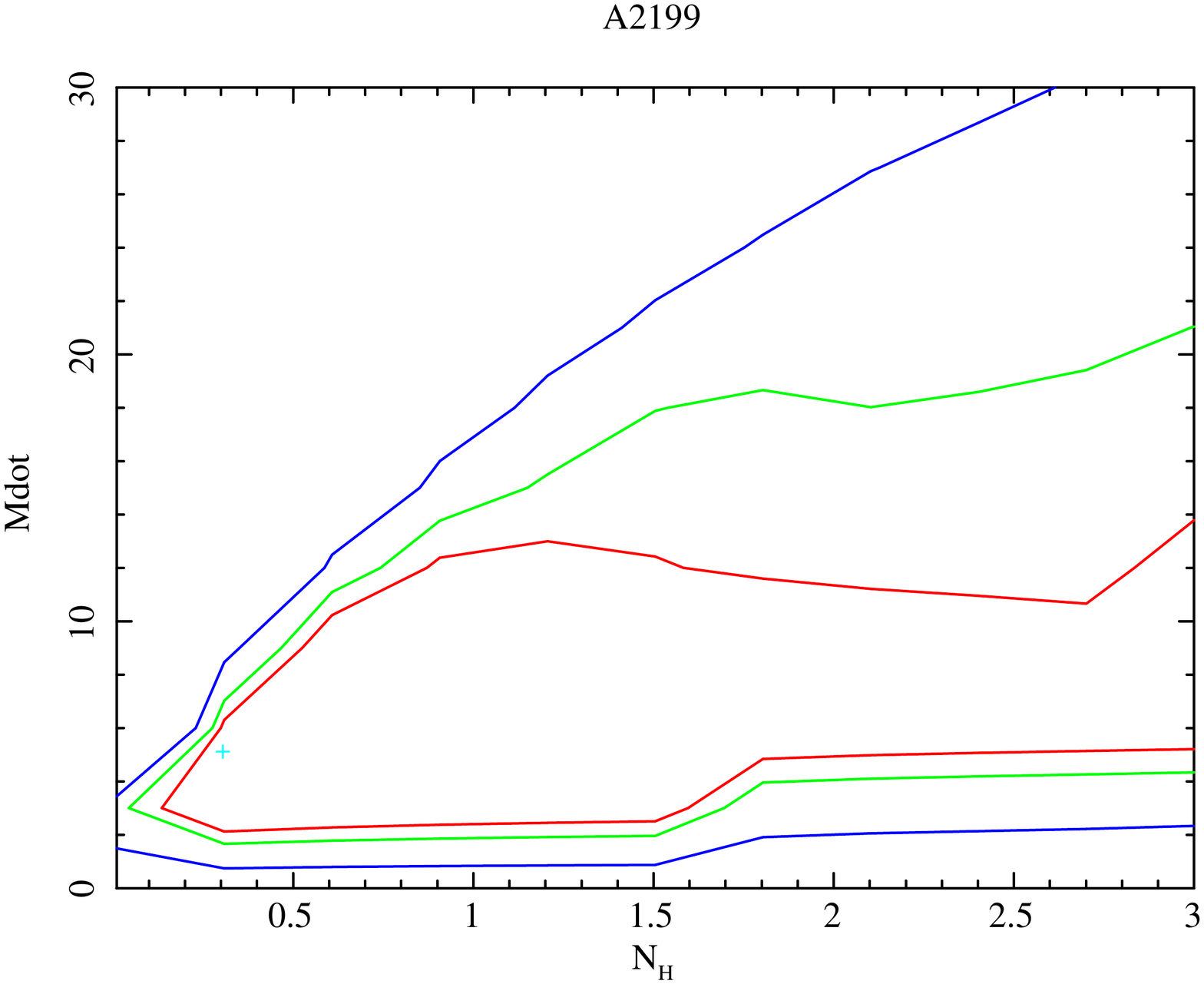}
      \hspace{-0.85cm} \includegraphics[width=0.48\textwidth]{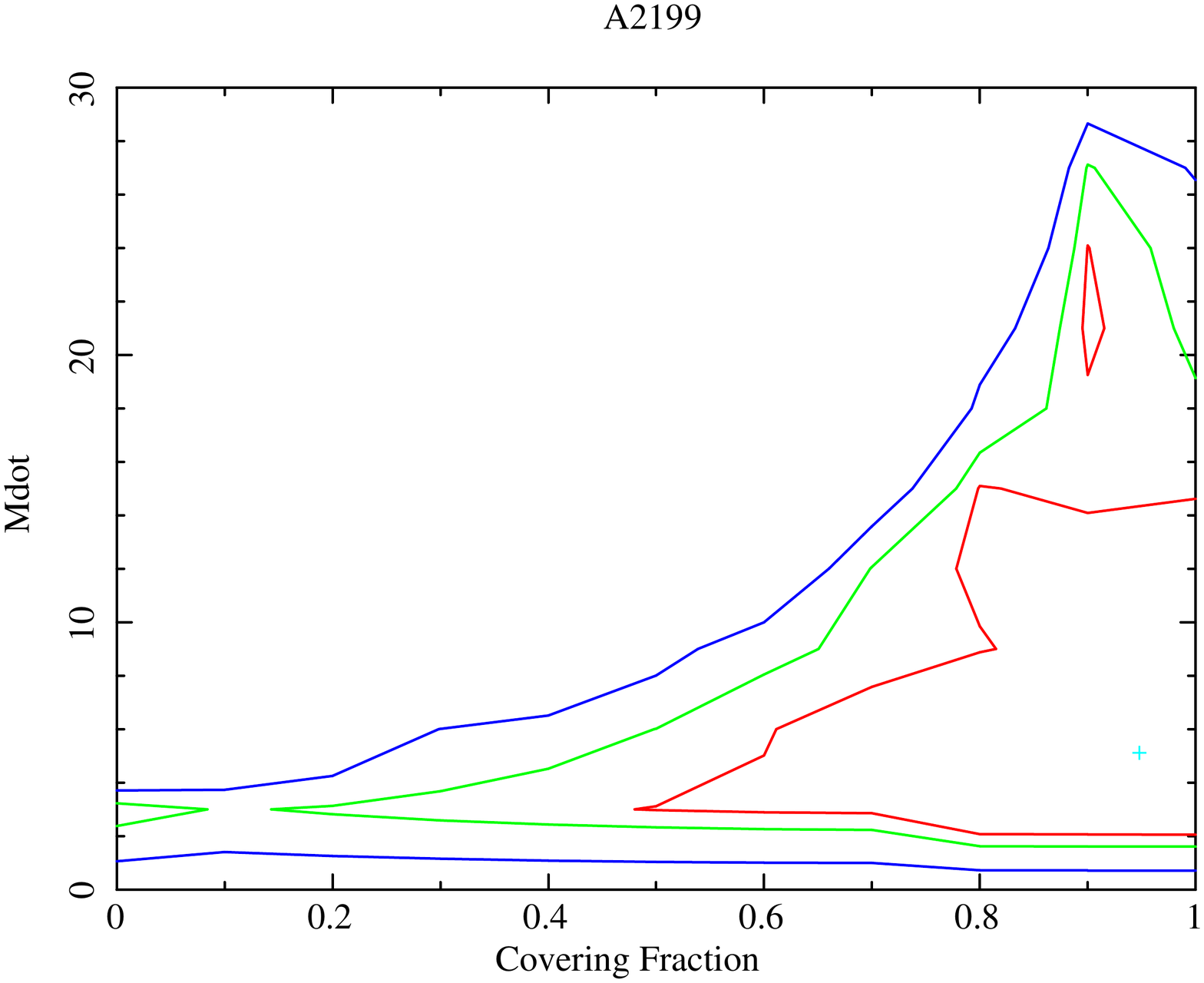}
    \caption{ A2199,  with details as in Fig B2.}
\end{figure}

\begin{figure}
    \centering    \includegraphics[width=0.48\textwidth]{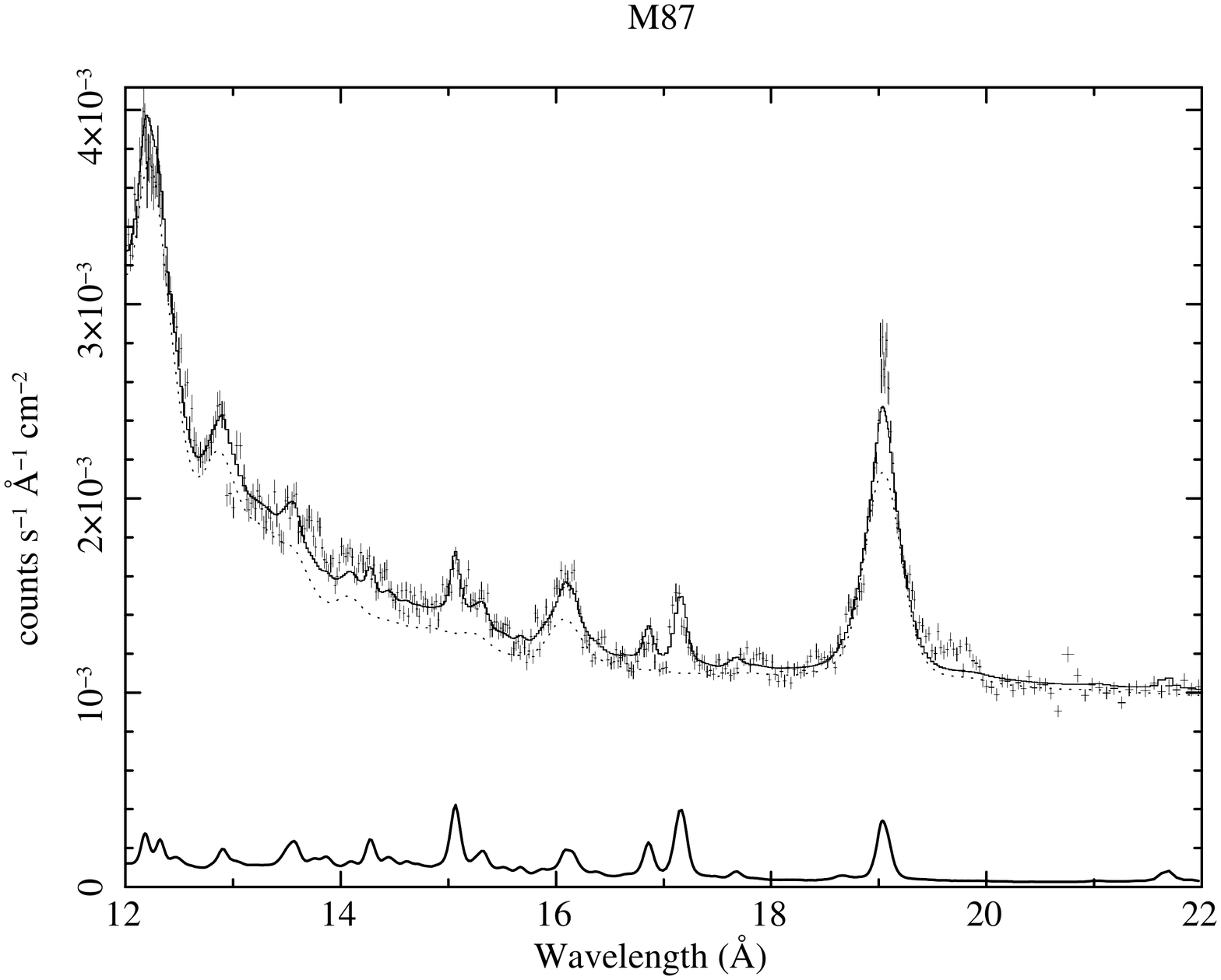}
    \hspace{-0.85cm}\includegraphics[width=0.48\textwidth]
    {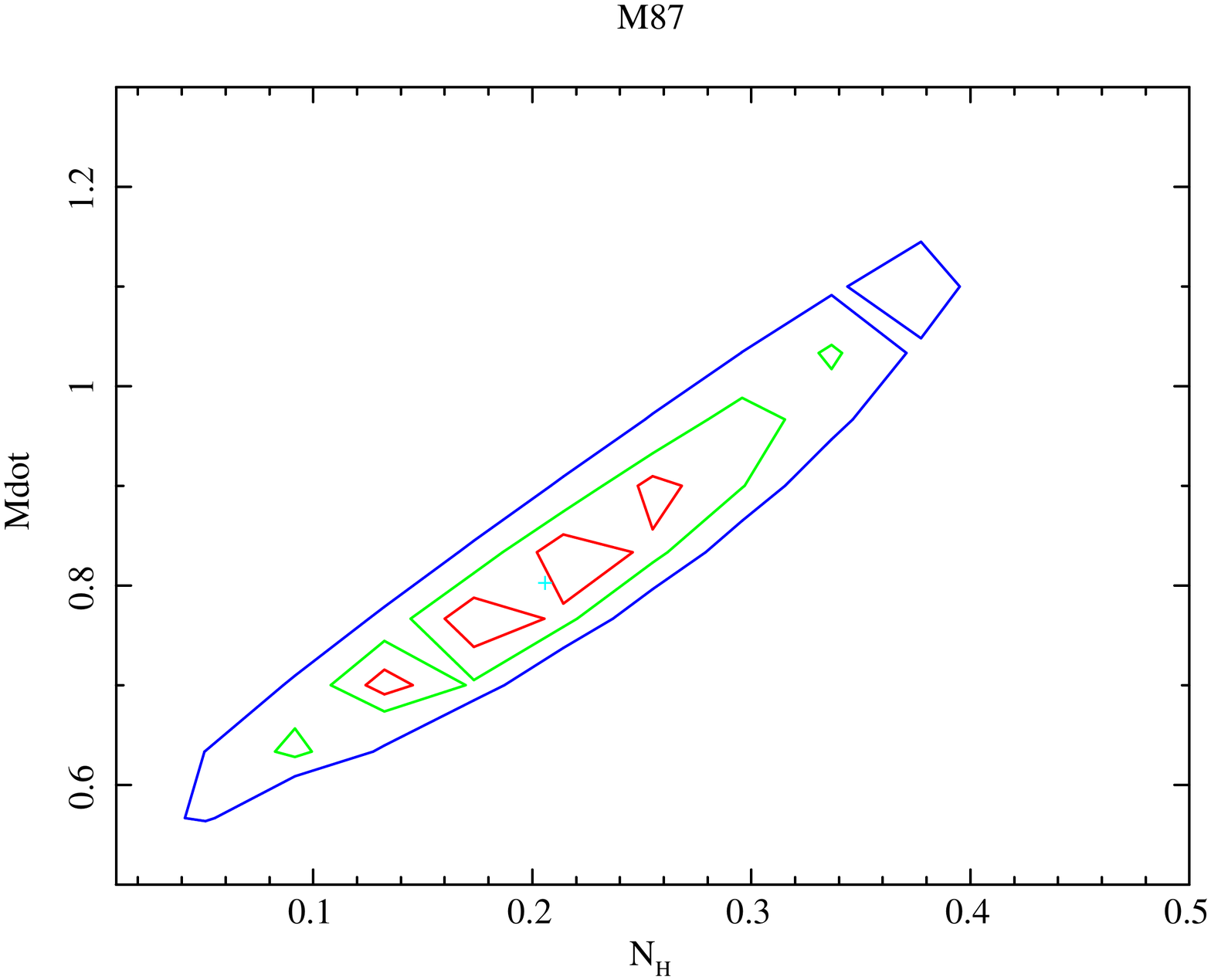}
      \hspace{-0.85cm} \includegraphics[width=0.48\textwidth]{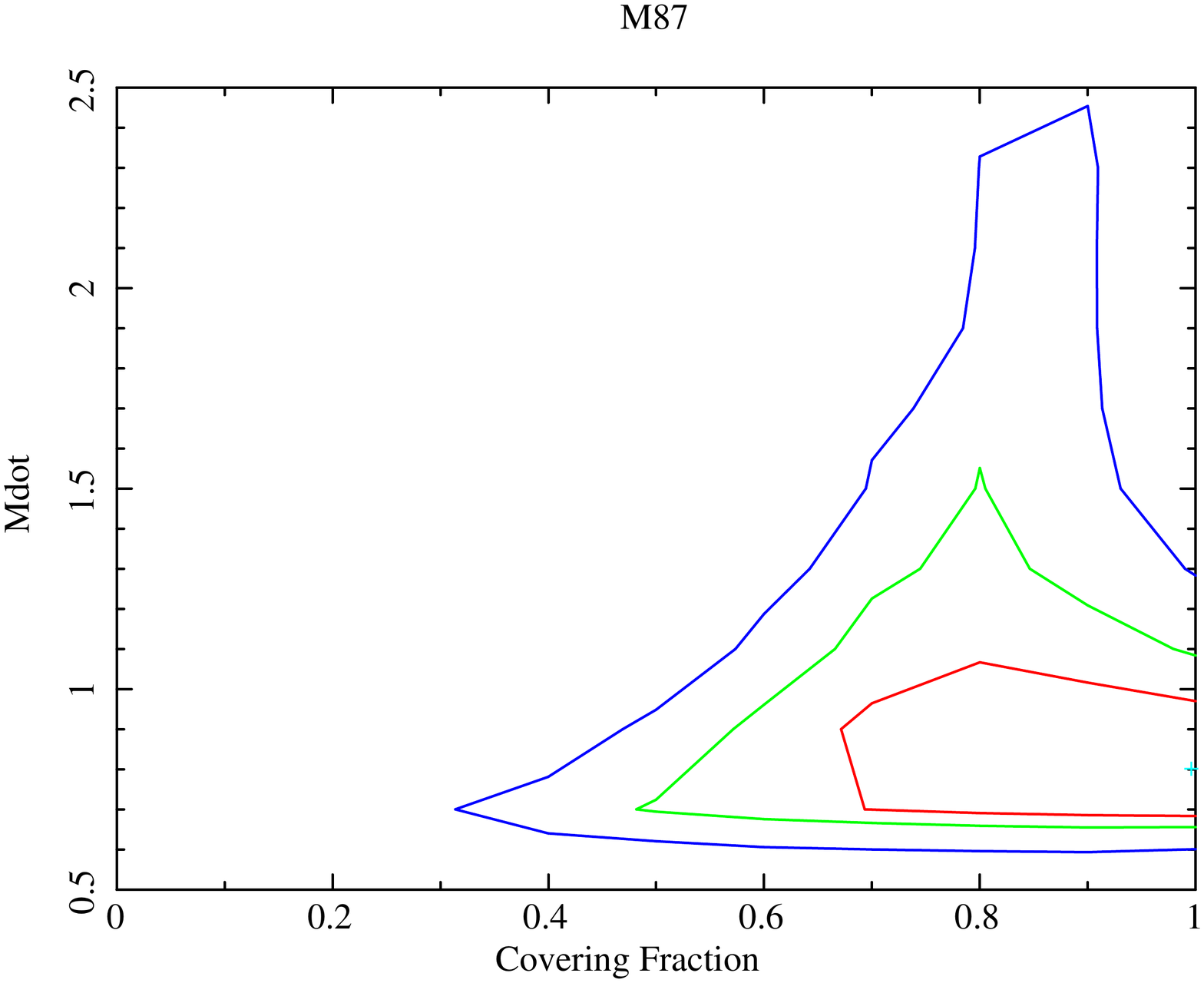}
    \caption{M87,  with details as in Fig B2.}
\end{figure}

\begin{figure}
    \centering    \includegraphics[width=0.48\textwidth]{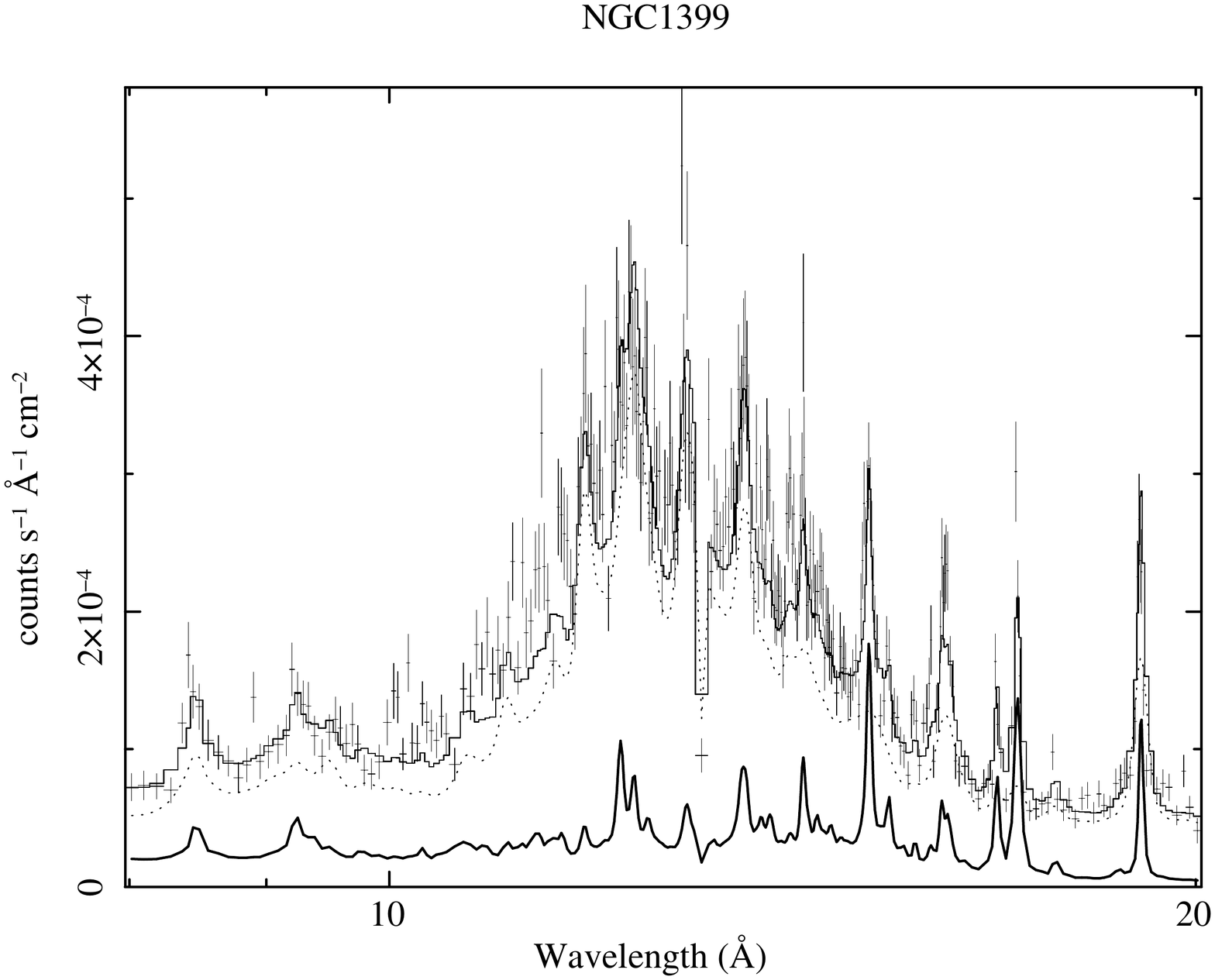}
    \hspace{-0.85cm}\includegraphics[width=0.48\textwidth]
    {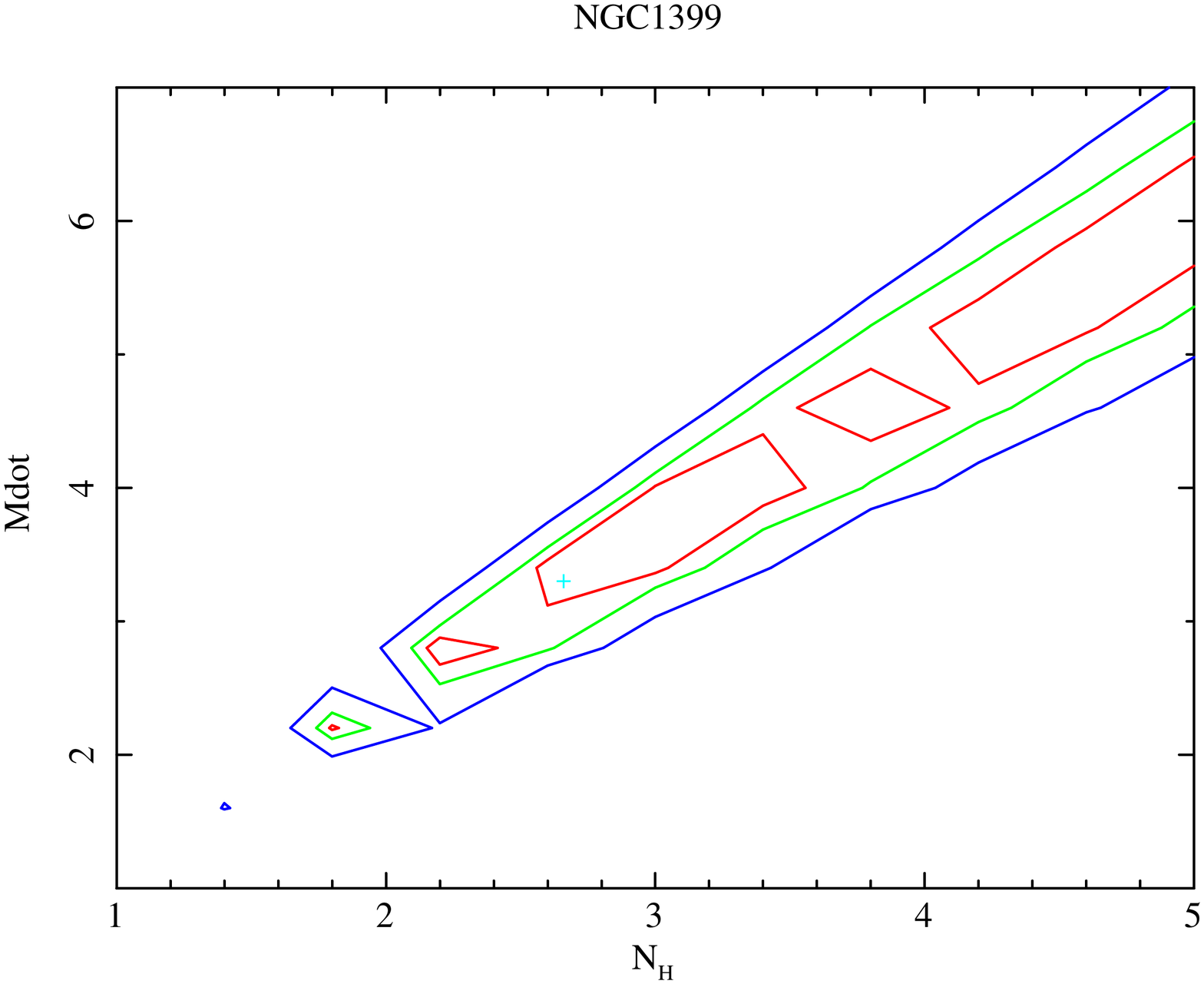}
      \hspace{-0.85cm} \includegraphics[width=0.48\textwidth]{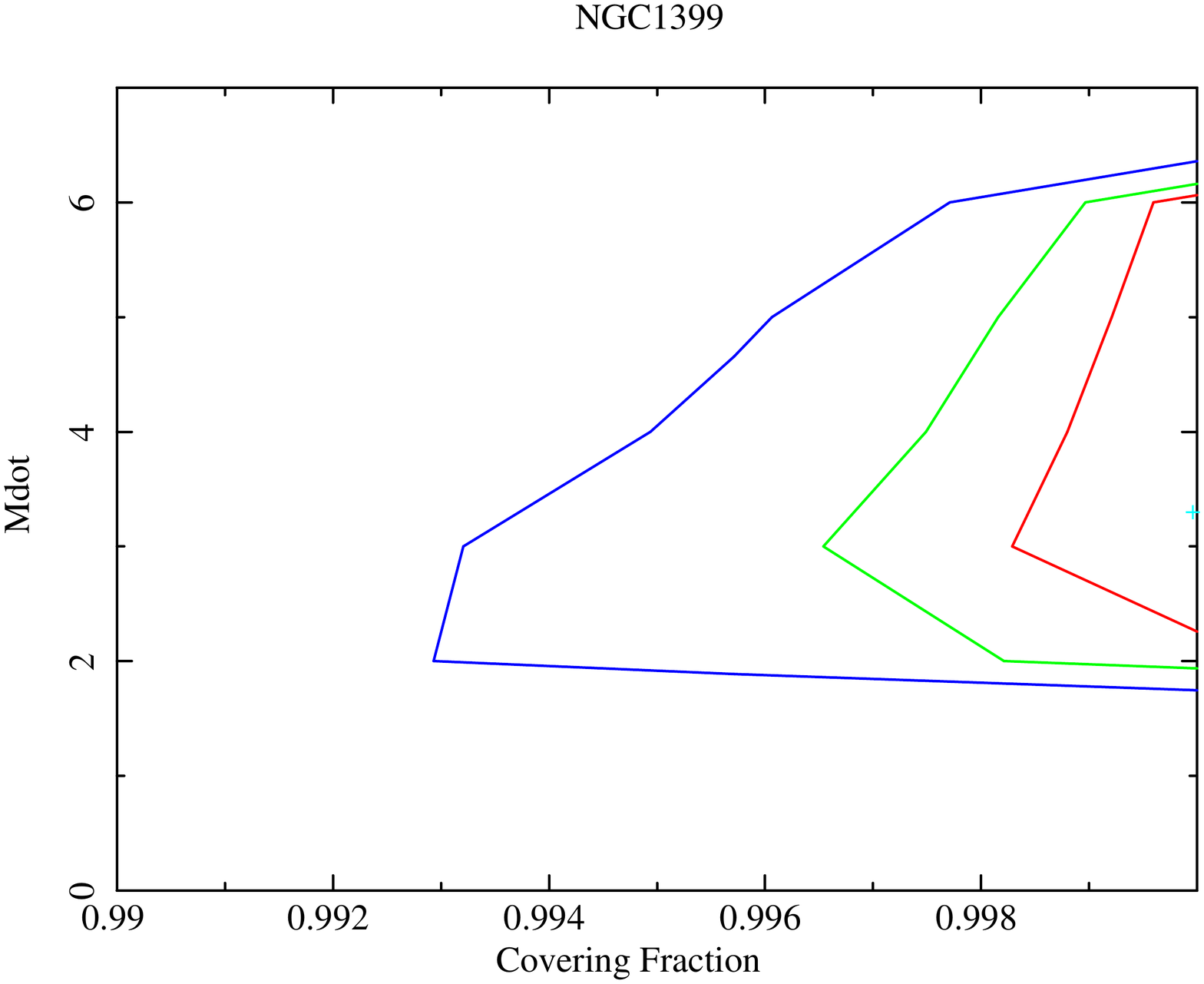}
    \caption{NGC1399,  with details as in Fig B2.}
\end{figure}

\begin{figure}
    \centering    \includegraphics[width=0.48\textwidth]{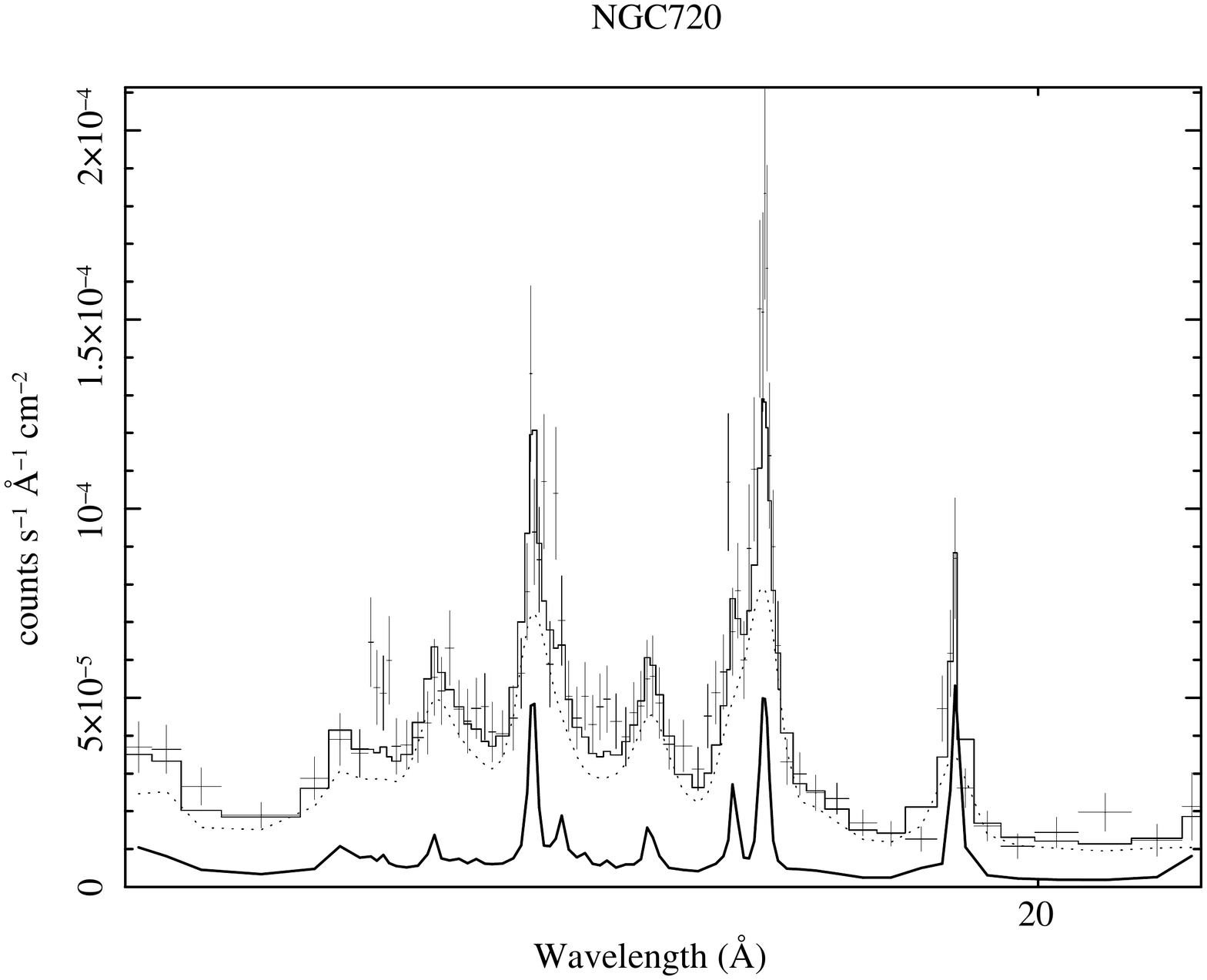}
   \hspace{-0.85cm}\includegraphics[width=0.48\textwidth]
  {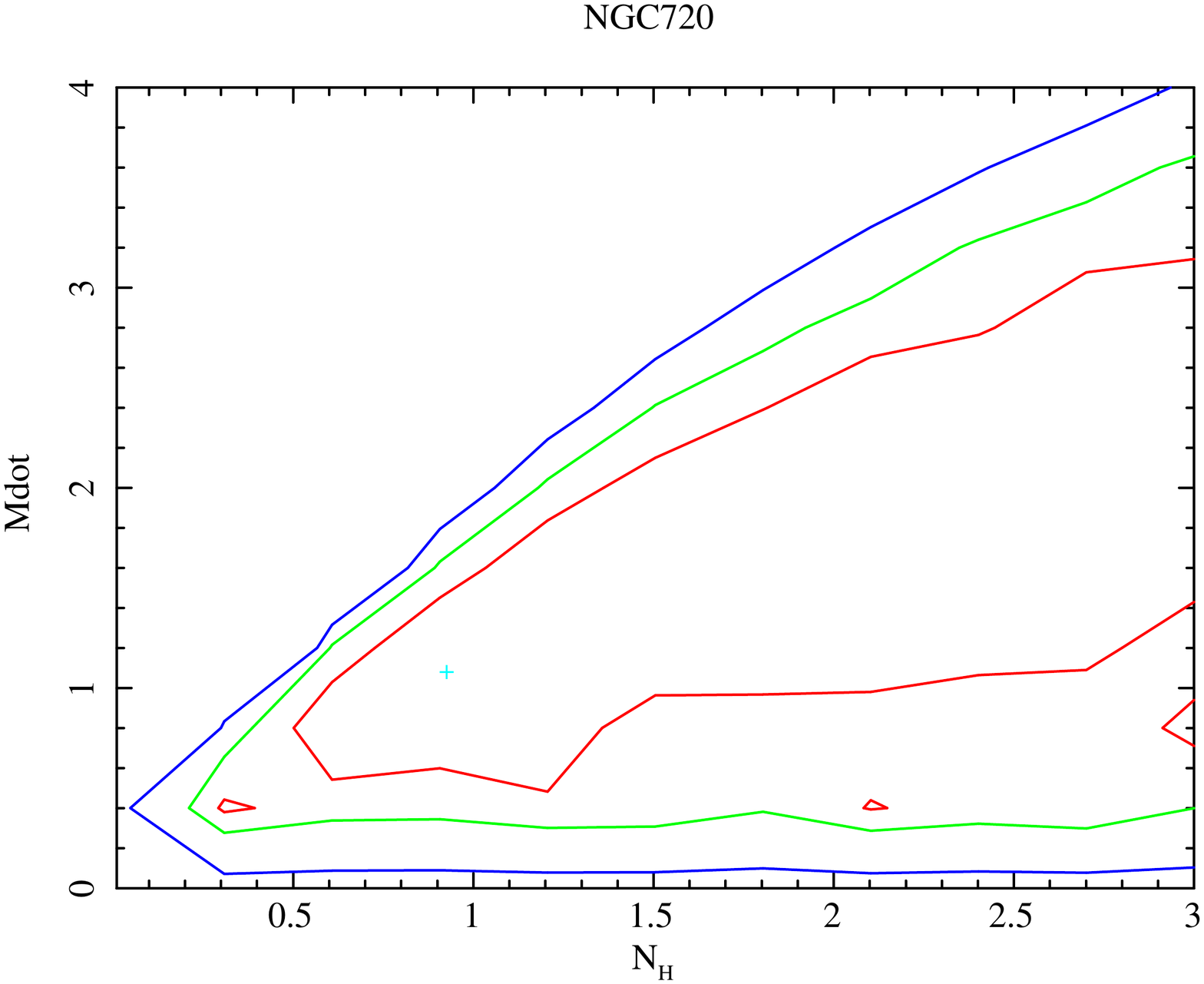}
    \hspace{-0.85cm} \includegraphics[width=0.48\textwidth]{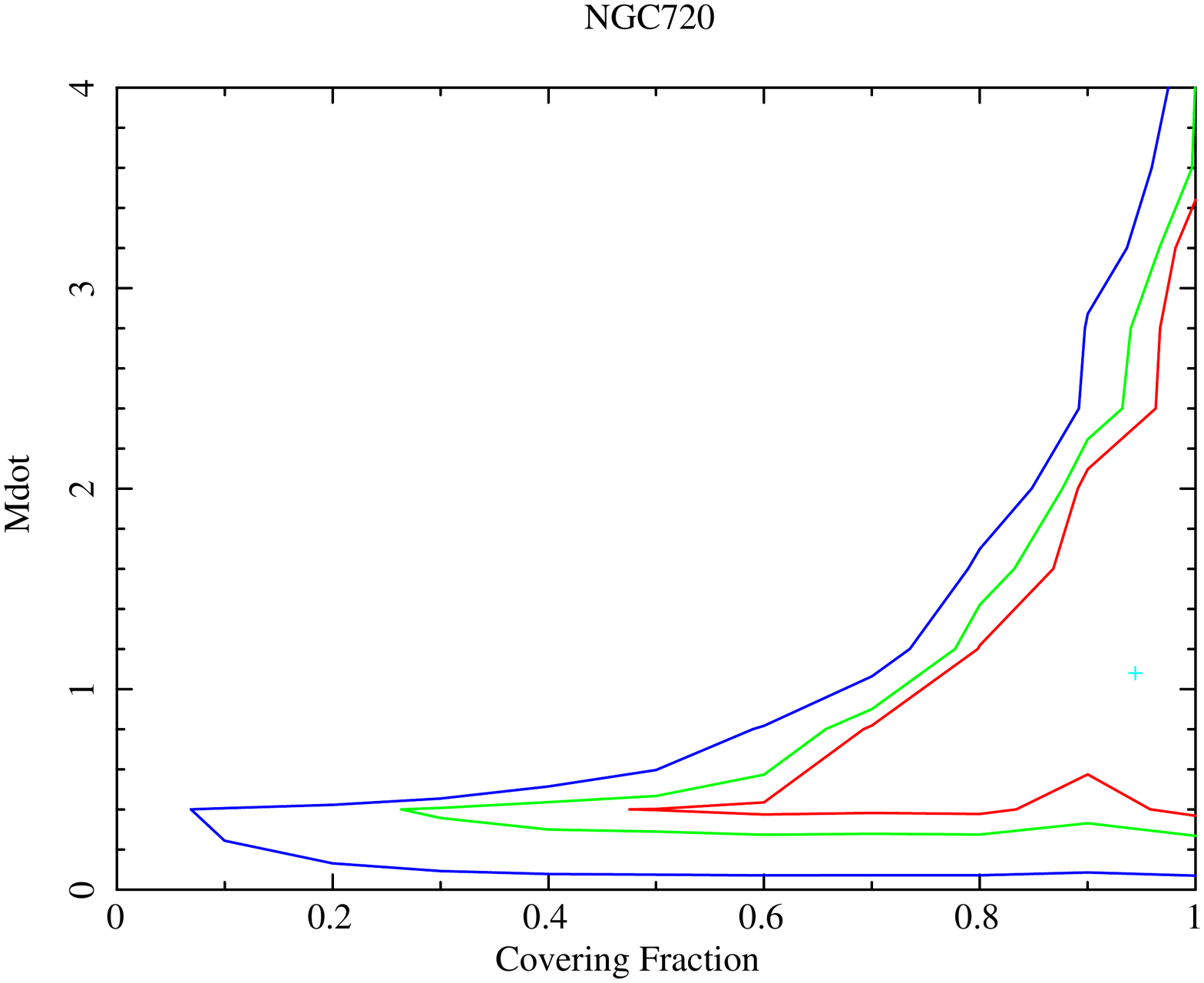}
    \caption{NGC720,  with details as in Fig B2.}
\end{figure}

\begin{figure}
    \centering    \includegraphics[width=0.48\textwidth]{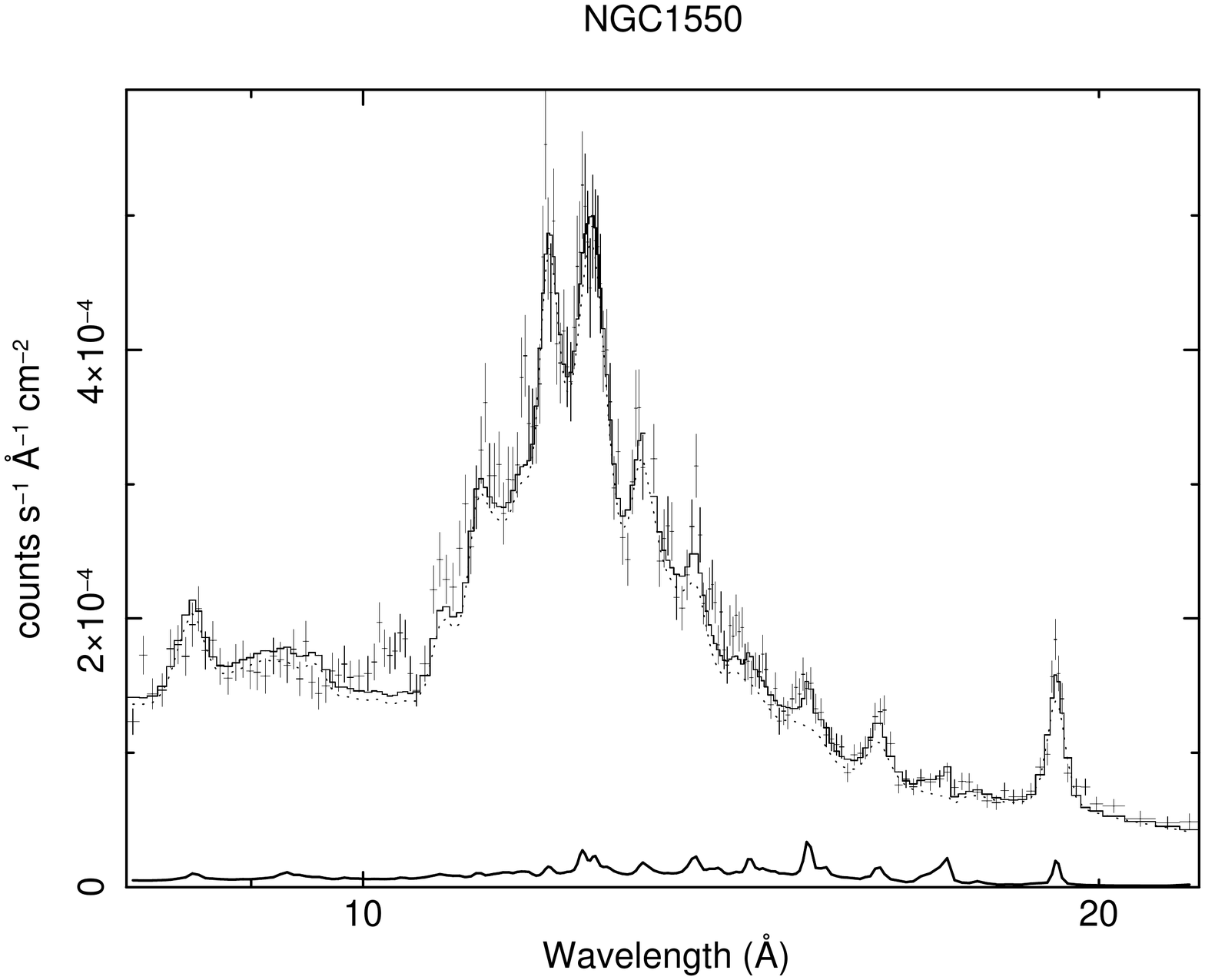}
    \hspace{-0.85cm}\includegraphics[width=0.48\textwidth]
    {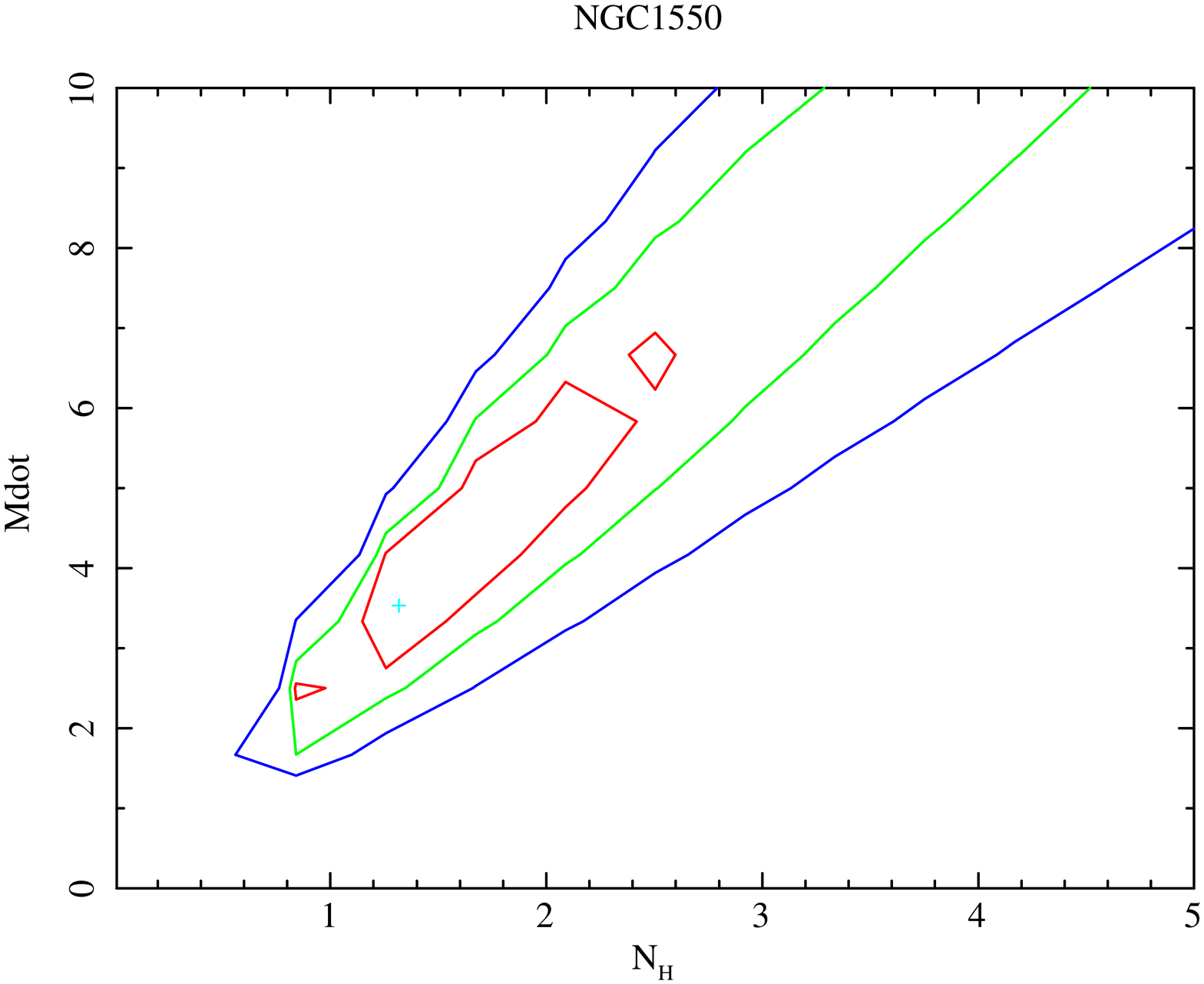}
      \hspace{-0.85cm} \includegraphics[width=0.48\textwidth]{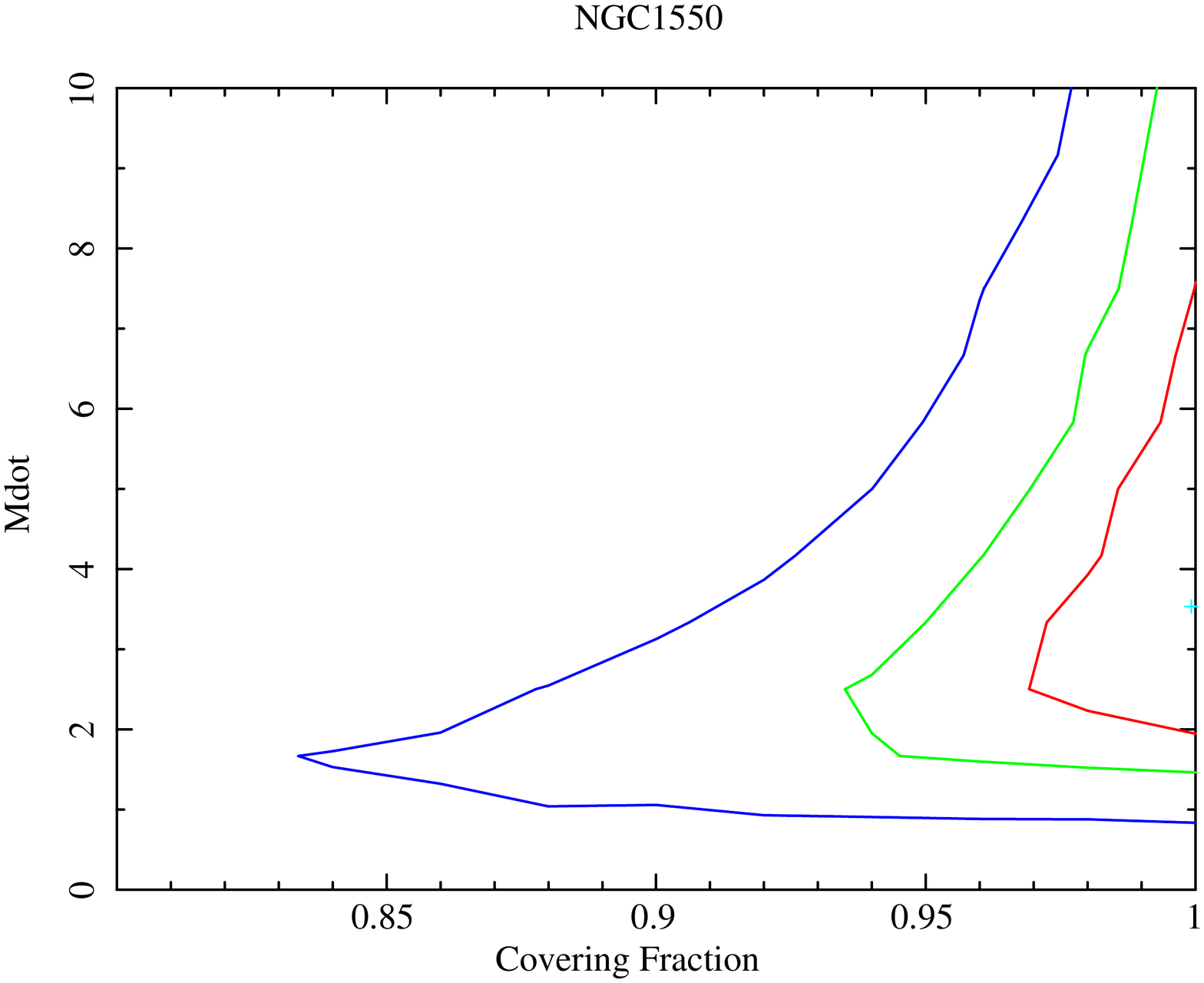}
    \caption{NGC1550,  with details as in Fig B2.}
\end{figure}

\begin{figure}
    \centering    \includegraphics[width=0.48\textwidth]{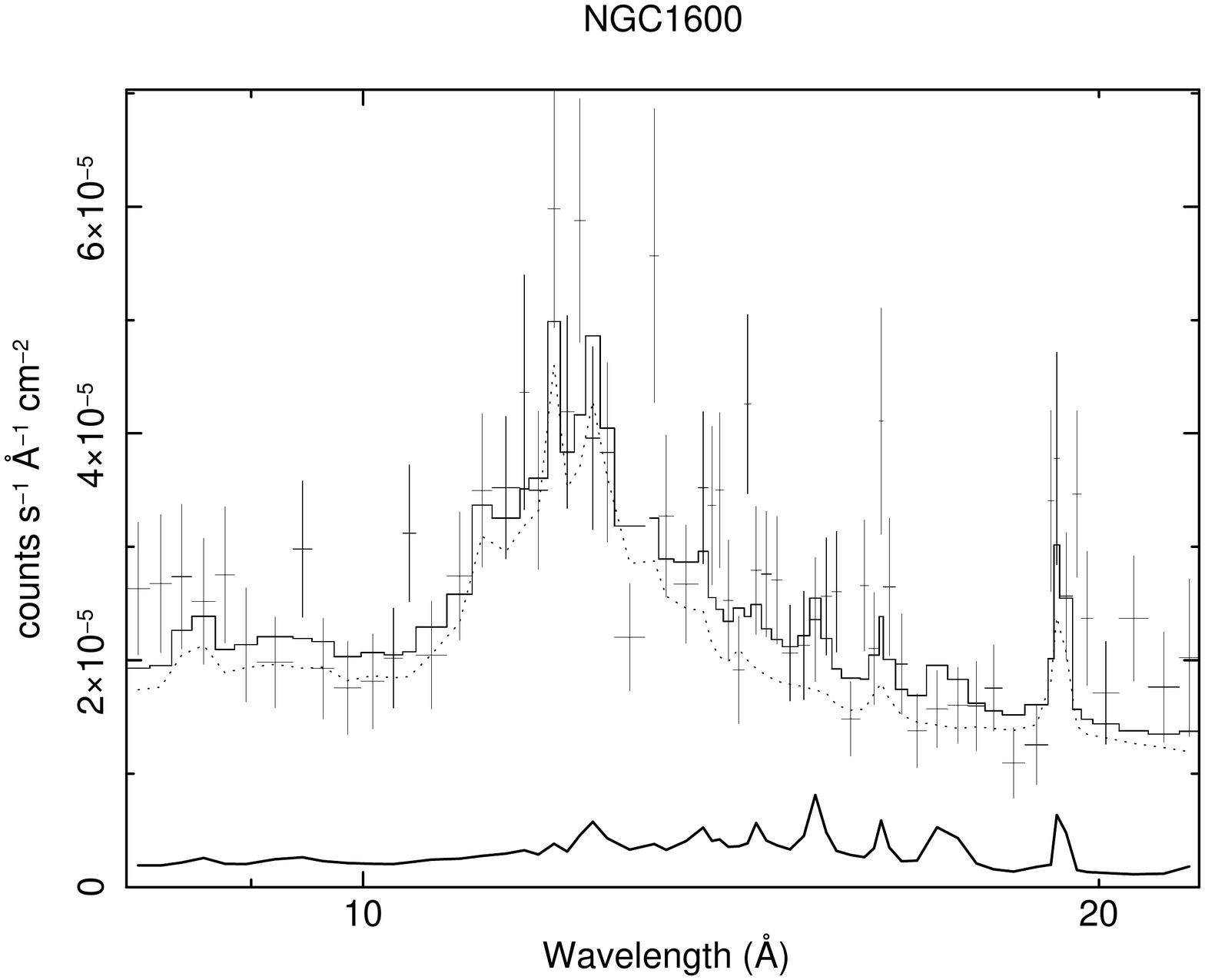}
    \hspace{-0.85cm}\includegraphics[width=0.48\textwidth]
    {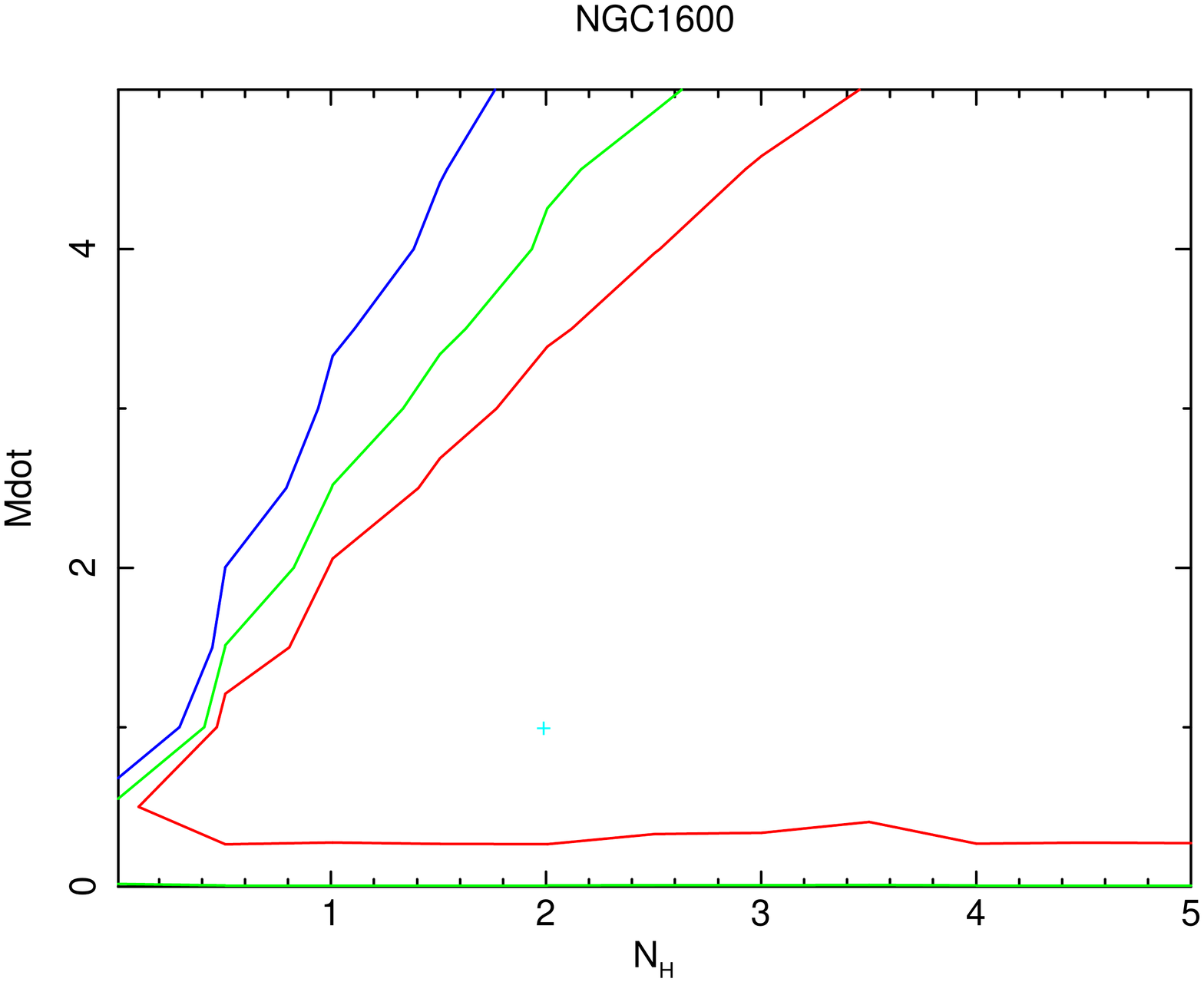}
      \hspace{-0.85cm} \includegraphics[width=0.48\textwidth]{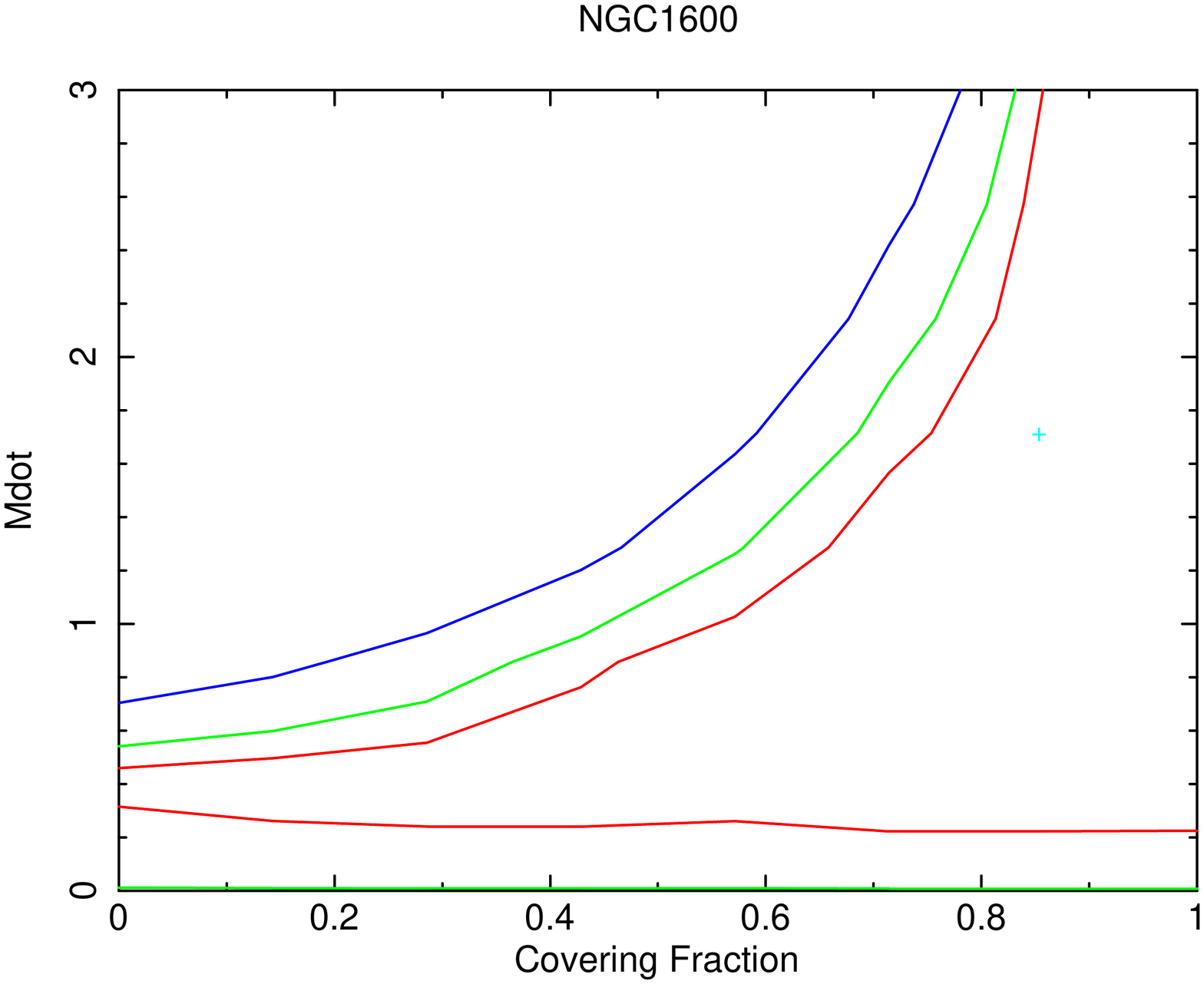}
    \caption{NGC1600,  with details as in Fig B2.}
\end{figure}

\begin{figure}
    \centering    \includegraphics[width=0.48\textwidth]{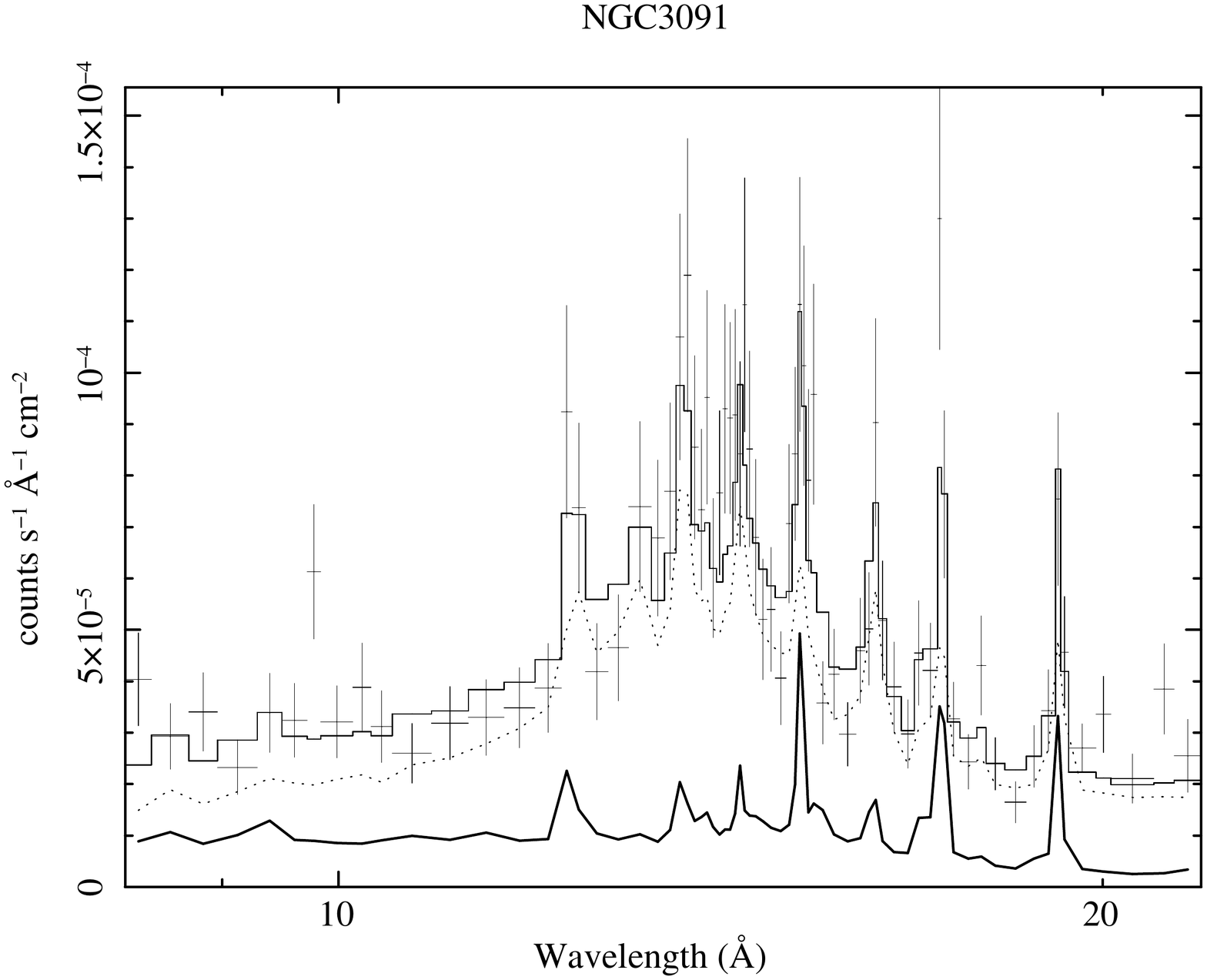}
    \hspace{-0.85cm}\includegraphics[width=0.48\textwidth]
    {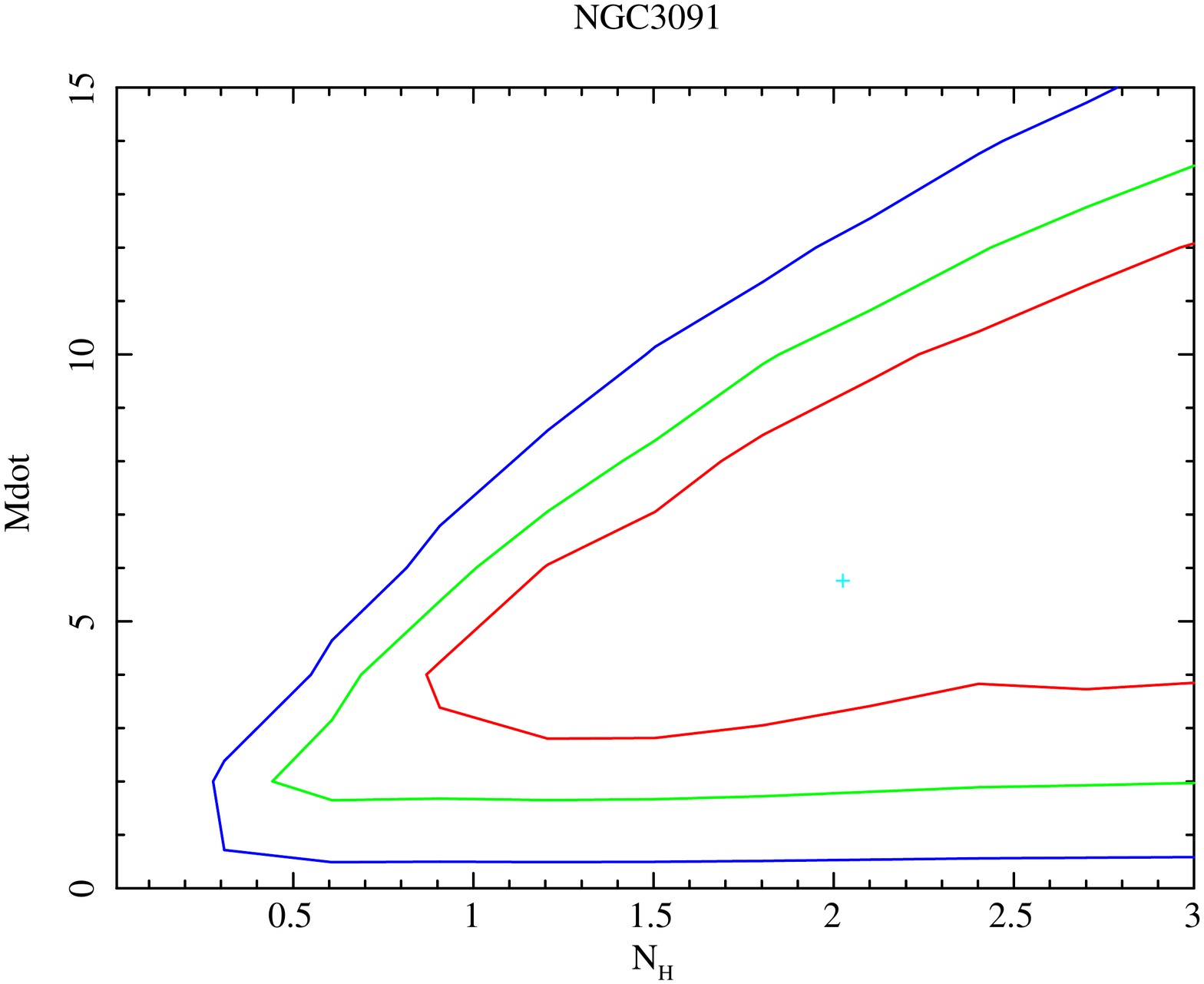}
      \hspace{-0.85cm} \includegraphics[width=0.48\textwidth]{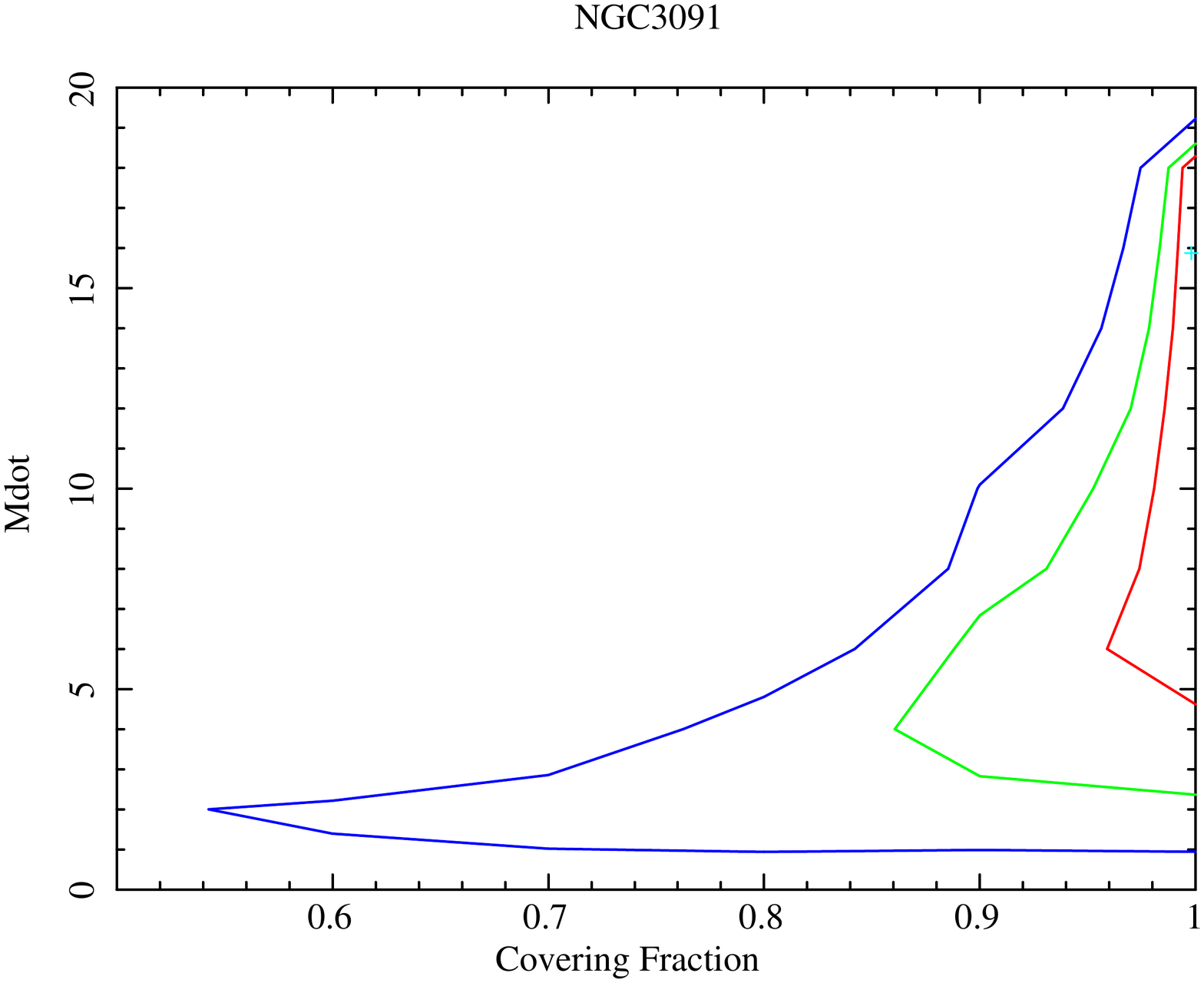}
    \caption{NGC3091,  with details as in Fig B2.}
\end{figure}

\begin{figure}
    \centering    \includegraphics[width=0.48\textwidth]{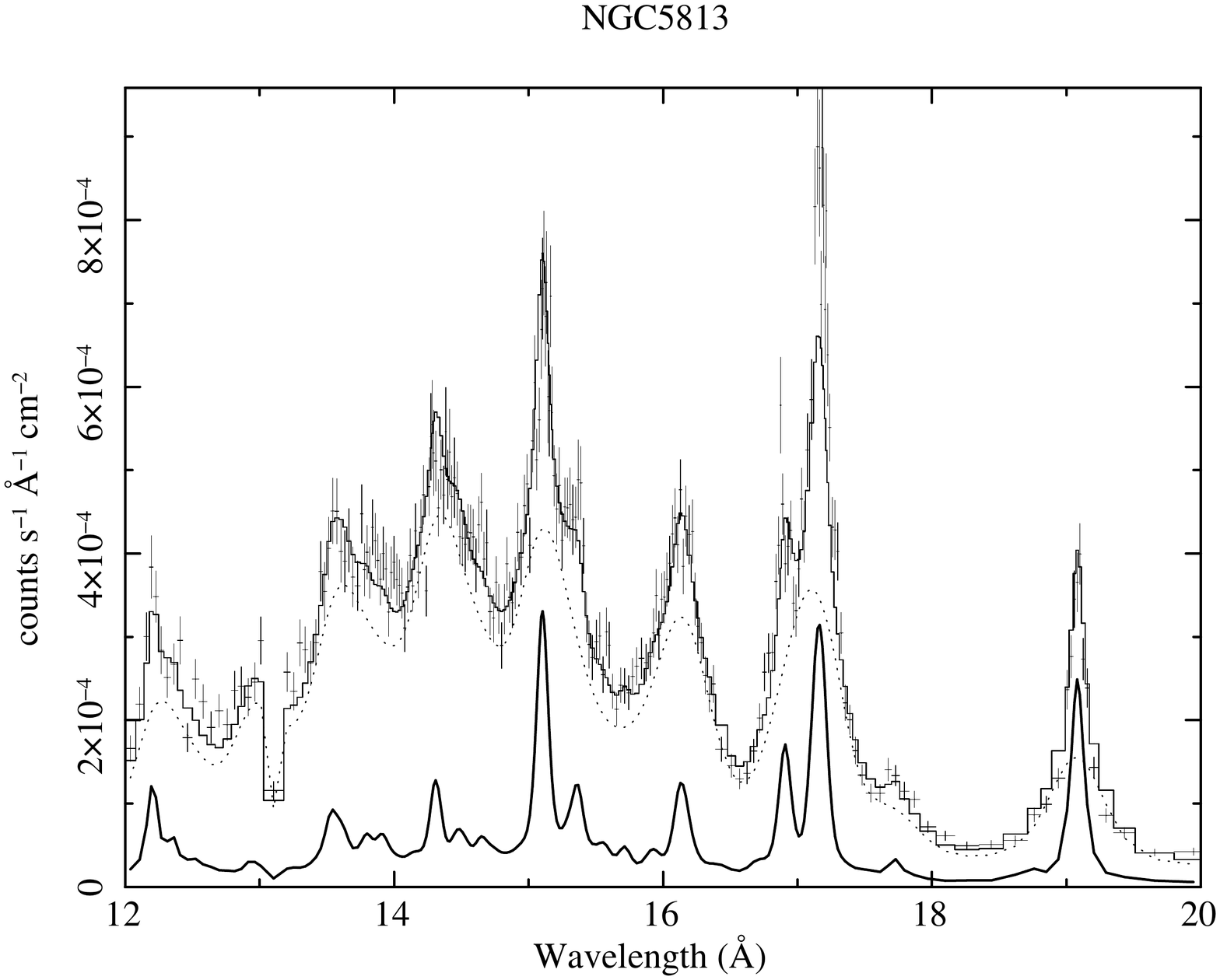}
    \hspace{-0.85cm}\includegraphics[width=0.48\textwidth]
    {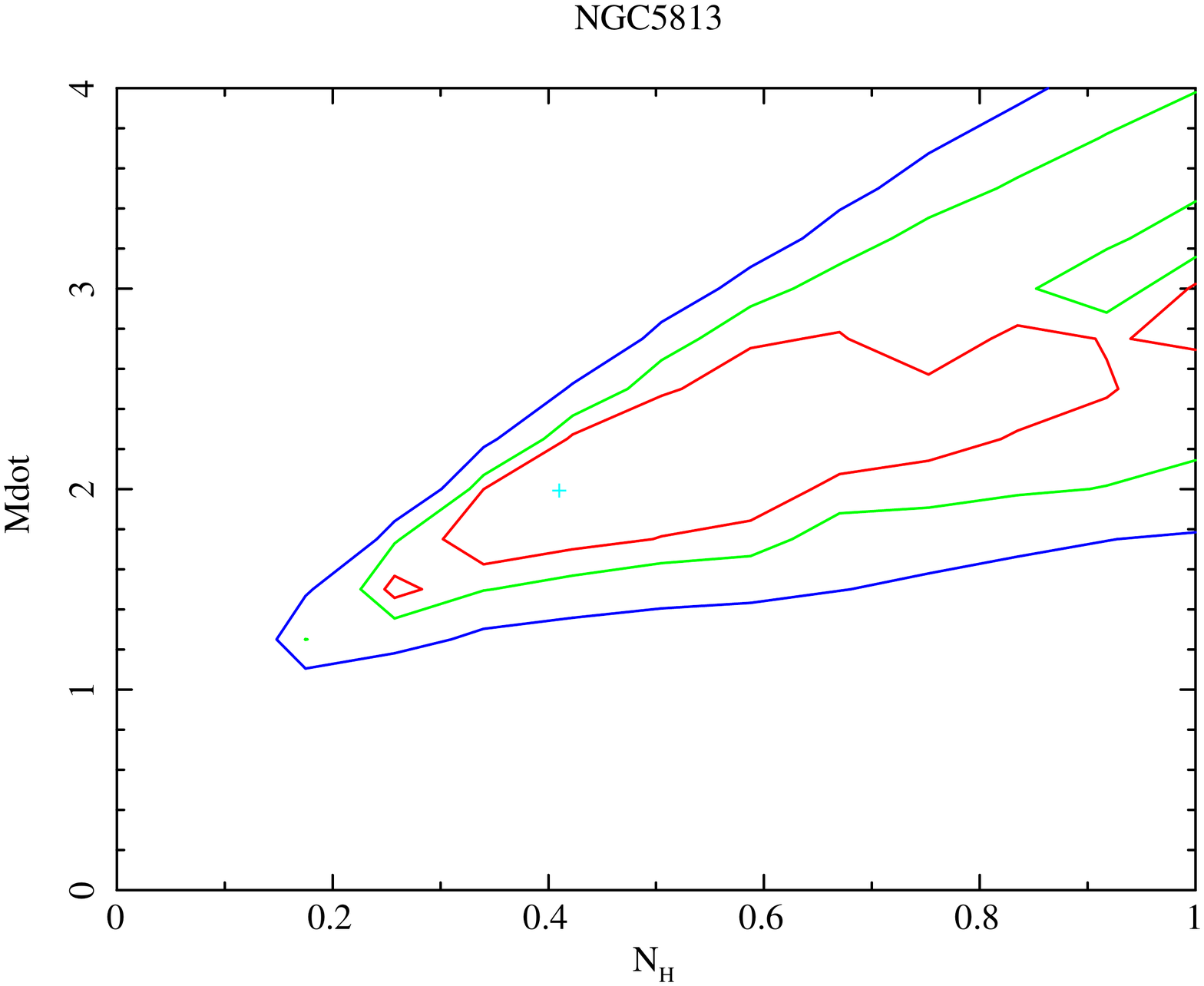}
      \hspace{-0.85cm} \includegraphics[width=0.48\textwidth]{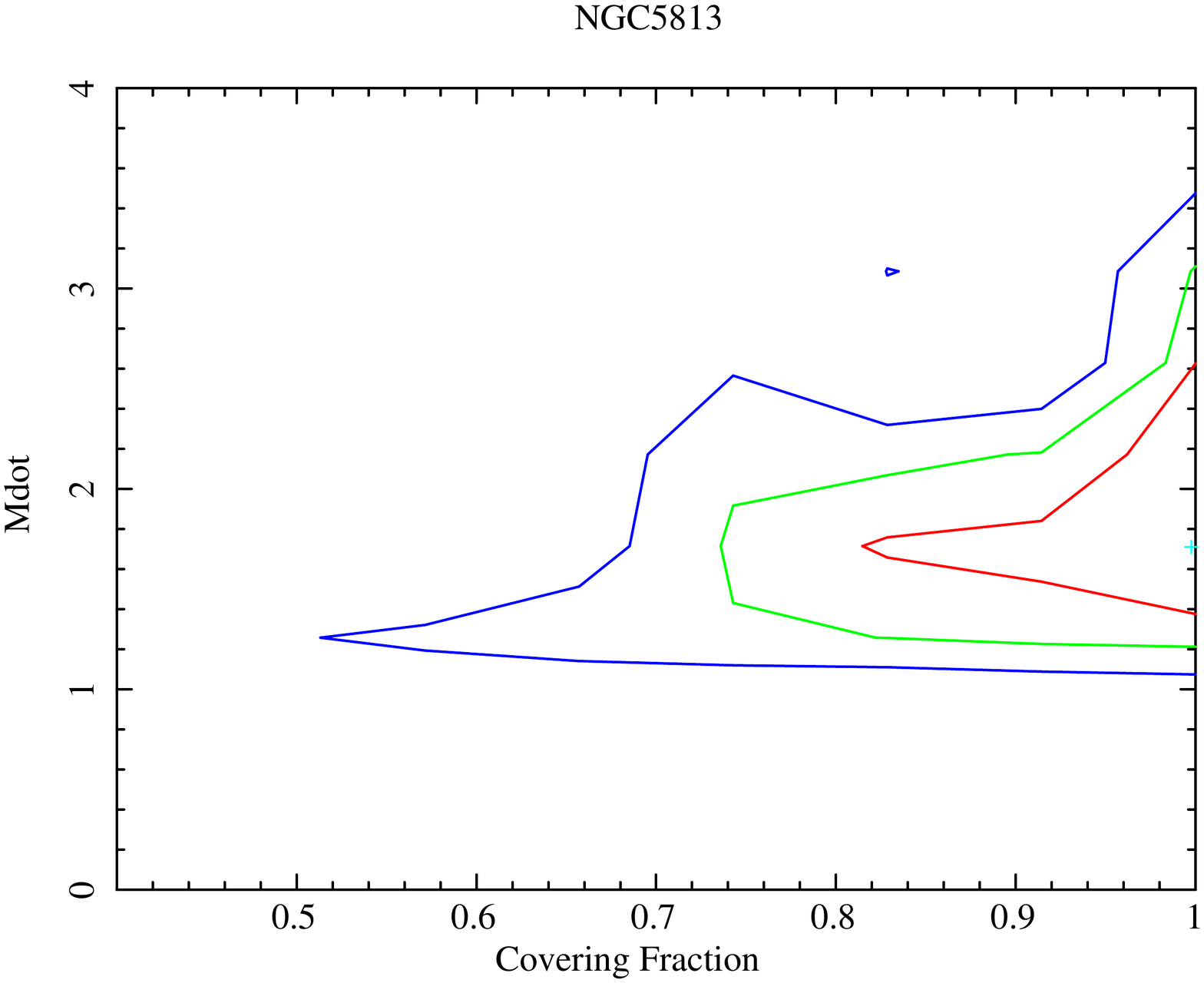}
    \caption{NGC5813,  with details as in Fig B2.}
\end{figure}

\begin{figure}
    \centering    \includegraphics[width=0.48\textwidth]{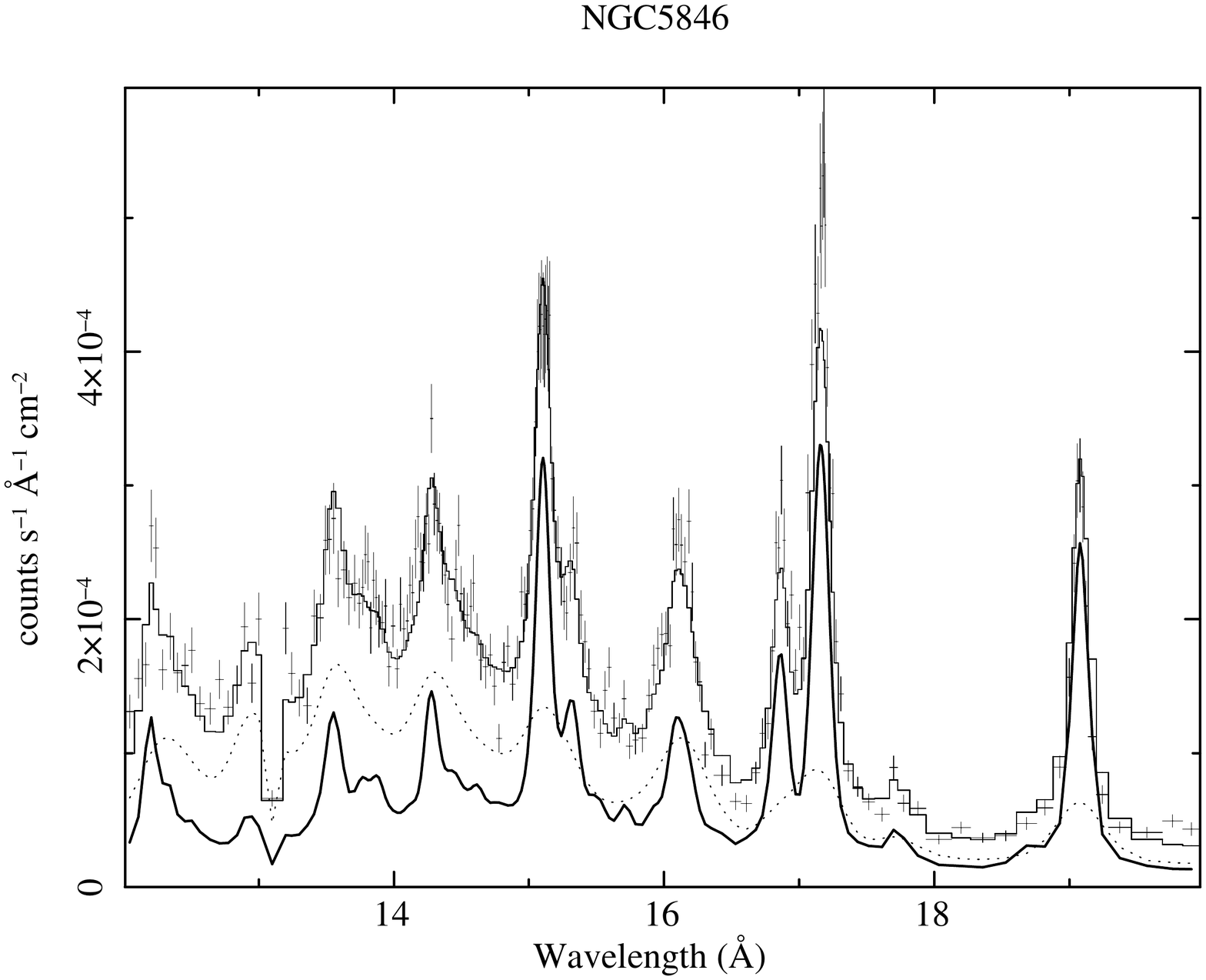}
   \hspace{-0.85cm}\includegraphics[width=0.48\textwidth]
  {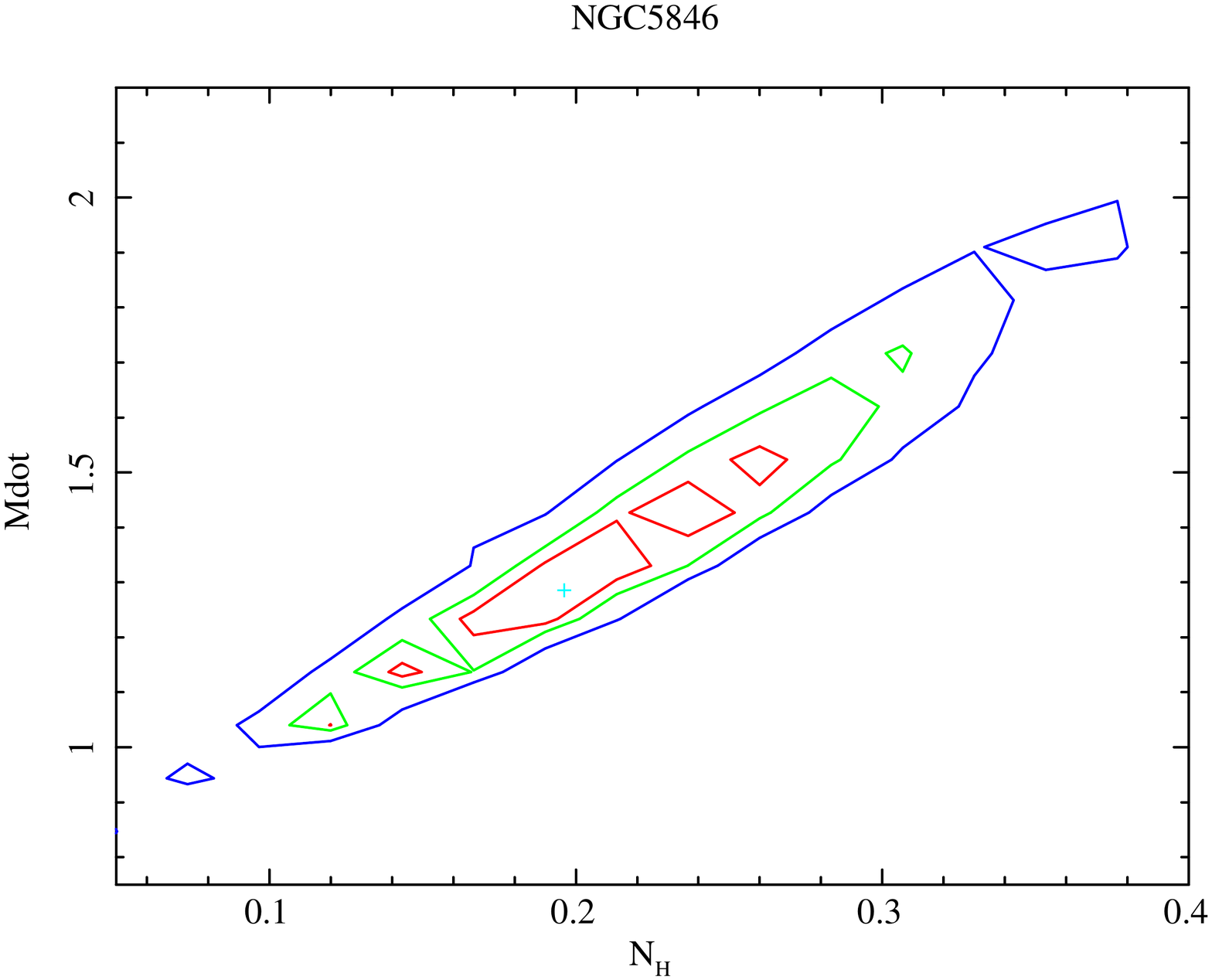}
    \hspace{-0.85cm} \includegraphics[width=0.48\textwidth]{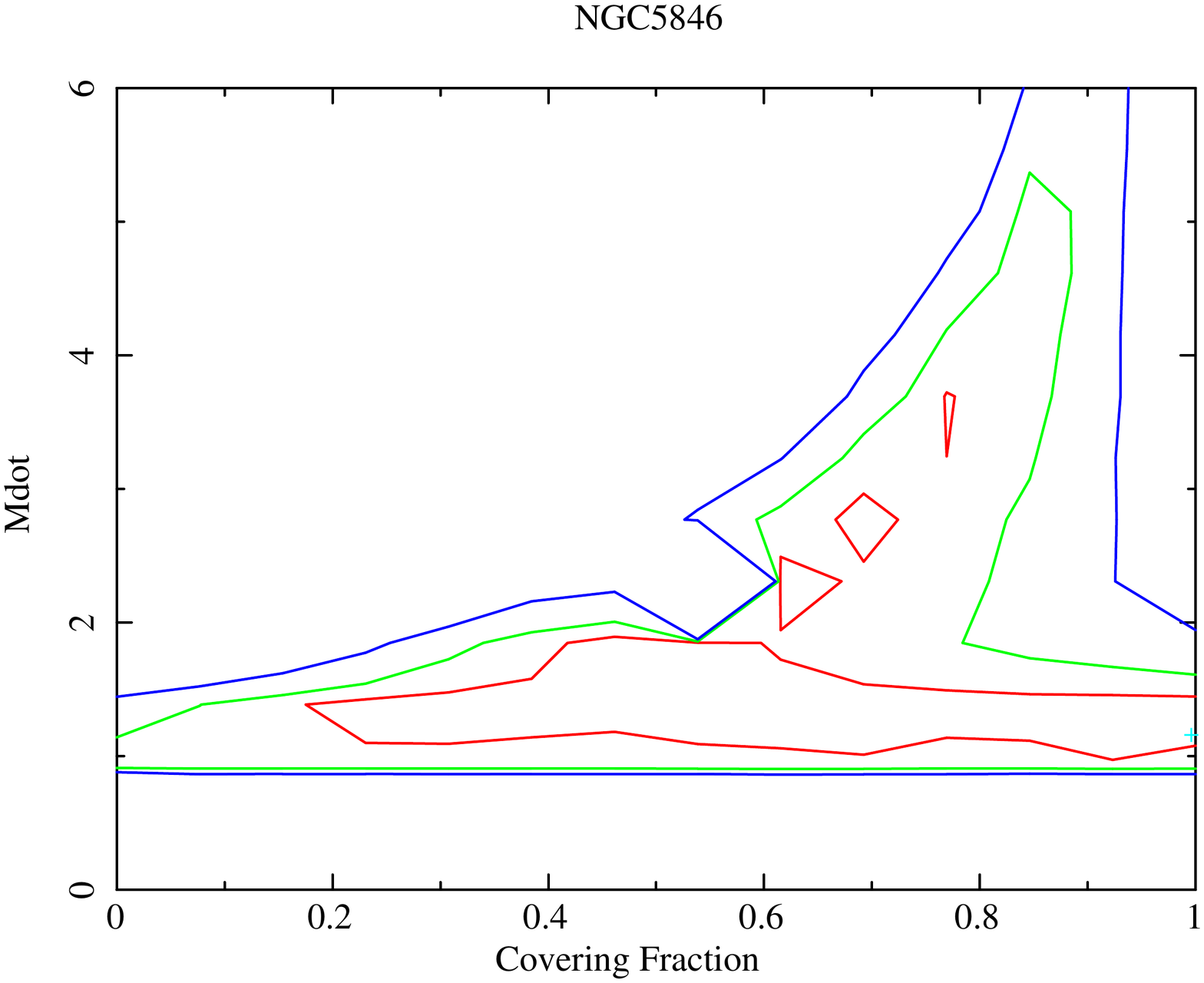}
    \caption{NGC5846,  with details as in Fig B2.}
\end{figure}

\begin{figure}
    \centering    \includegraphics[width=0.48\textwidth]{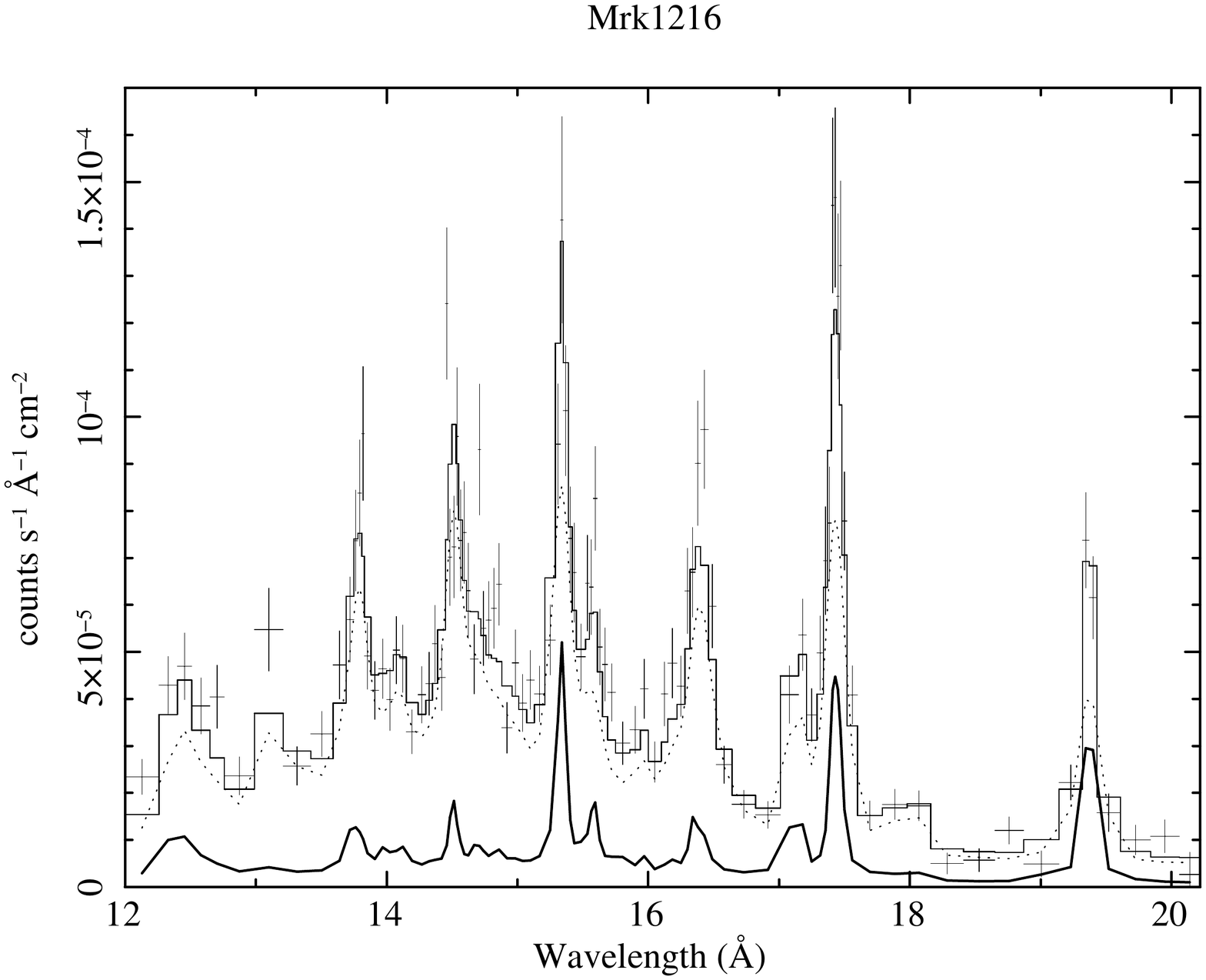}
   \hspace{-0.85cm}\includegraphics[width=0.48\textwidth]
  {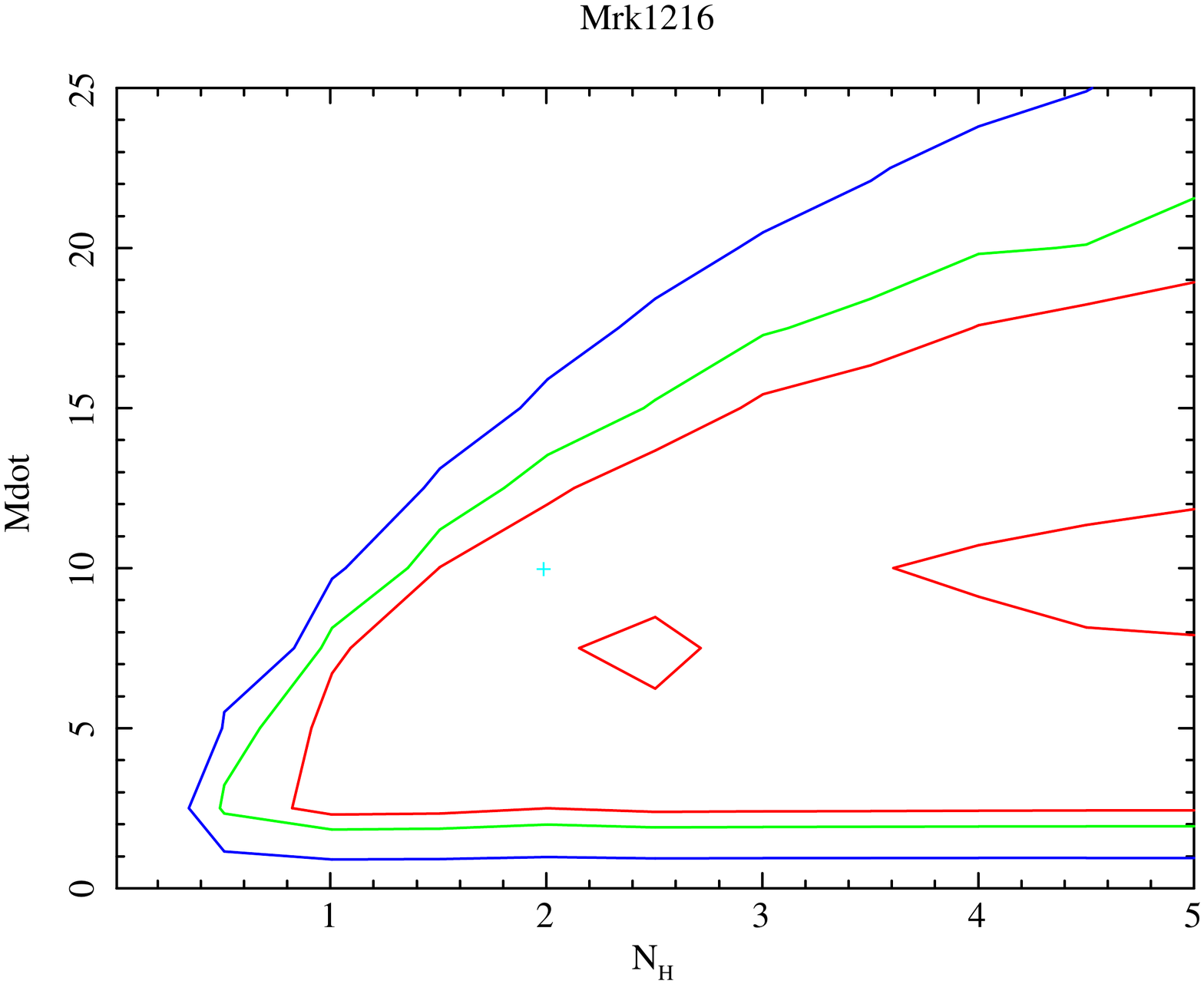}
    \hspace{-0.85cm} \includegraphics[width=0.48\textwidth]{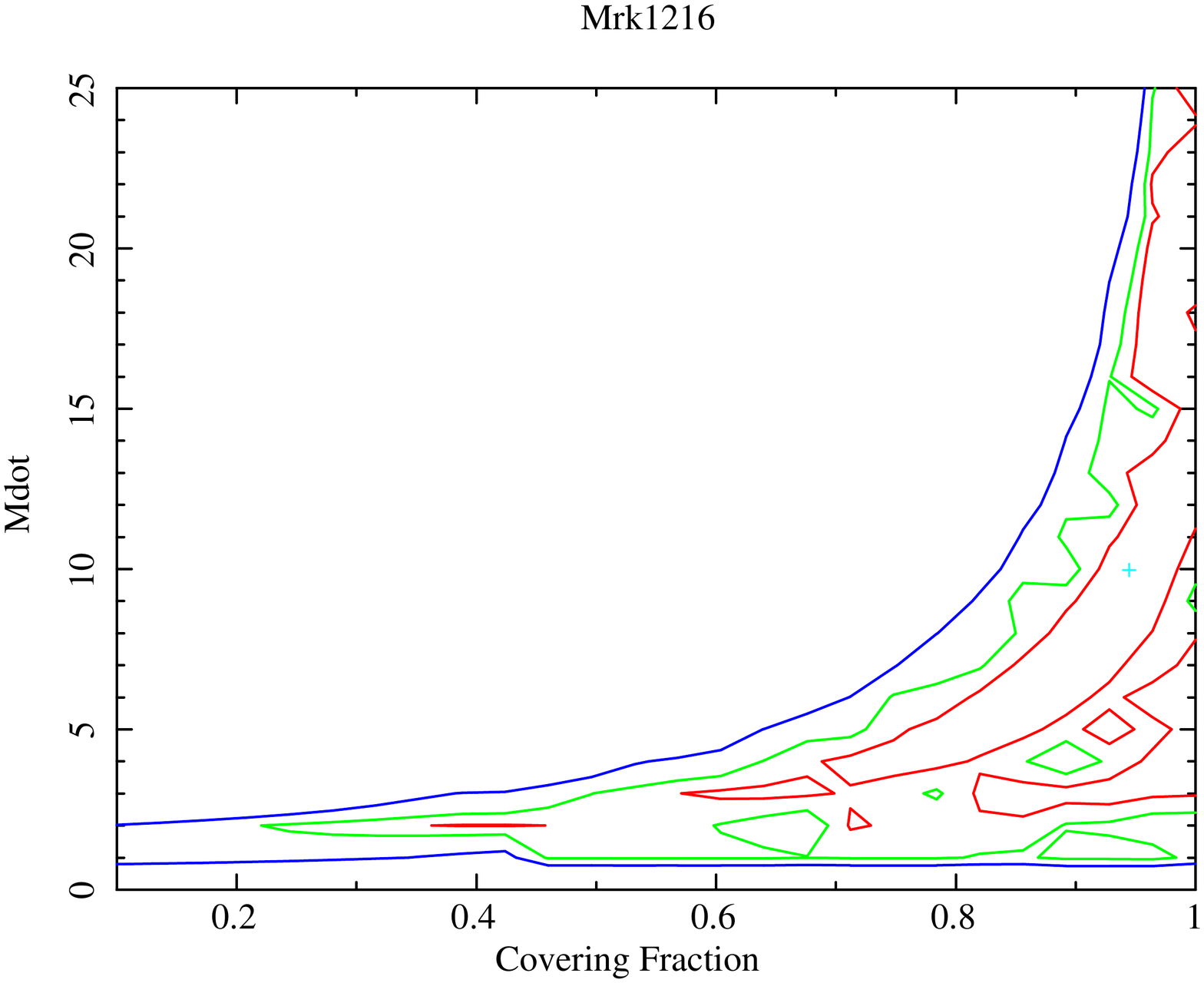}
    \caption{MRK 1216,  with details as in Fig B2.}
\end{figure}

\begin{figure}
    \centering    \includegraphics[width=0.48\textwidth]{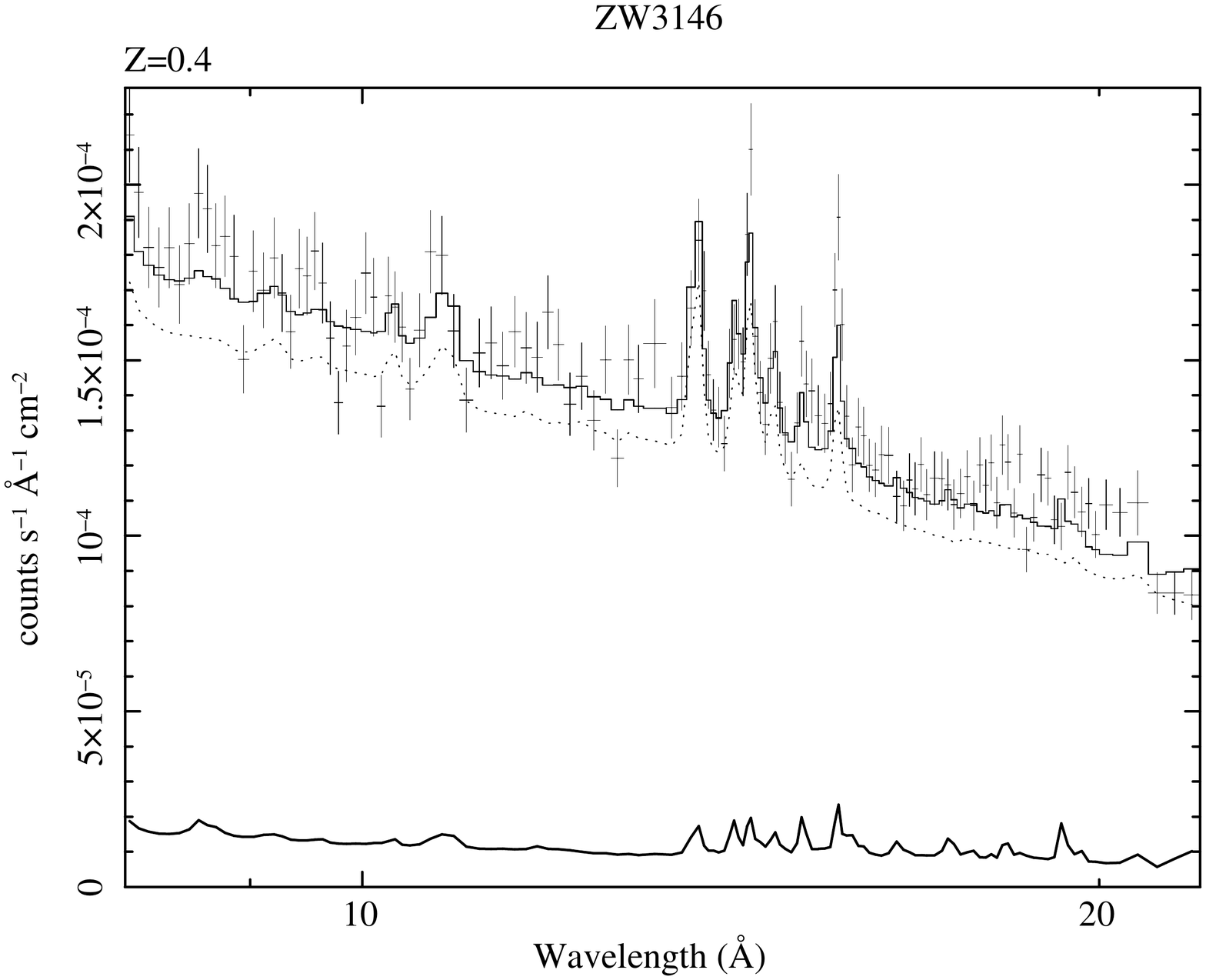}
    \hspace{-0.85cm}\includegraphics[width=0.48\textwidth]
    {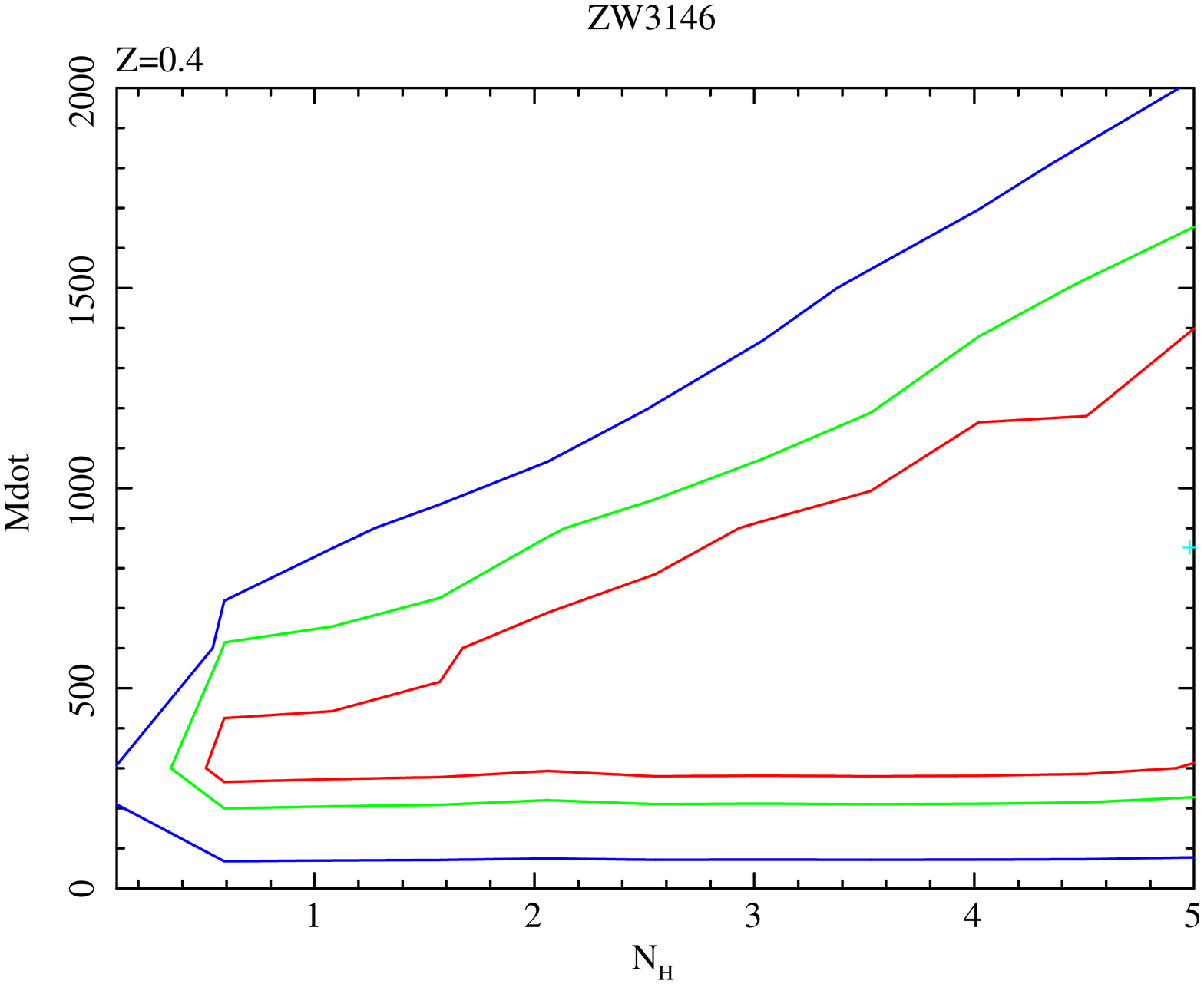}
      \hspace{-0.85cm} \includegraphics[width=0.48\textwidth]{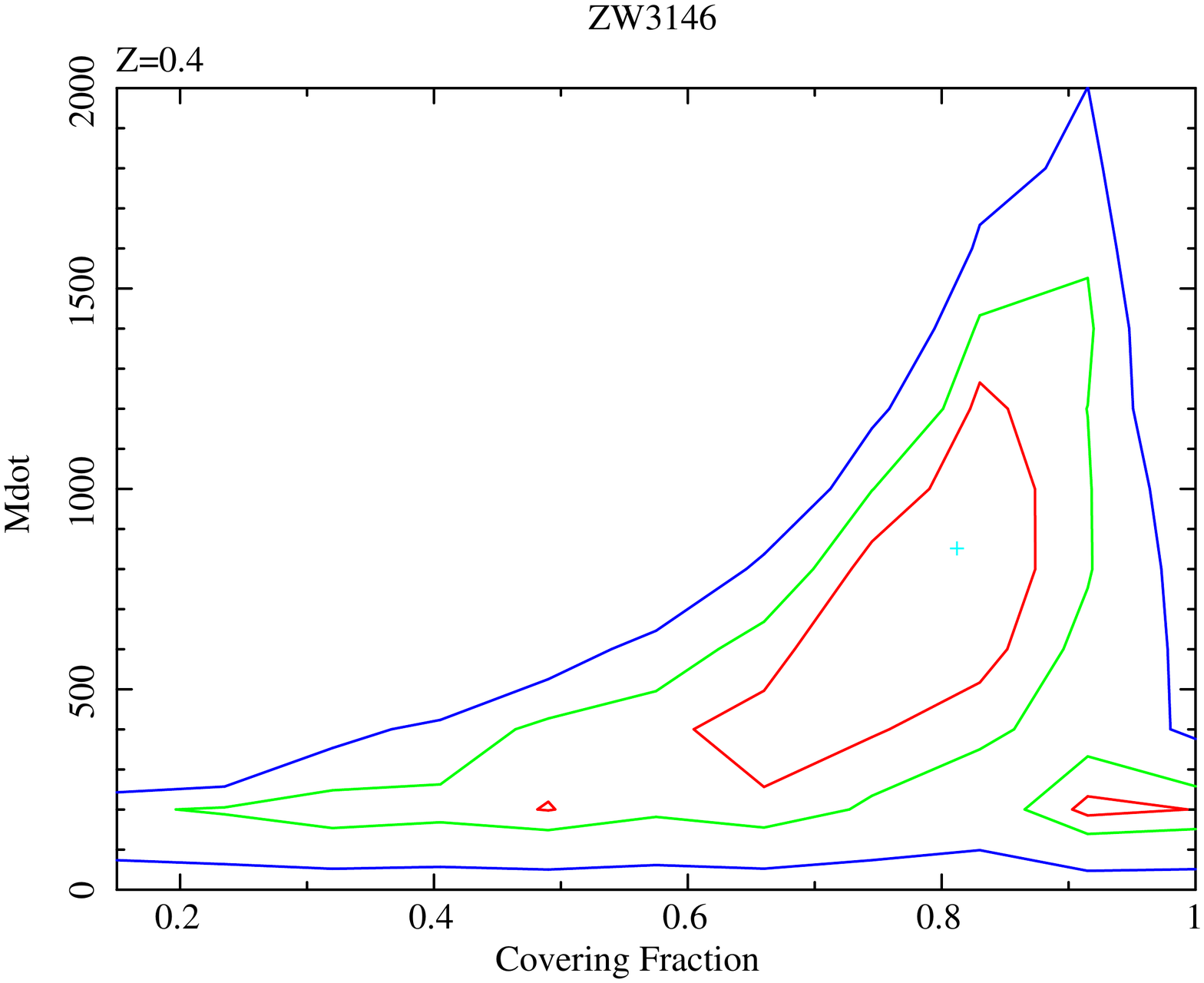}
    \caption{ZW3146,  with details as in Fig B2.}
\end{figure}

~~~~~~~~~~~~~~~~~~~~~~~~~~~~~~~~~~~~~~~~~~~~~~~~~~~~~~~~~~~~~~~~~~~~~~~~~~~~~~~~~~~~~~~~~~~~~~~~~~~~~~~~~~~
\end{document}